\documentclass[10pt]{article}
\pdfoutput=1
\usepackage[top=1cm,right=1cm,left=1cm,bottom=1.25cm]{geometry}
\usepackage{amsmath,amssymb}
\usepackage[utf8x]{inputenc}
\usepackage{textcomp,marvosym}
\usepackage{cite}
\usepackage{nameref,hyperref}
\usepackage[table]{xcolor}
\usepackage{array}
\makeatletter
\renewcommand{\@biblabel}[1]{\quad#1.}
\makeatother
\usepackage{lastpage,fancyhdr,graphicx}
\usepackage{epstopdf}
\usepackage[normalem]{ulem}
\usepackage{fancyhdr}
\usepackage{lastpage}
\usepackage{setspace}
\usepackage{lipsum}
\usepackage{geometry}
\usepackage{amsmath,amsthm,amsfonts}
\usepackage{graphicx}
\usepackage{algorithm}
\usepackage{algorithmic}
\usepackage{subfigure} 
\usepackage{multirow}
\usepackage{multicol}
\usepackage{appendix}
\usepackage{indentfirst}
\usepackage{amssymb}
\usepackage{pifont}
\usepackage{verbatim}
\usepackage{epstopdf}
\usepackage{mathtools}

\definecolor{LightGray}{rgb}{0.80,0.80,0.85}

\newcommand{\Real}{{\mathbb R}}

\newcommand{\row}{{\text{\tiny row}}}
\newcommand{\col}{{\text{\tiny col}}}
\newcommand{\cK}{{\cal K}}
\newcommand{\cJ}{{\cal J}}
\newcommand{\cG}{{\cal G}}
\newcommand{\cH}{{\cal H}}
\newcommand{\imin}{175}
\newcommand{\imax}{350}
\newcommand{\prs}{{\bf PRS}}
\newcommand{\prsorig}{{\bf PRS_{\text{\footnotesize wide}}}}
\newcommand{\prsbicl}{{\bf PRS_{\text{\footnotesize bicl}}}}
\newcommand{\auc}{{\bf AUC}}
\newcommand{\aucorig}{{\bf AUC_{\text{\footnotesize wide}}}}
\newcommand{\aucorigBDI}{{\bf AUC_{\text{\footnotesize wide\textbar BDI}}}}
\newcommand{\aucorigBDII}{{\bf AUC_{\text{\footnotesize wide\textbar BDII}}}}
\newcommand{\aucbicl}{{\bf AUC_{\text{\footnotesize bicl}}}}
\newcommand{\aucbiclBDI}{{\bf AUC_{\text{\footnotesize bicl\textbar BDI}}}}
\newcommand{\aucbiclBDII}{{\bf AUC_{\text{\footnotesize bicl\textbar BDII}}}}
\newcommand{\cRorig}{{\bf {\cal R}^{2}_{\text{\footnotesize wide}}}}
\newcommand{\cRbicl}{{\bf {\cal R}^{2}_{\text{\footnotesize bicl}}}}
\newcommand{\cRorigBDI}{{\bf {\cal R}^{2}_{\text{\footnotesize wide\textbar BDI}}}}
\newcommand{\cRorigBDII}{{\bf {\cal R}^{2}_{\text{\footnotesize wide\textbar BDII}}}}
\newcommand{\cRbiclBDI}{{\bf {\cal R}^{2}_{\text{\footnotesize bicl\textbar BDI}}}}
\newcommand{\cRbiclBDII}{{\bf {\cal R}^{2}_{\text{\footnotesize bicl\textbar BDII}}}}
\newcommand{\NSNP}{{N_{\text{\footnotesize SNP}}}}
\newcommand{\tp}{\tilde{p}}

\graphicspath{{dir_figs}{dir_figs_dist}{dir_figs_stack}{dir_figs_nagelkerke}{dir_lisa_xdropgenri}{dir_prs_collect}{dir_prs_collect/dir_jpg}{dir_Review_20240821}}

\begin{document}
\vspace*{0.2in}

\begin{flushleft}
  {\Large
\textbf\newline{Heterogeneity analysis provides evidence for a genetically homogeneous subtype of bipolar-disorder} 
  }
  \newline
  \\
Caroline C. McGrouther\textsuperscript{1},
Aaditya V. Rangan\textsuperscript{1*},
Arianna Di Florio\textsuperscript{2},
Jeremy A. Elman\textsuperscript{3},
Nicholas J. Schork\textsuperscript{4},
John Kelsoe\textsuperscript{5},
Bipolar Disorder Working Group of the Psychiatric Genomics Consortium\textsuperscript{6}
\\
\bigskip
\textbf{1} Courant Institute of Mathematical Sciences, New York University, New York, NY, United States of America
\\
\textbf{2} School of Medicine, Division of Psychological Medicine and Clinical Neurosciences, Cardiff University, Cardiff, United Kingdom
\\
\textbf{3} Department of Psychiatry, University of California San Diego, San Diego, CA, United States of America
\\
\textbf{4} The Translational Genomics Research Institute, Quantitative Medicine and Systems Biology, Phoenix, AZ, United States of America \\ 
\textbf{5} Department of Psychiatry, University of California San Diego, La Jolla, CA, United States of America \\ 
\textbf{6} see Acknowledgements section for full list \\ 
\bigskip
* avr209@nyu.edu

\end{flushleft}

\pagebreak

\section*{Abstract}
{\bf Background}: Bipolar Disorder (BD) is a complex disease. It is heterogeneous, both at the phenotypic and genetic level, although the extent and impact of this heterogeneity is not fully understood. One way to assess this heterogeneity is to look for patterns in the subphenotype data. Because of the variability in how phenotypic data was collected by the various BD studies over the years, homogenizing this subphenotypic data is a challenging task, and so is replication. An alternative methodology, taken here, is to set aside the intricacies of subphenotype and allow the genetic data itself to determine which subjects define a homogeneous genetic subgroup (termed ‘bicluster’ below).

{\bf Results}: In this paper, we leverage recent advances in heterogeneity analysis to look for genetically-driven subgroups (i.e., biclusters) within the broad phenotype of Bipolar Disorder. We first apply this covariate-corrected biclustering algorithm to a cohort of $2524$ BD cases and $4106$ controls from the Bipolar Disease Research Network (BDRN) within the Psychiatric Genomics Consortium (PGC). We find evidence of genetic heterogeneity delineating a statistically significant bicluster comprising a subset of BD cases which exhibits a disease-specific pattern of differential-expression across a subset of SNPs. This disease-specific genetic pattern (i.e., ‘genetic subgroup’) replicates across the remaining data-sets collected by the PGC containing $5781$/$8289$, $3581$/$7591$, and $6825$/$9752$ cases/controls, respectively. This genetic subgroup (discovered without using any BD subtype information) was more prevalent in Bipolar type-I than in Bipolar type-II.

{\bf Conclusions}: Our methodology has successfully identified a replicable homogeneous genetic subgroup of bipolar disorder. This subgroup may represent a collection of correlated genetic risk-factors for BDI.  By investigating the subgroup’s bicluster-informed polygenic-risk-scoring (PRS), we find that the disease-specific pattern highlighted by the bicluster can be leveraged to eliminate noise from our GWAS analyses and improve risk prediction. This improvement is particularly notable when using only a relatively small subset of the available SNPs, implying improved SNP replication. Though our primary focus is only the analysis of disease-related signal, we also identify replicable control-related heterogeneity.


\section*{Background}

{\bf Overview:\ }
Bipolar disorder (BD) is a brain disorder characterized by shifts in mood, energy and attention/focus \cite{AmericanPsychiatricAssociation_2013}.
BD affects roughly 50 million people across the world, with a mean age of onset of 20 years and an estimated lifetime prevalence of $\sim 1\%$ \cite{Angst_1998,Merikangas_2007,Merikangas_2011,GBD2016DiseaseandInjuryIncidenceandPrevalenceCollaborators_2017}.
BD is also highly heritable \cite{Bienvenu_2011}, with heritability estimates of $40\%$ or higher \cite{Craddock_2013,Merikangas_2002,Smoller_2003,Song_2015,Kendler_2020} and evidence of increased risk when family-members exhibit other psychiatric disorders \cite{Craddock_2013,Song_2015,Kendler_2020}.

There is growing consensus that BD is heterogeneous, both at the phenotypic and genetic level \cite{WorldHealthOrganization_1992,Grande_2016,Charney_2017,Allardyce_2018,Markota_2018,BipolarDisorderandSchizophreniaWorkingGroupofthePsychiatricGenomicsConsortium_2018,Lewis_2019,Charney_2019,Stahl_2019,Coombes_2020,Carvalho_2020,OConnel_2021}.
For example, diagnostic systems usually consider at least two subtypes of bipolar disorder: bipolar I and bipolar II. 
The diagnostic criteria for bipolar I require the presence of at least one manic episode, while those for bipolar II require at least one hypomanic and one major depressive episode \cite{AmericanPsychiatricAssociation_2013}.
Response to medication (such as lithium) is highly heterogeneous across patients, and genetic predictors of drug-response have been difficult to clearly determine and replicate \cite{InternationalConsortiumonLithiumGenetics[ConLi+Gen]_2018,Amare_2020,Nunes_2020,Gordovez_2020,Ho_2020}.

The high degree of heterogeneity for BD at the clinical and phenotypic level
may make it more difficult to identify genetic risk-factors for BD.
To briefly summarize: while the overall heritability of BD is estimated at $\sim 40\%$, the overall single-nucleotide-polymorphism (SNP) heritability is only $\sim 18.6\%$ \cite{Mullins_2020}, which is moderate when compared to many other psychiatric and neurological disorders \cite{Gatz_2006,Bienvenu_2011,Browne_2014,Zilhao_2017,Walters_2018,CrossDisorderGroupofthePsychiatricGenomicsConsortium_2019,Demontis_2019a,Faraone_2019,Jansen_2019,Purves_2020}.
Recent genome-wide association studies (GWASs) have been used to identify several (i.e. $\sim 100$) independent loci associated with BDI and BDII, with the overall variance explained by SNPs reaching $\sim 15-18\%$ \cite{Mullins_2020}.
However, many of the loci that seem promising in one cohort fail to replicate in other cohorts \cite{Djurovic_2010,Smith_2011,OConnel_2021}.
Studies attempting to uncover gene-environment interactions in BD have also encountered challenges finding replicable signals \cite{Winham_2014,Aas_2014,Oliveira_2016,Hosang_2017,Aas_2020}.

Rather than focusing on small sets of loci, one can also consider collections of SNPs which individually may not be of genome-wide significance.
Along this vein, Polygenic-risk-scores (PRSs), which are usually weighted sums of genetic variants, have been used to summarize the genome-wide risk for BD \cite{InternationalSchizophreniaConsortium_2009}.
These PRSs may provide an estimate of overall risk and/or severity: those individuals with PRSs in the top $90\%$ were $3.62$ times more likely to be a case than those with average PRSs.
These PRSs also contain information regarding multiple phenotypic traits, including the risk of other psychiatric disorders, psychopathology, educational attainment and more \cite{Wilcox_2017,Mistry_2018,Reginsson_2018,Mistry_2019a,Mistry_2019b,Musliner_2019,Musliner_2020,Mullins_2019,GrigoroiuSerbanescu_2020}.
Depite these successes, to the best of our knowledge, no individual PRS has yet been able to explain a large fraction of the variation between the main bipolar subtypes.

The high degree of heterogeneity within BD poses a challenge to understanding its etiology and developing new interventions.
Ultimately, a comprehensive depiction of the landscape of BD will involve clear descriptions of the heterogeneity at the phenotypic level, as well as at the genetic level.

To date, the main research efforts aimed at understanding the genetic heterogeneity underlying BD have focused on (i) increasing the power of BD meta-GWAS, (ii) running subphenotypic-specific meta-GWAS, and (iii) performing pathway-specific analyses \cite{Chen_2013,Joslyn_2016,Eser_2018}.
These research efforts are non-trivial and in some cases require insights we do not yet have.
Generally speaking, recruiting, assessing, and genotyping new subjects is expensive; there is often a trade-off between the quantity of subjects that can be recruited and the `quality' or accuracy with which their data is processed.
For example, one promising resource for genotyped data is 23andMe, but many of the data-sets available through this resource rely on self-reported diagnoses \cite{URL_23andMe}.
Consequently, any synchronization effort involves the integration and harmonization of data collected using different phenotypic instruments or genotyping methods and may inadvertently introduce non-disease-related signal. 
Furthermore, in many cases, the relevant subphenotypic information was not collected at all, forcing interested researchers to contact prior participants or lose those data points entirely. 
Finally, even when promising results are obtained, it is not always easy to find an appropriate replication sample \cite{McGrouther_2012}.
Since we do not yet know which trait or combination of subphenotypic traits (if any) is responsible for BD genetic heterogeneity, it is not always clear how best to proceed.

{\bf Contribution:}
Ultimately, we seek to investigate the genetic heterogeneity of BD by using an approach which does not require the user to provide pathways or subphenotypes. 
As described below, we introduce a methodology which first uses the genotyped data to identify a genetic subgroup within BD, and then uses that genetic subgroup for a downstream analyses (in this case risk prediction).
To briefly summarize: we use a covariate-corrected biclustering algorithm to search for statistically significant biclusters comprising subsets of BD cases which exhibit disease-specific patterns of differential-expression across subsets of SNPs.
In this study we find one statistically-significant disease-specific structure, which is limited to only a fraction of the case-subjects.
These case-subjects collectively exhibit a shared pattern of differential-expression -- i.e., a form of genetic homogeneity -- which is not shared by the other BD-cases nor by the control-subjects; we refer to this bicluster as a `genetic subgroup'.
We then demonstrate that this genetic subgroup is useful for risk-prediction.

In more detail, our analysis begins by collecting data within which to search for genetic subgroups of BD.
As members of the Psychiatric Genomics Consortium (PGC), we had access to the raw genotypes of $\sim 18K$ BD cases and $\sim 30K$ controls.
This data was generated by 27 studies and genotyped on a variety of platforms (OMEX, Affymetrix, Illumina).
When the PGC analyzed this data \cite{Stahl_2019}, they synchronized the data using imputation. 
We were not certain how imputation might impact the potentially subtle relationships between BD cases, and therefore decided to limit our analysis to the available raw genotyped data \cite{McGrouther_2012}.
This choice to limit ourselves to raw genotyped data placed constraints on our choices for the training and testing data sets, as the various genotyping platforms types emphasize different SNP sets (see Fig \ref{fig:arms}). 

In order to minimize batch-effects and reduce the chances of spurious false-positives, we chose to initially focus our primary analysis on a relatively large curated study from the Bipolar Disorder Research Network (BDRN) comprising raw genotyped data collected across $2524$ BD cases and $4106$ controls (OMEX platform) \cite{LRKetal_2020}.
We use this BDRN study as our training-arm, and set aside the remaining independent data for subsequent replication analyses (i.e., our replication-arms).
We grouped all the BD cases in our training-arm together and searched within the training-arm for any subsets of subjects which exhibited a distinct genetic signature (i.e., differential expression) across a subset of SNPs.
Any such subset of subjects along with the associated subset of differentially-expressed SNPs is referred to as a `bicluster', or a `genetic subgroup'. 

As described in \cite{Xie_2019,Dahl_2020}, many commonly used biclustering approaches suffer from two methodological issues.
First, a bicluster that is found within the case-population may not be disease-related, as a similar signal may be found within the control-population (e.g., a bicluster representing non-disease-specific heterogeneity).
Second, many biclustering algorithms proceed under the assumption that biclusters exist, often identifying `false-positive' structures that are not statistically-significant.

To address these issues we searched for biclusters using the `half-loop' algorithm of \cite{Rangan_2012,Rangan_2018}.
As described in \cite{Rangan_2018}, this algorithm ensures that the pattern of differential-expression within the bicluster is {\em not} similarly present within the control-population, reducing the likelihood that we highlight structures unrelated to disease status.
Second, the half-loop algorithm uses a permutation-test to estimate the p-value of each bicluster found, allowing us to test against the null hypothesis that no bicluster exists.
Finally, the half-loop algorithm also allows us to correct for other covariates, such as proxies for genetic-ancestry (see Methods).
While our approach is much simpler than some of the more recent machine-learning approaches, our biclusters are directly associated with subject- and SNP-subsets, which can be directly interpreted and assessed for homogeneity and/or used in downstream analyses. 

Using the relatively conservative half-loop method mentioned above, we found strong evidence for genetic heterogeneity.
We discovered one bicluster which is statistically significant and which replicates in all three other data-sets.
This primary bicluster was enriched for (but not completely driven by) BDI over BDII.
After removing this bicluster we saw further evidence of residual heterogeneity, but our training data-set was not sufficiently powered to clearly identify a secondary bicluster.

We then assessed the role of our bicluster in risk-prediction.
We found that the subset of case-subjects highlighted by the bicluster can be used to improve the performance of a PRS.
This advantage was more pronounced when (i) the SNPs included in the PRS were limited to those of high estimated significance, and (ii) the case-population was limited to those diagnosed with BDI.
These observations suggest that focusing on genetically identifiable subgroups of BD-subjects might improve overall risk-prediction and enhance replication across the top SNPs.

Finally, we also ran a simple gene-set over-representation analysis, revealing that the genetic subgroup identified above (i.e., the bicluster) is significantly enriched for many pathways associated with neuronal development and maintenance.

In summary, we find strong evidence for the genetic-heterogeneity of BD in the form of a bicluster.
Notably, BD subphenotype information was not required to identify this signature, nor were rare-variants (i.e., we relied only on common SNPs with maf greater than $25\%$ within the training-arm).
The signature of this bicluster has the potential to refine downstream analyses (e.g., improving genome-wide risk-prediction), and the associated gene-enrichment suggests an association with certain mechanisms of neuronal development. 


\section*{Methods}

In this section we describe several aspects of our methodology.
An even more detailed description, including an outline of the steps involved and the considerations we made along the way, is available in S1 Text.

\subsection*{Data}
We make use of data from $27$ of the cohorts described in \cite{Stahl_2019}.
These cohorts have been curated as described in \cite{Stahl_2019} and its supplementary information, and include de-identified subjects from several countries in Europe, North America and Australia, totaling over $18000$ cases and $29000$ controls of European descent.
Case-subjects were required to meet international consensus criteria (DSM-IV, ICD-9, or ICD-10) for a lifetime diagnosis of BD established using structured diagnostic instruments from assessments by trained interviewers, clinician-administered checklists, or medical record review.
Control-subjects in most samples were screened for the absence of lifetime psychiatric disorders, as indicated.
For each of the $27$ cohorts, we had access to both the raw genotypes and the imputed data generated by Stahl et al. using the 1000 Genomes (1KG) European reference-panel (see \cite{Stahl_2019}).


Due to the details of our heterogeneity analysis (described further below), we make three additional choices.
First, for our primary analysis we use only the raw genotyped data within each cohort, but not the imputed data.
This is because we want to avoid any concerns of spurious correlations that might arise from imputation \cite{McGrouther_2012}.
Second, when running our biclustering algorithm we do not explicitly correct for linkage-disequilibrium (LD) between genotyped SNPs at the level of the data-set itself (e.g., by eliminating SNPs in strong LD with other SNPs).
Instead, we implicitly correct for LD within our biclustering algorithm by contrasting cases against controls.
Third, it is typically quite difficult to reliably detect signal associated with rare variants (i.e., SNPs with a low minor-allele-frequency, a.k.a. `maf'), especially when the power of the data-set is low.
This difficulty is compounded when searching for heterogeneity, as the effective sample-size (e.g., the number of subjects in a bicluster) is further reduced -- often only a fraction of the total subject-population \cite{Rangan_2018}.
Thus, in order to avoid spurious results associated with rare-variants, we limit our analysis to common variants (i.e., SNPs with maf greater than $25\%$). This high maf-threshold has the added benefit that the signals that we do find are described in terms of common variants, which will hopefully be easier to access in future studies. 

\begin{figure}
  \centering
  \includegraphics[trim=350 250 300 250,clip,width=5.5in]{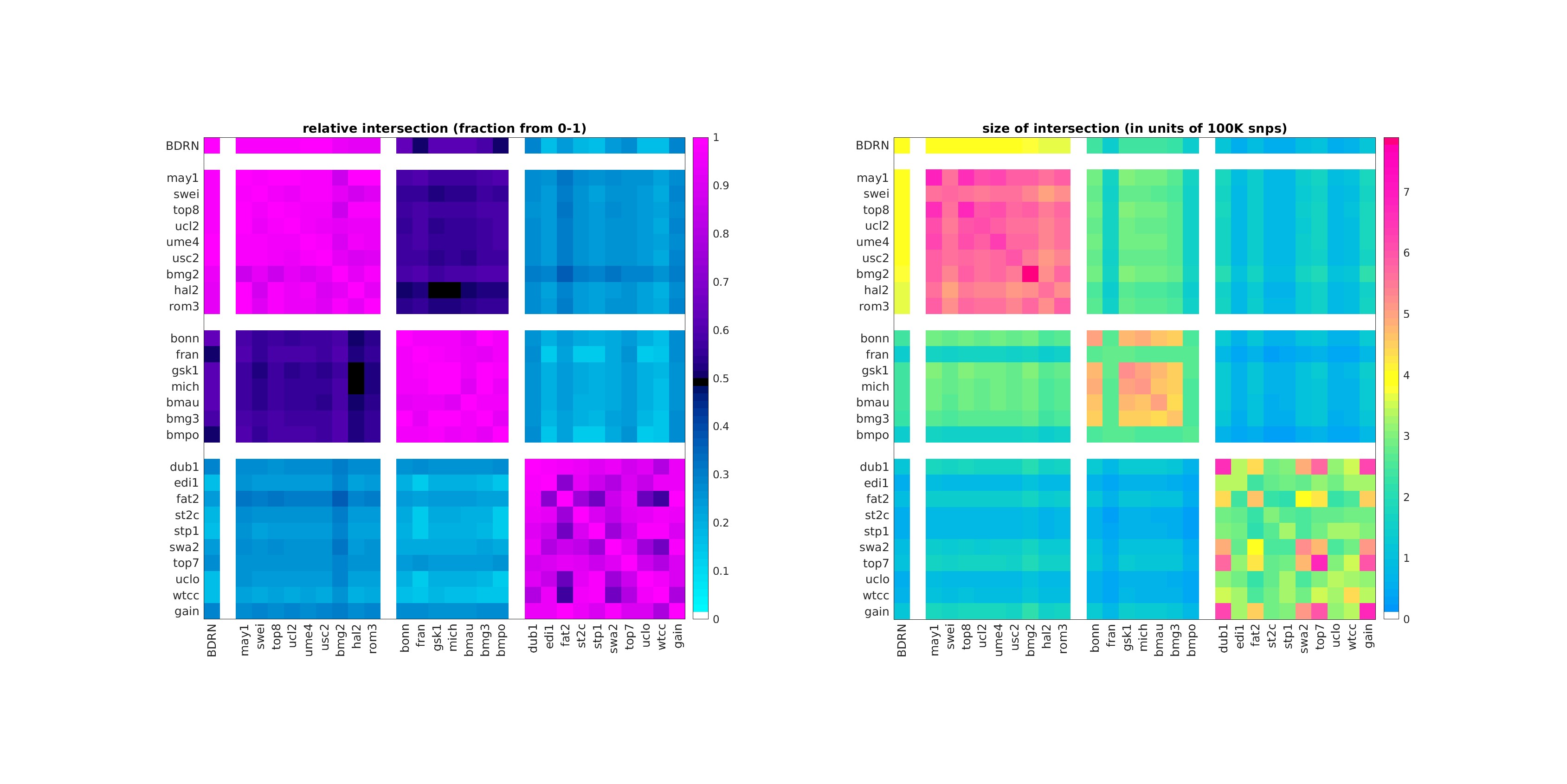}
  \caption{
    In this figure we illustrate the absolute (right) and relative (left) snp overlap between the studies available to us.
    The relative-overlap is calculated using the Szymkiewicz–Simpson coefficient (i.e., the overlap-coefficient between sets $X$ and $Y$ is $|X\cap Y|/\min(|X|,|Y|)$).
    Guided by the relative-overlap and genotyping platform used, we divided the studies into four arms (shown along the coordinate axes).
    The first arm contains only the single `BDRN' data-set, which we use as a training/discovery set to search for heterogeneity (see Methods).
    We reserve the remaining studies (organized into three arms) for replication.
    Note that the training-set overlaps strongly with arm-2, and less strongly with arm-3 and arm-4.
    The magnitude of this overlap will constrain how faithfully any patterns of differential-expression found in arm-1 can possibly manifest within the other arms (see Fig \ref{fig:AUC_trn4_tst1}, \ref{fig:AUC_trn4_tst2} and \ref{fig:AUC_trn4_tst3}).
  }
  \label{fig:arms}
\end{figure}

As shown in Fig \ref{fig:arms}, the common genotyped SNP-overlap between the cohorts varies significantly.
Cohorts that were genotyped using similar platforms tend to have large SNP-overlaps, while those genotyped on different platforms tend to have smaller SNP-overlaps.
After clustering the cohorts by platform (and removing any duplicate subjects across cohorts) we defined four `arms', as shown along the axes in Fig \ref{fig:arms}.
Arm-1 consists of the single cohort labeled `BDRN' (2524 cases, 4106 controls, OMEX).
Arm-2 includes cohorts `may1' through `rom3' (5781 cases, 8289 controls, OMEX).
Arm-3 includes cohorts `bonn' through `bmpo' (3581 cases, 7591 controls, Illumina).
Arm-4 includes cohorts `dub1' through `gain' (6825 cases, 9752 controls, Affymetrix).

The first arm (comprising the single cohort `BDRN') is relatively large and collected within the UK, comprising case-subjects of European descent over the age of 17 (see \cite{WellcomeTrustCaseControlConsortium_2007,JMGSetal_2015,LRKetal_2020} for details).
As a result, we expect this cohort to be less susceptible to spurious heterogeneity associated with batch-effects, and we use this cohort as a `training' or `discovery' arm, reserving the other three independent data-sets for validation (i.e., `replication' arms).
This training-arm has a large SNP-overlap of $\sim 85\%$ with arm-2, and a smaller SNP-overlap with arms 3 and 4 (i.e., $\sim 50\%$ and $\sim 30\%$, respectively).
Correspondingly, we expect that any signal involving a multi-SNP-pattern found in arm-1 will only have an opportunity to replicate strongly in arm-2, and will not have the opportunity to replicate as strongly in arms 3 and 4 (as we will have fewer SNPs to use for validation).

{\bf Ethics statement:\ }
We first obtained access to this data on 2013-11-25, and we have never had access to any information that could identify individual participants during or after data collection.
This study was approved by the institutional review boards (IRBs) at University of California, San Diego as well as New York University.
Because all the data we were working with had been de-identified, both IRBs certified our study as exempt from review and continuing review.
The metadata for the subjects included genome-wide principal-components, which we used as a proxy for ancestry and corrected for in our primary- and secondary analyses, as described below.
The metadata also included sex, whose effect we assessed a-posteriori (see Fig 17 in S1 Text).
We did not have access to confounding variables such as socioeconomic status, nutrition, environmental exposures, or other similar factors, and could not correct for these in our analysis.

\subsection*{Correcting for ancestry}

We use the genome-wide principal-components calculated by Stahl et al. to assess relatedness and correct for ancestry. 
Of the first 20 principal-components, denoted $\{U_{1},\ldots,U_{20}\}$, Stahl et al. determined that the first six principal-components ($U_{1}$-$U_{6}$) and $U_{19}$ showed significant correlation with the main phenotype across the studies considered in \cite{Stahl_2019}. 

In the discovery phase of our analysis we are restricted to the training-arm (arm-1).
For this sample we use an F-test applied to a nested logistic model, which selects only the first two principal-components (i.e., $U_{1}$ and $U_{2}$) as significantly related to case-control status in arm-1.
Therefore, to mimic the analyses one might conduct with access only to arm-1, we correct our biclustering algorithm for these two principal-components, under the assumption that they are a proxy for ancestry.

In all but this initial biclustering analysis on arm-1, we remain consistent with \cite{Stahl_2019} and correct for principal-components $U_{1}$ through $U_{6}$, as well as $U_{19}$.
This includes both the calculation of $\auc$ in the subsequent replication studies (e.g., $A(i)$ and $A'(i)$ in Fig \ref{fig:AUC_trn4_tst1}) as well as the PRS-analysis described below.

\subsection*{Biclustering}
For our initial biclustering of arm-1 we use the half-loop method of \cite{Rangan_2018}.
To briefly summarize the method, we first introduce some notation.
Assume that the data-set contains $M_{D}$ case-subjects, and $M_{X}$ control-subjects, each measured across $N$ allele-combinations (note, each SNP is associated with three allele combinations: heterozygous and homozygous dominant and recessive).
We denote the array of case-subjects by $D$, with $D(j_{D},k)$ referring to allele-combination-$k$ in case-subject-$j_{D}$.
Similarly, we denote the array of control-subjects by $X$, with $X(j_{X},k)$ referring to allele-combination-$k$ in control-subject-$j_{X}$.
We'll use the generic subject-index $j$ to refer to both the $j_{D}$ and the $j_{X}$.

In its most basic form, the half-loop algorithm proceeds as follows:
\begin{description}
\item[Step-0] First we load/initialize the data-arrays $D$ and $X$.
\item[Step-1] For each case $j_{D}$ and allele-combination $k$, we measure the fraction of other cases in $D$ which share that allele-combination, denoted by $[D\leftarrow D](j_{D},k)$. Similarly, we measure the fraction of controls in $X$ which share that allele-combination, denoted by $[D\leftarrow X](j_{D},k)$. The difference between these two values, denoted by $Q(j_{D},k) = [D\leftarrow D](j_{D},k) - [D\leftarrow X](j_{D},k)$ is a measure of differential-expression.
\item[Step-2] After calculating $Q(j_{D},k)$, we form the `row-scores' $Q^{\row}(j_{D}) = \sum_{k} Q(j_{D},k)$, as well as the `column-scores' $Q^{\col}(k) = \sum_{j_{D}} Q(j_{D},k)$ and the `trace' $\bar{Q}=\sum_{j_{D},k} Q(j_{D},k)$. The row- and column-scores measure how strongly each case-subject and allele-combination contribute to the trace, which is itself a measure of the overall differential-expression exhibited between $D$ and $X$.
\item[Step-3] We remove a small fraction of case-subjects and allele-combinations from $D$ with the lowest row- and column-scores.
\item[Step-4] We return to Step-1, iterating until there are no more case-subjects within $D$.
\end{description}

The algorithm proceeds iteratively; at each iteration $i$ we remove a small fraction $\gamma$ of the remaining case-subjects and allele-combinations. In this analysis we choose $\gamma=0.5^{8}\sim0.004$, which is sufficiently small that we expect statistical convergence of the algorithm's accuracy (see Fig 32 in supplementary section 7.3 in \cite{Rangan_2018}). After each iteration $i$, a subset $\cJ(i)$ comprising $M(i)$ case-subjects and a subset $\cK(i)$ comprising $N(i)$ allele-combinations remain, together forming an $M(i)\times N(i)$ sub-array $D(i)$ of the original $D$.
If the case-array $D$ were to contain a bicluster with a sufficiently strong signal, then the rows and columns of that bicluster would be retained until the end, with the other rows and columns eliminated earlier.

This half-loop method has detection-thresholds similar to spectral-clustering and message-passing \cite{Alon_1998,Deshpande_2015}, but has several additional useful features.
First, the half-loop method allows us to search for disease-specific heterogeneity by directly correcting for control-subjects.
This case-control correction also motivates the null-hypothesis H0 described below; the permutation-test allows us to avoid spurious structures that are unrelated to the disease-label.
Second, the half-loop scores in Step-1 allow us to (implicitly) correct for linkage-disequilibrium (LD).
More specifically, subsets of SNPs which are in equally strong LD in both the case- and control-populations will be excluded as the algorithm proceeds, unless some of those SNPs are involved in a pattern of differential-expression specific to the remaining case-subjects, in which case they will be retained (as desired).
Third, the method also allows us to correct for continuous covariates.
This covariate-correction is described in detail in \cite{Rangan_2018}, but essentially amounts to a reweighting of the $Q(j,k)$ in Step-1 to reduce the overall level of differential-expression contributed by structures which are not evenly distributed in covariate-space.
Finally, the method itself is rather straightforward and does not require the fine-tuning of parameters.

As mentioned in Step-2, the overall level of differential-expression between $D(i)$ and $X$ at each iteration is recorded as the trace $\bar{Q}(i)$.
The significance-level of $\bar{Q}(\cdot)$ is determined with respect to a null hypothesis (H0) which assumes that the heterogeneity is independent of case- and control-labels.
Samples from H0 are drawn by randomly permuting the case- and control-labels in arm-1 (i.e., randomly interchanging rows of $D$ and $X$) while respecting proximity in covariate-space.
By comparing the values of the $\bar{Q}(i)$ from the original data to the distribution of $\bar{Q}(i)$ associated with the null-hypothesis, we assign an (empirical) training-$p$-value to the individual $\bar{Q}(i)$ for each iteration $i$.
Similarly, we calculate an overall empirical training-$p$-value (across all iterations), which estimates the probability that the trace $\bar{Q}$ from the original data-set could be drawn from the null-hypothesis.

Within this context, the detection of a disease-specific bicluster corresponds to an elevated (i.e., statistically-significant) value of $\bar{Q}(i)$.
The case-subjects and allele-combinations comprising the bicluster can then be approximated by the subsets $\cJ(i)$ and $\cK(i)$ for those $i$.

\subsection*{Replication}
When discussing any particular replication-arm (e.g., arm-2), we will use primed indices (e.g., cases and controls will be indexed via $j_{D}'$ and $j_{X}'$).
To assess replication we first consider the set of allele-combinations $\cK'$ available within the replication-arm.
This subset will limit the alleles we can use from within the original training-arm (i.e., arm-1).
For any iteration $i$, we select the allele-subset $\cK(i)$ from the training-data-set, and then construct the intersection $\cK'(i):=\cK(i)\cap\cK'$.
For the replication-arm arm-2 the allele set $\cK'(i)$ will have a size $N'(i)$, which is typically around $85\%$ of $N(i)$ (i.e., $85\%$ of the full size of $\cK(i)$).
For the other replication-arms (i.e., arms 3 and 4) the overlap will be lower.
Using $\cK'(i)$ as well as the case-subject subset $\cJ(i)$, we define the $M(i)\times N'(i)$ submatrix $D'(i)$ within the training data (note $D'(i)$ is a submatrix of the $M(i)\times N(i)$ submatrix $D(i)$ defined above).
We then calculate the dominant SNP-wise principal-component $v(i)\in \Real^{N'(i)}$ of $D'(i)$.

We project each subject within the training-data-set onto $v(i)$, producing a `bicluster-score' (i.e., a single number) $u_{j_{D}}(i)$ for each case-subject in the training-data-set, and $u_{j_{X}}(i)$ for each control-subject in the training-data-set (recall that $j_{D}$ and $j_{X}$ index the case- and control-subjects in the training-data-set).
Based on the definition of the bicluster, we expect that the typical values of $u_{j_{D}}(i)$ will be larger than the typical values of $u_{j_{X}}(i)$.
We measure this difference by calculating the area under the receiver-operator-characteristic curve ($\auc$) between the sets $\{u_{j_{D}}(i)\}$ and $\{u_{j_{X}}(i)\}$; we refer to this $\auc$ as $A(i)$.
When calculating $A(i)$ we correct for the same ancestry-related covariates as in \cite{Stahl_2019} (see Methods and \cite{McGrouther_2012}).

We also project each subject in the replication-arm onto the same vector $v(i)$, producing bicluster-scores $u'_{j_{D}'}(i)$ for each case-subject in the replication-arm, and $u'_{j_{X}'}(i)$ for each control-subject in the replication-arm.
Once again, we expect that the typical values of $u'_{j_{D}'}(i)$ will be larger than the typical values of $u'_{j_{X}'}(i)$ in the replication-arm.
We measure this difference by calculating the $\auc$ $A'(i)$, once again correcting for the ancestry-related covariates.

We assess the overall significance of the replication by considering a null-hypothesis where the structure of the replication-arm is independent of disease-status.
We can draw a sample from this null-hypothesis (H0') by randomly permuting the case- and control-labels within the replication-arm (while respecting proximity in covariate-space).
In this manner we compare the original replication $\auc$ $A'(\cdot)$ (as a function of $i$) to the distribution of $A'(\cdot)$ obtained under H0'.

Later on below (e.g., Fig \ref{fig:AUC_trn4_tst1}) we calculate the average $\bar{A}'$ of $A'(\cdot)$ over a range of iterations, and then compare $\bar{A}'$ to the distribution of $\bar{A}'$ obtained under this label-shuffled null-hypothesis.
We define the range of iterations by taking an interval which is significant for both the trace $\bar{Q}(i)$ and the $\auc$ $A(i)$ defined using only the training-arm.
For example, in Fig \ref{fig:AUC_trn4_tst1} we consider the range of iterations $i\in[\imin,\imax]$.

\subsection*{Polygenic-Risk-Scores (PRSs)}
We calculate PRSs using the general strategy from \cite{Stahl_2019}, and further described in page $60$ of the Supplementary Information within that paper.
To briefly summarize: We use the genotype-level data from \cite{Stahl_2019}, which was imputed using the 1KG reference-panel. We then run a GWAS on this genotype-level data.
This GWAS produces summary-statistics defined by contrasting cases and controls from the training-arm, while correcting for ancestry-related covariates.
Once we have the summary-statistics defined by the GWAS, we run Plink's `clump' function to account for LD.
We perform this clumping step using the same parameters as in \cite{Stahl_2019} (e.g., info-score threshold of $0.9$, $R^{2}$-threshold of $0.1$, genomic window of $500$Kb, and minor-allele-frequency threshold of $0.05$.)
As a technical note: our ultimate goal is to analyze these PRS scores in the context of our heterogeneity analysis, which can be influenced by subtle relationships between SNPs.
Consequently, we wanted to use the most accurate available information regarding LD.
After the initial data-sets described in \cite{Stahl_2019} were published, the Haplotype Reference Consortium European Reference Panel (HRC EUR panel) became available through the Wellcome Trust Sanger Institute \cite{HRCERP_2016}.
This HRC EUR panel dramatically increased the amount of information available for approximating LD, and we use this panel when clumping our summary statistics.
Finally, after clumping, we use the assigned weights for each SNP to form a PRS. 
We test the performance of this PRS on our replication-arms.

For any subject $j'$ within a particular replication-arm, we denote by $\prsorig(j')$ the `population-wide' PRS defined by contrasting {\em all} the cases in the training-arm with the controls in the training-arm (when generating the summary-statistics).
We further denote by $\prsorig(j';\tp)$ the population-wide PRS constructed after restricting the SNP-weight-vector to include only those SNPs with individual GWAS $p$-values that are more significant than the threshold $\tp$ (when forming the PRS).

We also define a `bicluster-informed' PRS, denoted by $\prsbicl(j';i)$, by contrasting {\em only} the cases in $D(i)$ with the controls from the training-arm (when generating the summary-statistics).
We further denote by $\prsbicl(j';i,\tp)$ the bicluster-informed PRS constructed after restricting the SNP-weight-vector to include only those SNPs with individual GWAS $p$-values that are more significant than the threshold $\tp$ (when forming the PRS).
With this notation $\prsorig(j')$ and $\prsorig(j';\tp)$ are equivalent to $\prsbicl(j';1)$ and $\prsbicl(j';1,\tp)$, respectively.
However, we will typically consider $\prsbicl$ for iterations $i\in[175,350]$; in this range $\prsorig(j';\tp)$ and $\prsbicl(j';i,\tp)$ will differ.

We measure the performance of the population-wide $\prsorig(j')$ by calculating the $\aucorig$ between the case-values $\{\prsorig(j_{D}')\}$ and the control-values $\{\prsorig(j_{X}')\}$, once again correcting for the ancestry-related covariates.
Similarly, we measure the performance of $\prsorig(j';\tp)$, $\prsbicl(j';i)$ and $\prsbicl(j';i,\tp)$ by calculating the associated $\auc$s, denoted by $\aucorig(\tp)$, $\aucbicl(i)$ and $\aucbicl(i,\tp)$, respectively.

\subsection*{Gene-enrichment analysis}
We perform a simple over-representation analysis using the {\tt go\_bp} ontology from Seek \cite{Zhu_2015}.
We restrict our attention to the $132$ neuronally-related pathways (i.e., those referencing neurons, synapses or axons).
For any given iteration $i$ we consider the remaining allele-combinations within $\cK(i)$, retaining those genes which have more than half their originally associated alleles remaining.
These retained genes form a gene-set $\cG(i)$ which we then overlap with each pathway $\cH_{l}$ to obtain the intersection $\cG(i)\cap\cH_{l}$.
From this intersection we obtain the gene-count $\kappa(i,l) = |\cG(i)\cap\cH_{l}|$ for pathway $l$ at iteration $i$.

We assess the significance of the gene-counts by considering the same null-hypothesis H0 used when biclustering.
We compare each of the $\kappa(i,l)$ to the distribution of $\kappa(i,l)$ obtained under the label-shuffled null-hypothesis.
Later on below we calculate the average z-score $\bar{z}$ of the $\kappa(i,l)$ over a range of iterations and all the neuronally-related pathways, and then compare that $\bar{z}$ to the distribution of $\bar{z}$ obtained under H0.


\section*{Results}

\begin{figure}
  \centering
  \includegraphics[width=5.5in]{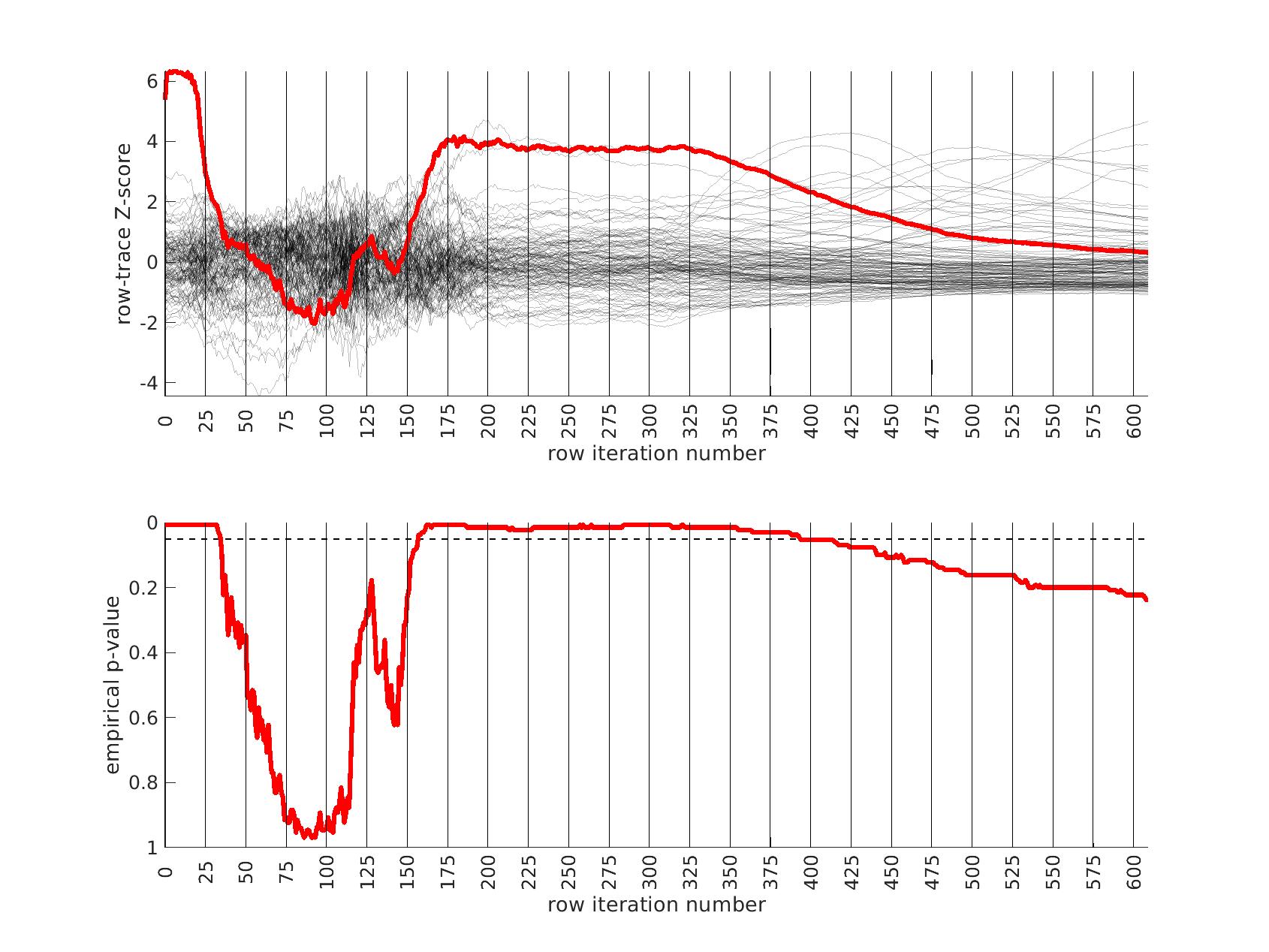}
  \caption{
    In this figure we show the output of the half-loop biclustering algorithm applied to the BDRN cohort in arm-1 (limited to those SNPs with maf $\geq 0.25$).
    As described in the main text, the algorithm proceeds iteratively, eliminating rows and columns from the case-subject-array $D$ until all have been removed.
    At each iteration $i$, the remaining submatrix $D(i)$ comprises case-subjects $\cJ(i)$  and allele-combinations $\cK(i)$.
    At each iteration we record the `row-trace' $\bar{Q}(i)$, which is the covariate-corrected average level of differential-expression between $D(i)$ and the control-subjects $X$.
    In the top row of subplots we show the row-trace for the data (red) as well as for $128$ label-shuffled trials (black).
    Each of the row-traces has been transformed into an iteration-dependent z-score (estimated using the distribution of label-shuffled trials at that iteration).
    In the bottom row we show the corresponding empirical p-value, as estimated for each iteration using the label-shuffled trials.
    The dashed black-line corresponds to the 95th percentile (i.e., a significance value of 0.05 if each iteration were considered independently).
    If the signal were homogeneous we would expect to see the red trace begin at a high value and decay relatively monotonically.
    By contrast, we see strong evidence for heterogeneity; the red trace is far from monotonic.
    The overall p-value for the data (red-trace), estimated using the strategy in \cite{Rangan_2018}, is $p\lesssim 1/64$.
    Note that the trace is significant over a range of iterations, including $i\in\left[\imin,\imax\right]$.
  }
  \label{fig:trace}
\end{figure}

We apply the half-loop-counting algorithm (see Methods) to the `BDRN' cohort used as the training arm.
The trace $\bar{Q}(\cdot)$ associated with the original data is shown in red in Fig \ref{fig:trace}.
Were the signal homogeneous, we would expect to see a trace that starts out high and gradually decreases in magnitude. 
Instead, we see a trace that behaves non-monotonically, and is statistically insignificant for a range of iterations.
The trace from the original data (in red) attains values that are significantly higher than the majority of the traces one would expect under the null-hypothesis (black) near iteration $i\sim \imin$.
This is an indicator that the data is heterogeneous, and that a bicluster has been detected near iteration $i\sim \imin$; the identity of the bicluster can be approximated by one of the submatrices $D(i)$ where the training-$p$-value is large.
We can calculate the empirical $p$-value associated with the entire trace $\bar{Q}(\cdot)$ by comparing the red curve (across all iterations) to the black curves, estimating an overall p-value of $p\lesssim 1/64$.

In idealized scenarios where the `true' bicluster is sharply defined, the trace typically has a sharp peak near the $D(i)$ that most closely corresponds to the bicluster \cite{Rangan_2012,Rangan_2018}.
However, in this case while the trace has a peak at around $i\sim \imin$, this peak is not particularly sharp, and the trace is nearly as significant across a range of iterations $i\in[\imin,\imax]$.
The largest of these submatrices (i.e., $D(\imin)$) corresponds to $\sim 47\%$ of the case-subjects and $\sim 31\%$ of the allele-combinations.
The smallest of these submatrices (i.e., $D(\imax)$) corresponds to $\sim 21\%$ of the case-subjects and $\sim 9\%$ of the allele-combinations. 

This `plateau' of significance indicates that the true signal is not a perfectly crisp and well-delineated bicluster.
Instead, this plateau suggests that, while there are certain `core' case-subjects that exhibit a strong similarity across certain allele-combinations, there are additional case-subjects that are `adjacent' to those in the core.
These adjacent subjects exhibit a slightly weaker similarity involving a slightly expanded set of allele-combinations.
Consequently, we expect iterations in the interval $i\in[\imin,\imax]$ to provide a range of approximations to the true `core' signal (which is still unknown).
One could certainly select the iteration with the highest training-$p$-value to approximate the bicluster, but as nearby iterations have nearly the same training-$p$-value, we expect them to also provide reasonable estimates of the true signal.

Given our approximation to the signal described above from the training-data-set, we test for replication in each of the replication-arms 2, 3 and 4.
We are interested in how strongly our approximate signal replicates, as well as whether our approximation has been compromised by overfitting.
Because the signal spans a range of iterations in arm-1, we assess the extent of replication across the plateau $i\in[\imin,\imax]$. 
This interval corresponds to significant values of the trace $\bar{Q}(i)$ as well as the $\auc$ $A(i)$ defined only using the training-data.

\begin{figure}
  \centering
  \includegraphics[width=6.5in]{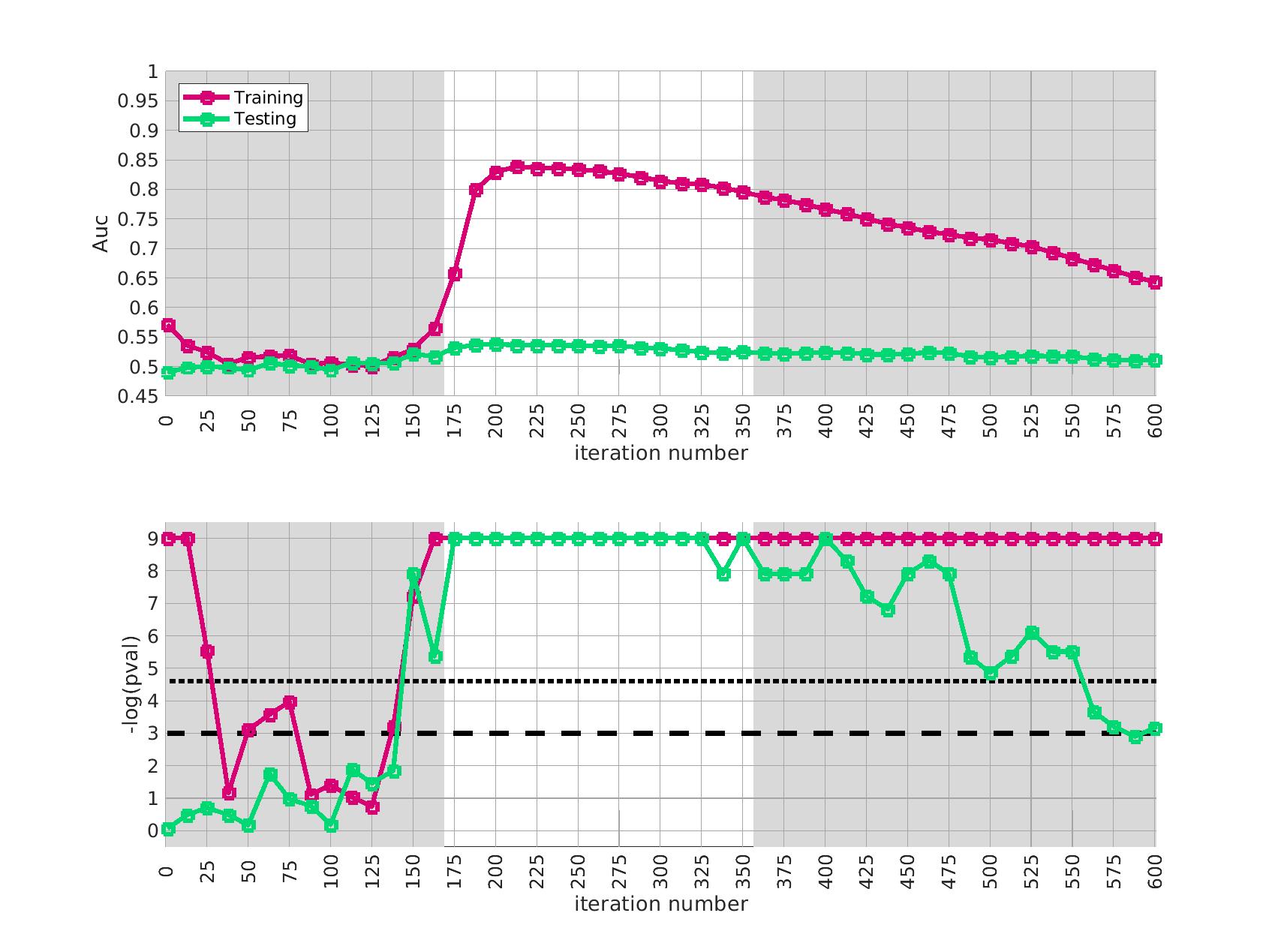}
  \caption{
    In this figure we illustrate the replication of the bicluster in arm-2.
    Note that the SNP-overlap between arm-1 and arm-2 is $\sim 85\%$.
    On the top we show $A(i)$ in red and $A'(i)$ in green.
    On the bottom we show the associated p-values for $A(i)$ and $A'(i)$, calculated with respect to $H0$ and $H0'$ for each iteration individually.
    Standard significance-levels $0.05$ and $0.01$ are shown in dashed- and dotted-lines, respectively.
    The interval $i\in[\imin,\imax]$ is highlighted in white.
    Note that both $A(i)$ and $A'(i)$ have peaks within the range that the trace was significant (c.f. Fig \ref{fig:trace}).
    The overall replication for arm-2 within the interval $i\in[\imin,\imax]$ is estimated at $p\lesssim 10^{-12}$.
  }
  \label{fig:AUC_trn4_tst1}
\end{figure}

\begin{figure}
  \centering
  \includegraphics[width=6.5in]{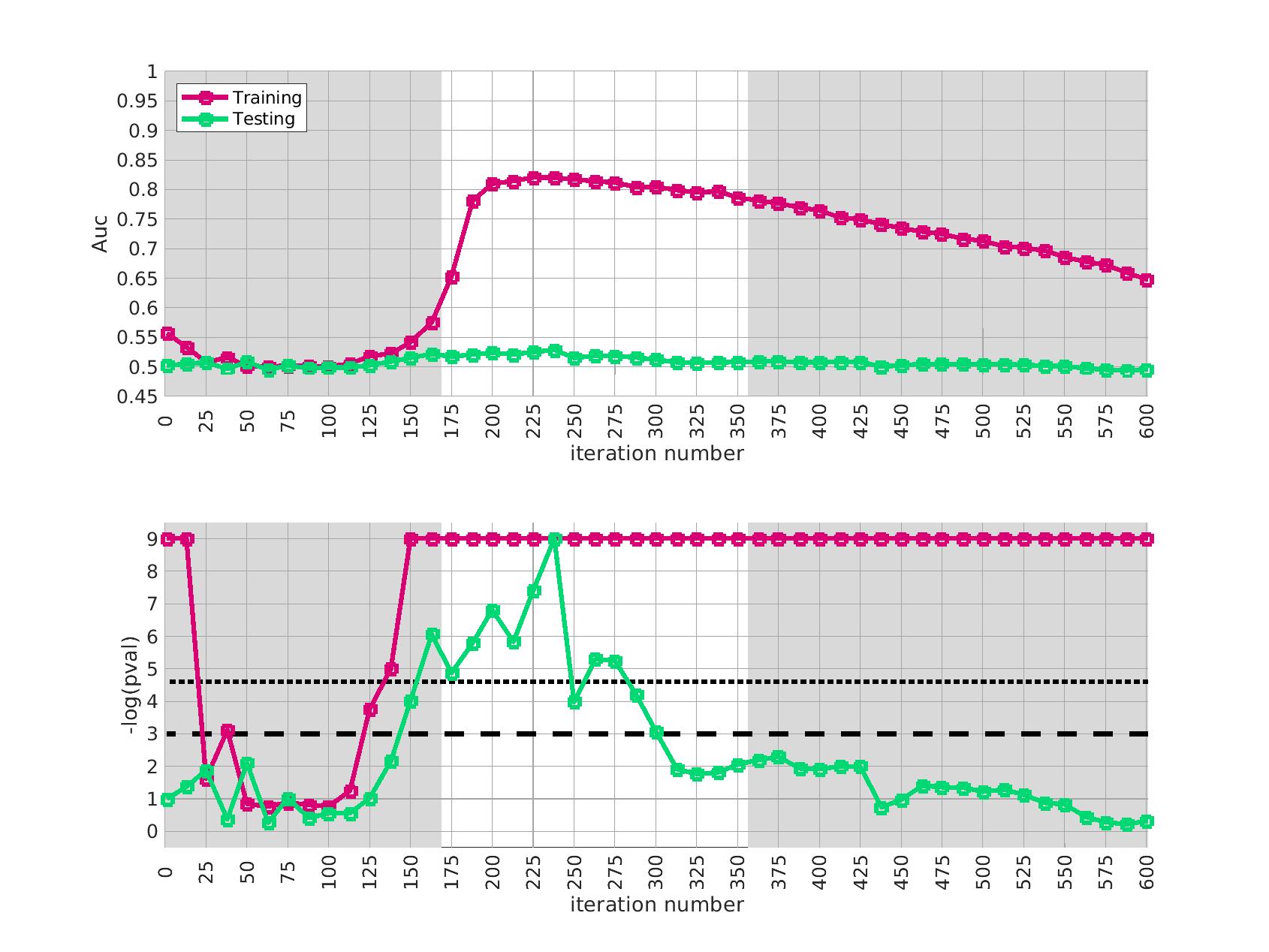}
  \caption{
    This figure is similar to Fig \ref{fig:AUC_trn4_tst1}, except that we use arm-3 instead of arm-2.
    The overall replication for arm-3 within the interval $i\in[\imin,\imax]$ is estimated at $p\lesssim 10^{-3}$.
    Note that the SNP-overlap between arm-1 and arm-3 is only $\sim 50\%$. 
  }
  \label{fig:AUC_trn4_tst2}
\end{figure}

\begin{figure}
  \centering
  \includegraphics[width=6.5in]{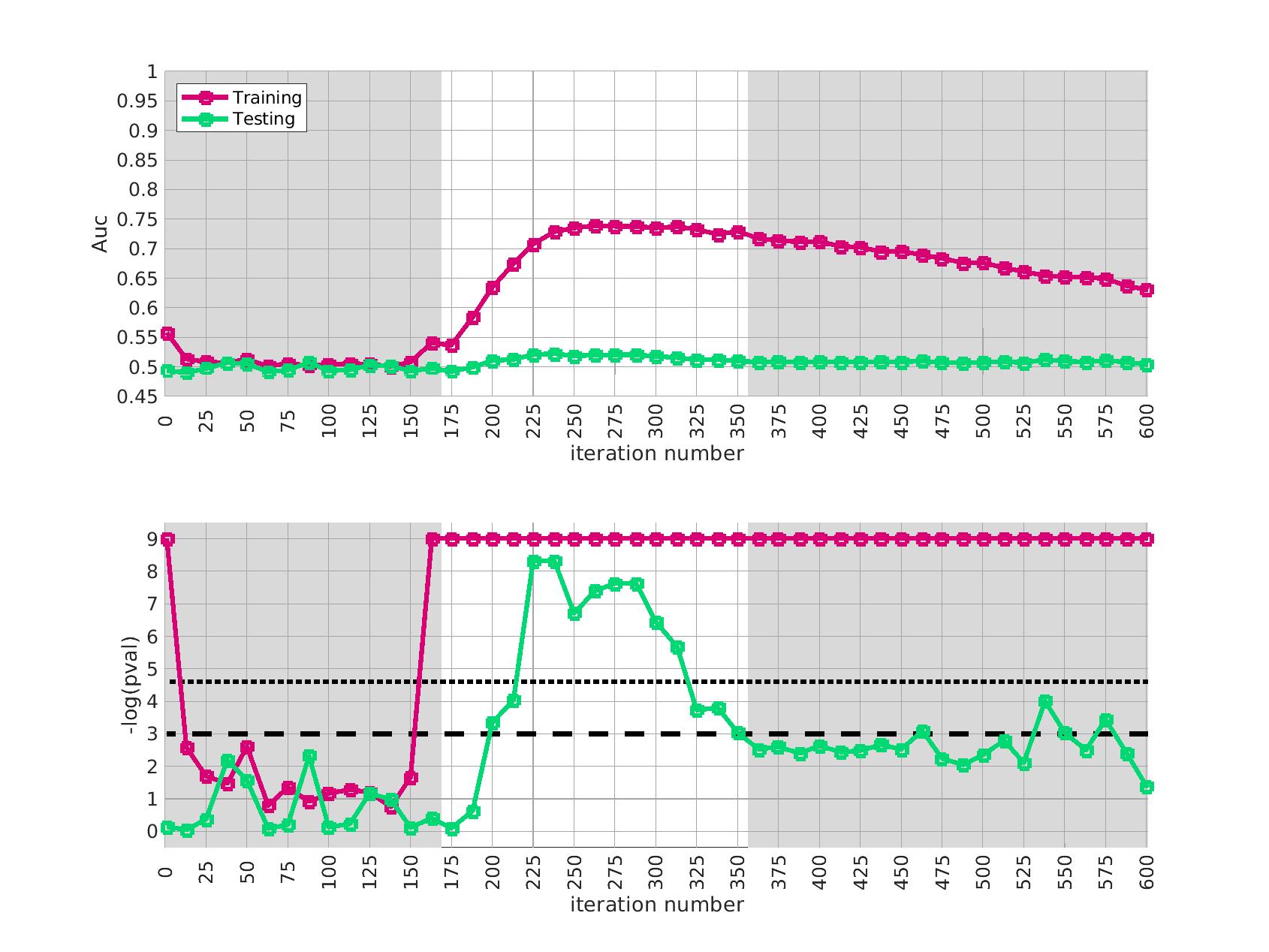}
  \caption{
    This figure is similar to Fig \ref{fig:AUC_trn4_tst1}, except that we use arm-4 instead of arm-2.
    The overall replication for arm-3 within the interval $i\in[\imin,\imax]$ is estimated at $p\lesssim 10^{-3}$.
    Note that the SNP-overlap between arm-1 and arm-4 is only $\sim 30\%$.
  }
  \label{fig:AUC_trn4_tst3}
\end{figure}

The results of this replication study for arm-2 are shown in Fig \ref{fig:AUC_trn4_tst1}.
The top subplot illustrates the $\auc$ $A(i)$ (red) and $A'(i)$  (green) as a function of $i$.
The bottom subplot shows the associated $p$-value for each $i$ (under a label-shuffled null-hypothesis).
Note that the training-$\auc$ $A(i)$ is high over the range of iterations $i\in[\imin,\imax]$ for which the training-$p$ value is significant.
Note also that the peak of $A(i)$ occurs within a few iterations of the peak of the training $p$-value.
This correspondence corroborates the claims made above: we believe we have detected a disease-related signal within the training-data-set that involves only a subset of subjects and alleles.
While the magnitude of the replication-$\auc$ $A'(i)$ is lower than the training-$\auc$ $A(i)$, the value of $A'(i)$ is also statistically significant over the range of iterations $i\in[\imin,\imax]$, with a peak at roughly the same point.

Similar results for arm-3 and arm-4 are shown in Figs \ref{fig:AUC_trn4_tst2} and \ref{fig:AUC_trn4_tst3}.
Note that the SNP-overlap between these arms and the training-data-set is quite a bit lower than that for arm-2.
Recall that arm-2 has a overlap of $\sim85\%$ with the SNPs in arm-1, while arm-3 and arm-4 have overlaps of $\sim 50\%$ and $\sim 30\%$, respectively.

We believe that this reduction in SNP-overlap is partially responsible for the reduction in the magnitude of replication-$\auc$s observed in these arms.
To test this hypothesis, we randomly eliminate SNPs from arm-2 until the SNP-overlap between the training-data-set and arm-2 is equal to the SNP-overlap between the training-data-set and arm-3.
The results of this replication-study are shown in Fig 13 in S1 Text: note that the amplitude of $A'(i)$ has degraded in comparison to the values shown in Fig \ref{fig:AUC_trn4_tst1}.
We then randomly eliminate even more SNPs, until the SNP-overlap between the training-data-set and arm-1 is equal to the SNP-overlap between the training-data-set and arm-4 (see Fig 14 in S1 Text), and the amplitude $A'(i)$ degrades even further.
More generally, by reducing the number of SNPs we include in the replication-arm, we can cause the values of $A'(i)$ to drop; depending on the subset of SNPs retained, the values of $A'(i)$ for arm-2 can be reduced to values similar to those observed in arm-3 and arm-4.

In summary, the $\auc$ associated with the genotype-based bicluster score discovered in the training-data-set replicates to varying degrees across all 3 replication arms.
In each case the average $A'(i)$ calculated over the interval $i\in[\imin,\imax]$ was significantly larger than what one would expect were the case- and control-labels in the replication-arm randomly permuted ($p\lesssim 1/1000$).
Consequently, we are fairly certain that -- while our approximation of the bicluster is far from perfect -- we have indeed identified a robust disease-related signal which generalizes across a variety of different BD studies.

\section*{Discussion}

\subsection*{Interaction with covariates}

Given the observations above, it is natural to ask what might be driving the signal associated with this bicluster.
We first checked to see if the bicluster was driven by the ancestry-related covariates in our data-set.
As shown in Figs 15 and 16 in S1 Text, the subjects in the bicluster have a distribution of ancestries similar to the remainder of arm-1 (recall that we corrected for ancestry as a covariate).
By considering the subjects remaining in $D(i)$, we also determined that the bicluster does not seem to be associated with sex (see Fig 17 in S1 Text). 

\subsection*{Interaction with BD subtype}

\begin{figure}
  \centering
  \includegraphics[width=5.5in]{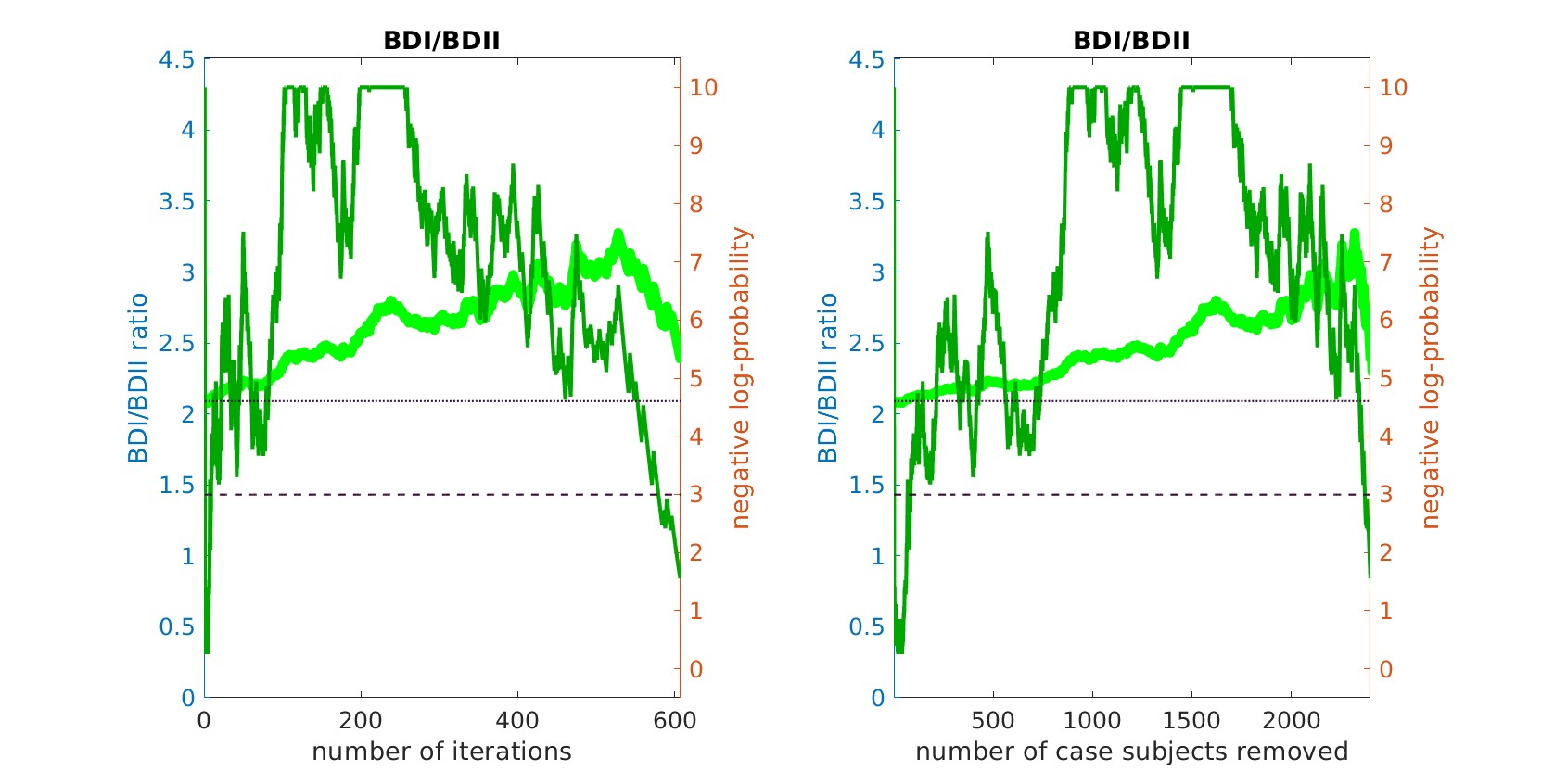}
  \caption{
    This figure plots the ratio of BDI to BDII subjects within $\cJ(i)$ (light-green, left y-axis) as a function of the iteration $i$ (left) and the number of removed case-subjects (right). The dark-green line corresponds to the negative-log-probability (right y-axis) of observing a ratio at least as large by chance.
    The dashed and dotted horizontal lines indicate $0.05$ and $0.01$ significance values, respectively.
    Note that the BDI population is over-represented across a range of iterations including $i\in[\imin,\imax]$, implying that the bicluster we observe is significantly enriched for BDI subjects.
  }
  \label{fig:pheno_stack}
\end{figure}

We then checked to see if the bicluster was associated with bipolar subtype. We measured the fraction of subjects classified as bipolar-type-1 versus bipolar-type-2 as our algorithm proceeded.
Specifically, we measured the fraction of case-subjects in $\cJ(i)$ that were classified as BDI and BDII.
If the bicluster were driven by BDII subjects, then we would expect the proportion of remaining BDII case-subjects to increase with the iteration-index $i$.
Conversely, if the bicluster were driven by BDI subjects, then we would expect the proportion of remaining BDI case-subjects to increase with iteration-index.
As shown in Fig \ref{fig:pheno_stack}, we found that this latter scenario holds; the bicluster was significantly enriched for BDI relative to BDII. 
This enrichment for BDI also impacts our risk-prediction results (see below). 
Note that, when determining this enrichment, we compare the proportion of BDI and BDII case-subjects at each iteration to the proportion at iteration $i=1$ (i.e., across all case-subjects in arm-1). In this manner our enrichment is defined relative to the starting proportion of BDI and BDII subjects in our training-arm, and is not influenced by the recruitment rates for BDI and BDII (which can differ across studies).

While significant, this BDI-enrichment was not completely overwhelming:
the initial fraction of BDII participants in arm-1 was $\sim 31\%$, which dropped to $\sim 26\%$ at iteration $i=240$.
Thus, while the majority of the case-subjects in the bicluster are classified as BDI, those classified with BDII do still contribute to the overall signal.
It is possible that this BDI-enrichment is due to a true difference between the BD-subtypes at the genetic level.
However, it is also possible that this enrichment is partially driven by inaccuracies associated with classification \cite{Charney_2017}. 

\subsection*{Bicluster-informed PRS Performance}
\begin{figure}
  \centering
  \includegraphics[trim=300 0 200 0,clip,width=7.5in]{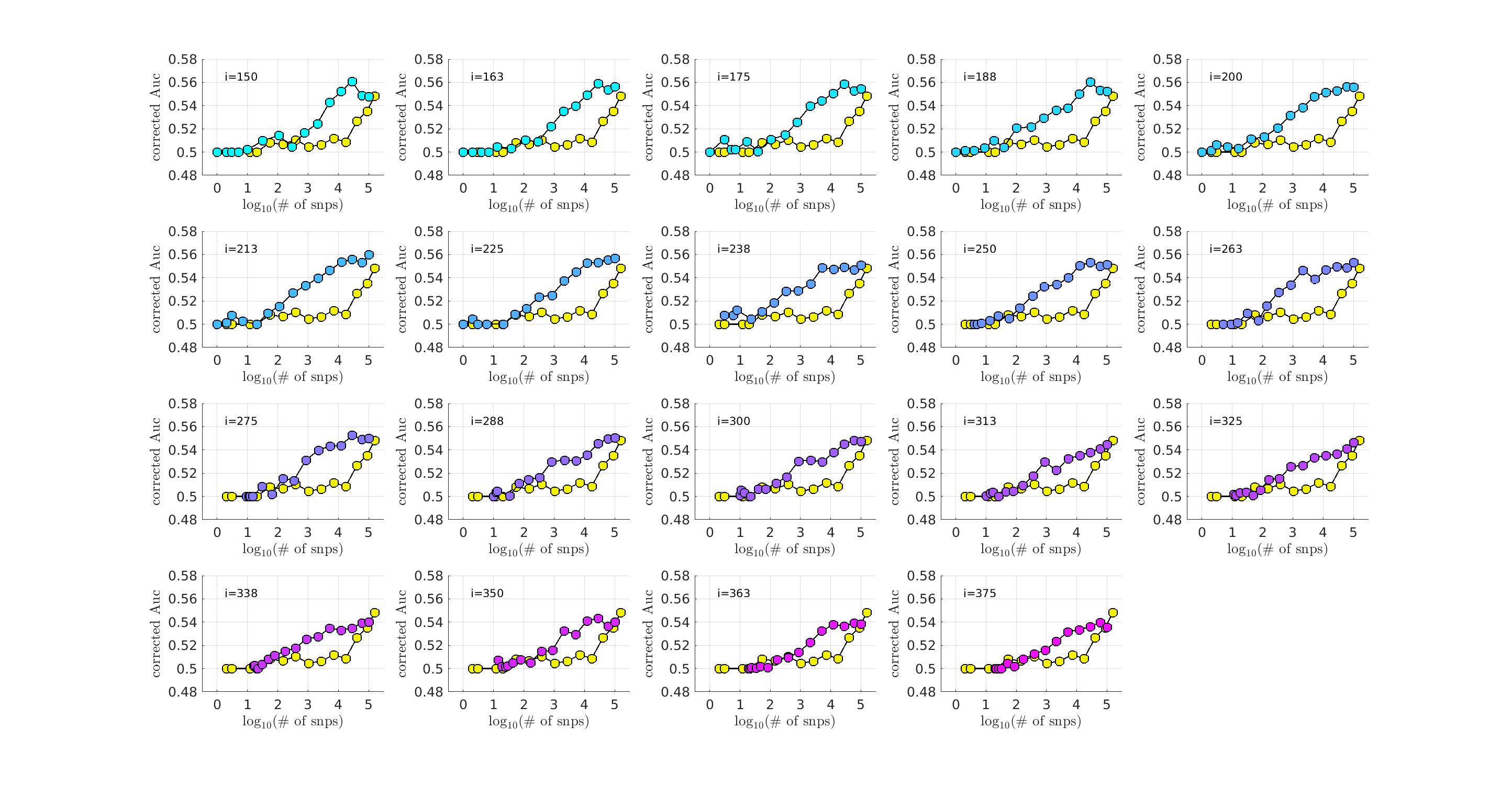}
  \caption{
    In each subplot we show in yellow the $\aucorig(\tp)$ (vertical) for arm-2 as a function of the number of SNPs corresponding to each $\tp$-threshold (horizontal, log-scale).
    Additionally, we show $\aucbicl(i,\tp)$ for a particular iteration $i$ (with $i$ varying across subplots). The color-code used for $\aucbicl(i,\tp)$ ranges from blue to pink, corresponding to the iteration index $i$.
    Note that, by using the bicluster to inform the PRS, the performance typically improves. This improvement in performance becomes marked when the number of SNPs is limited to a relatively small fraction of the total (e.g., $\sim 1\%$ of the total, corresponding to a $\log_{10}(\#)$ of $\sim 3$).
  }
  \label{fig:prs_comparison_trn4_tst1_nixxx}
\end{figure}

\begin{figure}
  \centering
  \includegraphics[trim=300 0 200 0,clip,width=7.5in]{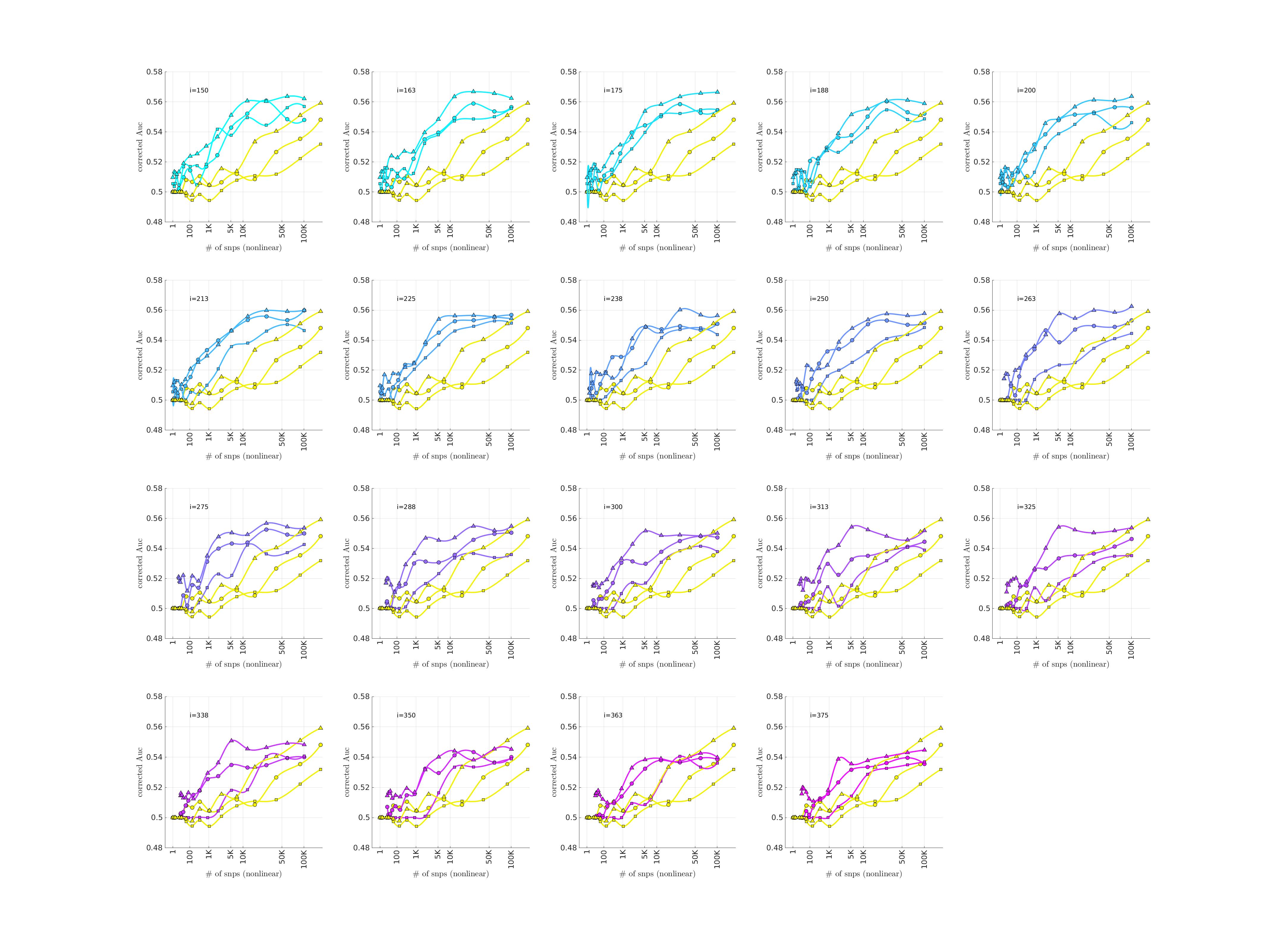}
  \caption{
    This figure uses circles to displays the same information as Fig \ref{fig:prs_comparison_trn4_tst1_nixxx} (corresponding to replication arm-2). 
    In this figure we use an algebraic-scale for the horizontal-axis (rather than a log-scale) in order to better emphasize the interval where the number of SNPs used is between $1K$ and $10K$. The results for replication arm-3 and arm-4 are shown using squares and triangles, respectively. 
    }
  \label{fig:prs_comparison_trn4_tsty_nixxx}
\end{figure}

As described in the Methods section, we calculated the population-wide $\prsorig(j';\tp)$ and the bicluster-informed $\prsbicl(j';i,\tp)$ across a variety of iterations $i$ and $\tp$-thresholds. We compared the bicluster-informed $\prsbicl(j';i,\tp)$ performance to the one generated by the population-wide $\prsorig(j';\tp)$ across a variety of $\tp$-thresholds.
Results for arm-2 are shown in Fig \ref{fig:prs_comparison_trn4_tst1_nixxx}.
Results for arm-3 and arm-4 are shown alongside arm-2 in Fig \ref{fig:prs_comparison_trn4_tsty_nixxx}, and individually in Figs 23 and 24 in S1 Text.

Note that, when constructing $\prsbicl(j';i,\tp)$, we restrict ourselves to a subset of case-subjects within the training-arm determined by $\cJ(i)$. In this case, when $i\in[\imin,\imax]$ the case-subset $\cJ(i)$ retains only $\sim 50\%-20\%$ of the original case-subjects in arm-1. Typically, one might expect a reduction in the number of case-subjects to yield a corresponding reduction in power, giving rise to a reduced discriminability in the testing-arms 2,3 and 4. However, as we see in Fig \ref{fig:prs_comparison_trn4_tsty_nixxx}, the discriminability for $\prsbicl(j';i,\tp)$ is typically {\em higher} than $\prsorig(j',\tp)$ when $i\in[\imin,\imax]$. This suggests that the case-subjects in $\cJ(i)$ identified by the bicluster correspond to a stronger genetic signal, likely arising from the increased homogeneity within $\cJ(i)$.

Note that $\prsbicl$ and $\prsorig$ are not capturing identical signals (see the Nagelkerke $R^{2}$ analysis in the Supporting information). It is useful to compare the performance of $\prsbicl$ with $\prsorig$ as there are features of $\prsbicl$  which indicate that it is more robust than $\prsorig$.
As one example, we point out that $\aucbicl(i,\tp)$ is markedly higher than $\aucorig(\tp)$ when the number of SNPs used (denoted by $\NSNP$) is  fewer; one begins to see the effect between $1K$ and $10K$. This suggests that the bicluster-informed $\prsbicl(j';i,\tp)$ is not only outperforming the population-wide $\prsorig(j';\tp)$, but also correctly attributing the largest PRS-weights to those SNPs that truly carry the signal (and which are most important for replication). As one illustration, by comparing the values of $\aucbicl$ to $\aucorig$ in Fig \ref{fig:prs_comparison_trn4_tsty_nixxx}, we can directly see that the bicluster-informed PRS would replicate across arms 2,3 and 4 for values of $i=225$ and $\NSNP\in[10^{3},10^{4}]$, while the population-wide PRS would not. 

Motivated by the significant BDI-enrichment seen within the training-arm (see Fig \ref{fig:pheno_stack}), we repeated these assessments for the BDI- and BDII-populations within the testing-arms.
More specifically, recall that, for any particular testing-arm, the $\aucorig(\tp)$ and $\aucbicl(i,\tp)$ values shown in Figs \ref{fig:prs_comparison_trn4_tst1_nixxx}-\ref{fig:prs_comparison_trn4_tsty_nixxx} are defined using the values of $\prsbicl(j';i,\tp)$ across all case- and control-subjects $j'$ for that testing-arm.
We can now use the same values of $\prsorig(j';\tp)$ and $\prsbicl(j';i,\tp)$, but only compare the BDI-case-subjects to the control-subjects in the testing-arm.
This produces `restricted' $\auc$-values, which we denote by $\aucorigBDI(\tp)$ and $\aucbiclBDI(i,\tp)$, respectively.
In a similar fashion we can restrict the case-subjects in the testing-arm to the BDII-case-subjects, and calculate $\aucorigBDII(\tp)$ and $\aucbiclBDII(i,\tp)$.

\begin{figure}
  \centering
  \includegraphics[trim=300 0 200 0,clip,width=7.5in]{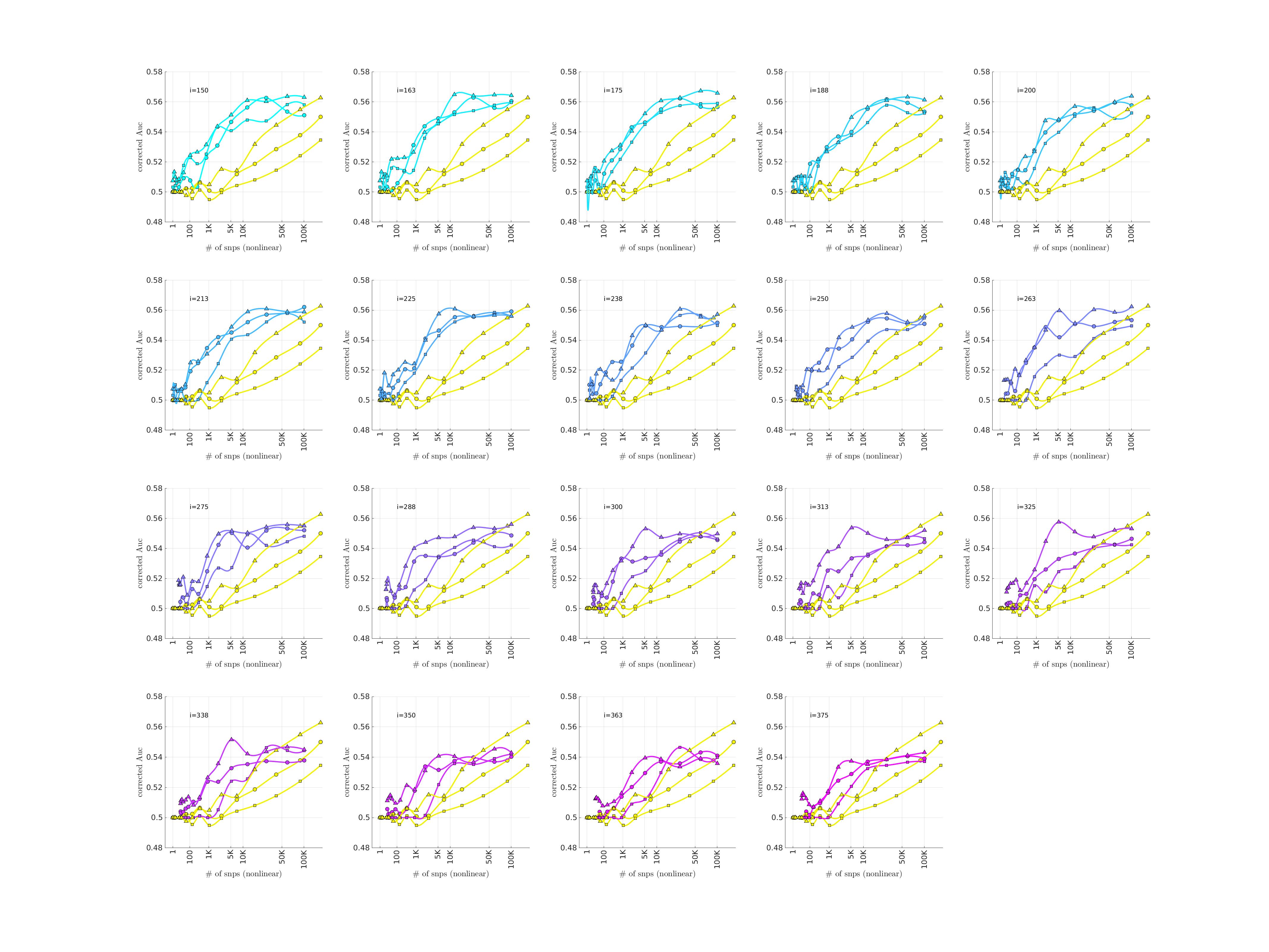}
  \caption{
    This figure is similar to Fig \ref{fig:prs_comparison_trn4_tsty_nixxx}, except that we limit ourselves only to those case-subjects in the replication-arms which are classified as BDI. This subset corresponded to $66\%$ ($M=3834$), $84\%$ ($M=2995$) and $75\%$ (M=5107) of the case-population for arms 2, 3 and 4, respectively. The corresponding $\auc$-values are denoted by $\aucorigBDI(\tp)$ and $\aucbiclBDI(\tp)$ in the main text. 
    For reference the training-arm had $M=1645$ BDI case-subjects, corresponding to $65\%$ of the case-population in arm-1.
    }
  \label{fig:prs_BD1_comparison_trn4_tsty_nixxx}
\end{figure}

\begin{figure}
  \centering
  \includegraphics[trim=300 0 200 0,clip,width=7.5in]{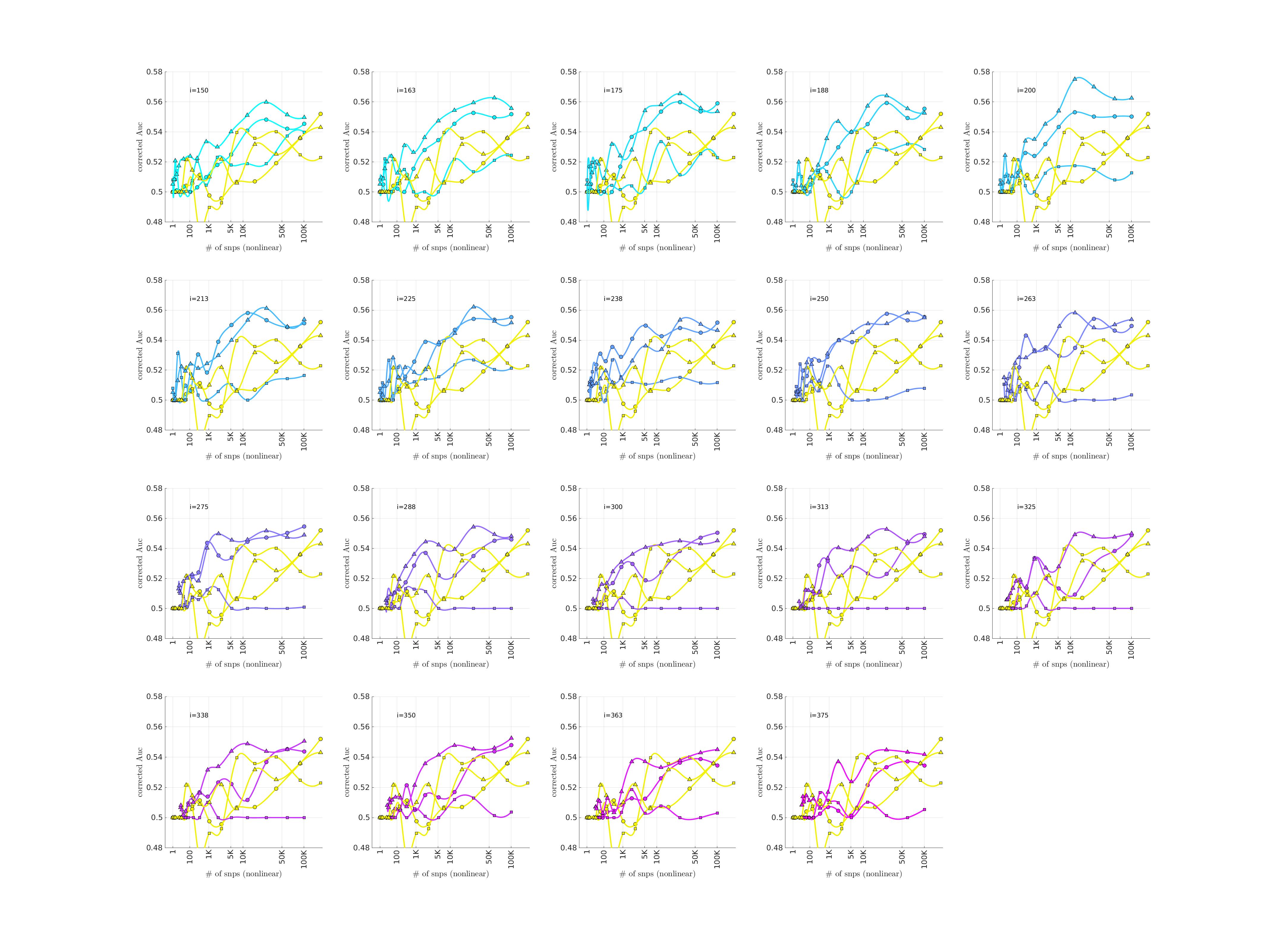}
  \caption{
    This figure is similar to Fig \ref{fig:prs_comparison_trn4_tsty_nixxx}, except that we limit ourselves only to those case-subjects in the replication-arms which are classified as BDII. This subset corresponded to $19\%$ ($M=1082$), $12\%$ ($M=435$) and $16\%$ (M=1060) of the case-population for arms 2, 3 and 4, respectively. The corresponding $\auc$-values are denoted by $\aucorigBDII(\tp)$ and $\aucbiclBDII(\tp)$ in the main text. 
    For reference the training-arm had $M=788$ BDII case-subjects, corresponding to $31\%$ of the case-population in arm-1.
    }
  \label{fig:prs_BD2_comparison_trn4_tsty_nixxx}
\end{figure}

The results are shown in Figs \ref{fig:prs_BD1_comparison_trn4_tsty_nixxx} and \ref{fig:prs_BD2_comparison_trn4_tsty_nixxx}, respectively.
Note that the improvement to risk-prediction persists for the BDI-population, but is not as robust for the BDII-population.
The performance of $\aucbiclBDII(i,\tp)$ is particularly poor for the BDII-population in arm-3, for which there were only $M=435$ BDII-subjects (i.e., the fewest out of all the arms).
It is possible that the variation in the performance of $\aucbiclBDII(i,\tp)$ for the BDII-population across the replication-arms has to do with these differences in power.
It is also possible that there are other systematic issues affecting the BDII-population, including variation in the life history of the subjects or the metrics used for their clinical diagnosis \cite{Charney_2017}.

\begin{figure}
  \centering
  \includegraphics[trim=300 0 200 0,clip,width=7.5in]{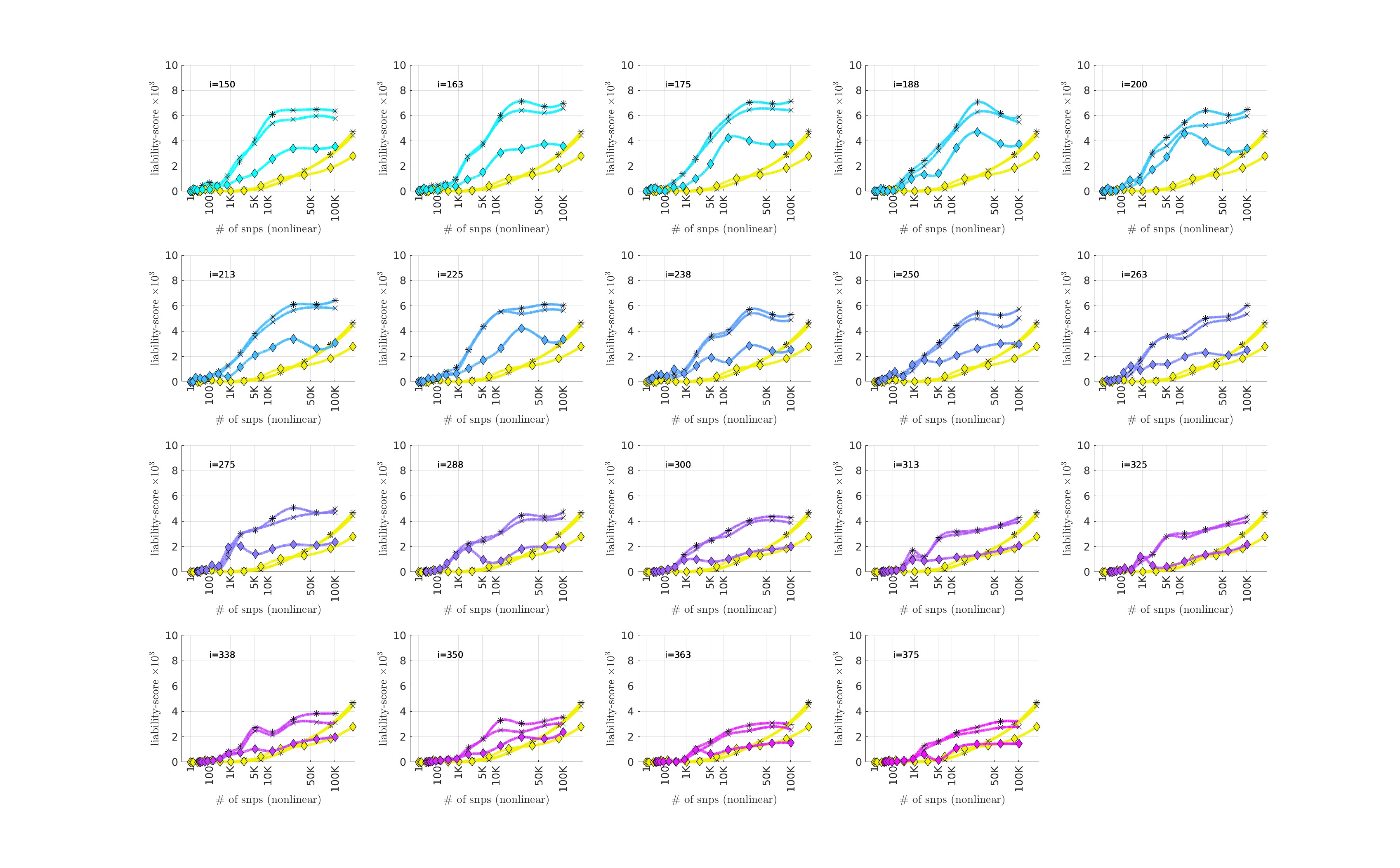}
  \caption{
    This figure is similar to Fig \ref{fig:prs_comparison_trn4_tsty_nixxx}, and uses the data from Figs \ref{fig:prs_comparison_trn4_tsty_nixxx}, \ref{fig:prs_BD1_comparison_trn4_tsty_nixxx} and \ref{fig:prs_BD2_comparison_trn4_tsty_nixxx}.
    This time we combine the information across all three replication-arms, and calculate replication $\auc$-values for this combined data-set. We then convert these $\auc$-values into liability-scores (see \cite{Lee_2012}).
    The results for all the cases ($\cRorig$ and $\cRbicl$) are shown with an asterisk `*', whereas the results for only the BD1-cases ($\cRorigBDI$ and $\cRbiclBDI$) are shown with an `$\times$', and the results for only the BD2-cases ($\cRorigBDII$ and $\cRbiclBDII$) are shown with a diamond.
    In each case the yellow curves correspond to the liability-scores derived from the population-wide PRS, whereas the cyan-magenta curves correspond to the liability-scores derived from the bicluster-informed PRS.
    Note that our overall results are closely matched by the BD1-cases, but not by the BD2-cases.
    }
  \label{fig:prs_BDX_liability_trn4_tsty_nixxx}
\end{figure}

To summarize the overall relationship between BD-subtype, the bicluster-informed PRS and the population-wide PRS, we pool the subjects across the replication-arms and convert the combined $\auc$-values into $R^{2}$-values on a liability-scale \cite{Lee_2012} using prevalences of $2\%$ for BD, and $1\%$ for BDI and BDII \cite{OConnell_etal_2023}.
Using notation analogous to the $\auc$-values, we denote these liability-scores as $\cRorig(\tp)$, $\cRorigBDI(\tp)$ and $\cRorigBDII(\tp)$, as well as $\cRbicl(i,\tp)$, $\cRbiclBDI(i,\tp)$ and $\cRbiclBDII(i,\tp)$, respectively.
The resulting liability-scores are shown in Fig \ref{fig:prs_BDX_liability_trn4_tsty_nixxx}.

We believe that Fig \ref{fig:prs_BDX_liability_trn4_tsty_nixxx} hints at the potential our methodology offers to researchers of complex disease.
By limiting our definition of a case to those with a more genetically homogeneous BD signature, we were able to generate a $\prsbicl$ which outperforms $\prsorig$ in the following ways: 
\begin{itemize}
\item The maximum $\cRbicl$ is $20$-$40\%$ higher than the maximum $\cRorig$, depending on the iteration-index $i$.  
\item This increase in liability-score occurs despite the fact that the $\prsbicl$ is generated using $\sim50\%$ to $80\%$ fewer cases than the $\prsorig$. For example, we considered only between $1191$-$526$ cases in arm-1 to generate the $\prsbicl(j';i,\tp)$ values for $i=[175,350]$, whereas $2524$ cases were used to generate the $\prsorig(j';\tp)$ values.
\item The p-values assigned to the SNPs via the bicluster-informed GWAS were less noisy than those p-values assigned using the population-wide GWAS. For example, Fig \ref{fig:prs_BDX_liability_trn4_tsty_nixxx} indicates that the first $10$K SNPs of highest significance from the population-wide GWAS contained almost no disease-related information. By contrast, the first $10$K SNPs of highest significance from the bicluster-informed GWAS typically contain most of the available disease-related information. The bicluster-informed GWAS produces a $\cRbicl$ with only $\sim 5$K to $10$K SNPs that surpasses the maximum of $\cRorig$ (e.g., within the $i=225$ subplot the value of $\cRbicl$ at $5$K SNPs is comparable to the value of $\cRorig$ at $\sim 150$K SNPs).
\item Furthermore, the values of $\cRbicl(i,\tp)$ typically plateau somewhere between $10$K--$35$K SNPs (corresponding to $\tp\in[0.22,0.36]$). Meanwhile, the values of $\cRorig(\tp)$ continue to increase until $\tp=1$ (including all $\sim 150$K SNPs).
\end{itemize}
To summarize: The $\prsbicl$ outperforms $\prsorig$ overall, and when restricted to either BDI and BDII, achieving a higher maximum for each subtype. The $\prsbicl$ also achieves its peak performance with far fewer SNPs, consistent with a far less noisy signal.
Put another way, the values of $\cRbicl$, $\cRbiclBDI$ and $\cRbiclBDII$ all plateau earlier than $\cRorig$, $\cRorigBDI$ and $\cRorigBDII$, indicating that the SNPs which are most relevant to the bicluster-informed PRS performance have indeed been identified by the bicluster-informed GWAS as having low individual $p$-values. 

Additionally, we note that there is a close relationship between $\cRbicl$ and $\cRbiclBDI$, but a discrepancy between $\cRbicl$ and $\cRbiclBDII$.
The values of $\cRbiclBDII$ indicate that some subset of BDII cases share the bicluster signature, but the maximum for $\cRbiclBDII$ is only $50\%$ of the maximum for $\cRbiclBDI$. 
This could imply that the bicluster has focused on a signature that is associated with BDI, perhaps serving as a risk factor for manic episodes in the presence of the necessary epigenetic or environmental influences.

\subsection*{Gene-enrichment:\ }

We also perform a simple over-representation analysis, measuring the overlap $\kappa(i,l)$ between the bicluster $D(i)$ at iteration $i$ and the various neuronally-related pathways $\cH_{l}$ from the {\tt go\_bp} ontology (see Methods).
The average z-score for the enrichment-values $\kappa(i,l)$, averaged over the interval $i\in[\imin,\imax]$ and all neuronally-related pathways, is quite significant, with $p\lesssim 1e-4$ (as determined by a permutation-test).
Examples of some of the more significantly over-represented pathways are shown in Table \ref{tab:genri_neu}.

\begin{table}
\resizebox{\textwidth}{!}{
  \begin{tabular}{ |r|c|c|c|c|c|c|c|c|c|} 
 \hline
                                         annotation &   175 &   200 &   225 &   250 &   275 &   300 &   325 &   350 &   375 \\
                       synaptic vesicle endocytosis & {\textcolor{red}{ 3.49}} & {\textcolor{red}{ 2.10}} &  1.28 & {\textcolor{red}{ 1.94}} &  0.22 &  0.24 &  0.24 &  0.26 &  0.26 \\
                positive regulation of neurogenesis & {\textcolor{red}{ 3.29}} & {\textcolor{red}{ 2.36}} &  0.36 &  0.39 &  0.75 &  0.08 &  0.11 &  0.13 &  0.15 \\
 neurological system process involved in regulation & {\textcolor{red}{ 3.10}} & {\textcolor{red}{ 2.88}} & {\textcolor{red}{ 2.10}} &  1.07 & {\textcolor{red}{ 1.46}} & {\textcolor{red}{ 1.74}} &  0.21 &  0.22 &  0.23 \\
    positive regulation of neuroblast proliferation & {\textcolor{red}{ 2.97}} &  0.71 &  0.18 &  0.20 &  0.22 &  0.23 &  0.24 &  0.26 &  0.30 \\
                        neurological system process & {\textcolor{red}{ 2.94}} &  1.10 & {\textcolor{red}{ 1.78}} & {\textcolor{red}{ 2.14}} &  1.11 &  0.84 &  0.43 &  0.98 &  1.20 \\
                        synaptic vesicle exocytosis & {\textcolor{red}{ 2.86}} & {\textcolor{red}{ 3.03}} & {\textcolor{red}{ 3.62}} & {\textcolor{red}{ 4.31}} & {\textcolor{red}{ 2.22}} &  0.17 &  0.19 &  0.19 &  0.19 \\
                         regulation of neurogenesis & {\textcolor{red}{ 2.81}} & {\textcolor{red}{ 1.57}} &  0.85 &  0.60 &  0.96 &  0.36 &  0.68 &  0.53 &  0.89 \\
     establishment of synaptic vesicle localization & {\textcolor{red}{ 2.78}} & {\textcolor{red}{ 2.03}} & {\textcolor{red}{ 2.70}} & {\textcolor{red}{ 3.66}} & {\textcolor{red}{ 1.85}} &  0.15 &  0.17 &  0.18 &  0.18 \\
                      synaptic vesicle localization & {\textcolor{red}{ 2.78}} & {\textcolor{red}{ 2.03}} & {\textcolor{red}{ 2.70}} & {\textcolor{red}{ 3.66}} & {\textcolor{red}{ 1.85}} &  0.15 &  0.17 &  0.18 &  0.18 \\
                         synaptic vesicle transport & {\textcolor{red}{ 2.78}} & {\textcolor{red}{ 2.03}} & {\textcolor{red}{ 2.70}} & {\textcolor{red}{ 3.66}} & {\textcolor{red}{ 1.85}} &  0.15 &  0.17 &  0.18 &  0.18 \\
      positive regulation of neuron differentiation & {\textcolor{red}{ 2.63}} &  0.99 &  0.47 &  0.34 &  0.54 &  0.87 &  1.27 &  0.16 &  0.20 \\
                                       axonogenesis & {\textcolor{red}{ 2.53}} & {\textcolor{red}{ 1.83}} & {\textcolor{red}{ 2.41}} &  1.03 & {\textcolor{red}{ 1.73}} &  1.30 &  1.03 &  1.03 &  0.85 \\
 cell morphogenesis involved in neuron differentiat & {\textcolor{red}{ 2.52}} & {\textcolor{red}{ 2.18}} & {\textcolor{red}{ 3.19}} & {\textcolor{red}{ 1.44}} & {\textcolor{red}{ 2.42}} & {\textcolor{red}{ 1.67}} & {\textcolor{red}{ 1.45}} & {\textcolor{red}{ 1.53}} & {\textcolor{red}{ 1.51}} \\
                              generation of neurons & {\textcolor{red}{ 2.31}} & {\textcolor{red}{ 1.98}} & {\textcolor{red}{ 2.54}} & {\textcolor{red}{ 1.46}} & {\textcolor{red}{ 2.06}} &  1.07 &  0.98 &  1.06 & {\textcolor{red}{ 1.36}} \\
                                   axon development & {\textcolor{red}{ 2.25}} & {\textcolor{red}{ 1.70}} & {\textcolor{red}{ 2.29}} &  0.99 & {\textcolor{red}{ 1.68}} &  1.26 &  0.99 &  1.00 &  0.83 \\
                positive regulation of axonogenesis & {\textcolor{red}{ 2.17}} & {\textcolor{red}{ 3.36}} &  0.64 &  0.43 &  0.64 &  0.13 &  0.15 &  0.18 &  0.19 \\
                               axonal fasciculation & {\textcolor{red}{ 2.07}} &  0.17 &  0.21 &  0.24 &  0.30 &  0.30 &  0.30 &  0.30 &  0.30 \\
                                 neuron development & {\textcolor{red}{ 2.04}} & {\textcolor{red}{ 1.73}} & {\textcolor{red}{ 2.73}} & {\textcolor{red}{ 2.08}} & {\textcolor{red}{ 2.62}} & {\textcolor{red}{ 1.32}} &  0.95 &  1.25 &  1.30 \\
 central nervous system projection neuron axonogene & {\textcolor{red}{ 2.04}} & {\textcolor{red}{ 1.76}} & {\textcolor{red}{ 2.94}} &  0.16 &  0.20 &  0.21 &  0.23 &  0.25 &  0.25 \\
                                       neurogenesis & {\textcolor{red}{ 2.01}} & {\textcolor{red}{ 1.64}} & {\textcolor{red}{ 2.38}} & {\textcolor{red}{ 1.57}} & {\textcolor{red}{ 2.42}} &  0.90 &  0.86 &  0.92 &  1.16 \\
                    neuron projection morphogenesis & {\textcolor{red}{ 1.91}} & {\textcolor{red}{ 1.53}} & {\textcolor{red}{ 2.23}} &  1.13 & {\textcolor{red}{ 1.96}} &  1.18 &  0.92 &  0.90 &  0.72 \\
         central nervous system neuron axonogenesis & {\textcolor{red}{ 1.88}} & {\textcolor{red}{ 1.85}} & {\textcolor{red}{ 3.40}} &  0.12 &  0.16 &  0.17 &  0.19 &  0.22 &  0.23 \\
                            neurotransmitter uptake & {\textcolor{red}{ 1.87}} &  0.41 &  0.12 &  0.15 &  0.17 &  0.18 &  0.20 &  0.22 &  0.24 \\
                      neuron projection development & {\textcolor{red}{ 1.81}} &  1.17 & {\textcolor{red}{ 2.33}} & {\textcolor{red}{ 1.57}} & {\textcolor{red}{ 1.84}} &  1.02 &  0.90 &  1.02 &  0.93 \\
                                      axon guidance & {\textcolor{red}{ 1.81}} &  0.83 & {\textcolor{red}{ 1.64}} &  1.11 & {\textcolor{red}{ 1.63}} & {\textcolor{red}{ 2.05}} & {\textcolor{red}{ 2.18}} & {\textcolor{red}{ 1.80}} & {\textcolor{red}{ 1.58}} \\
                         neuron projection guidance & {\textcolor{red}{ 1.81}} &  0.83 & {\textcolor{red}{ 1.64}} &  1.11 & {\textcolor{red}{ 1.63}} & {\textcolor{red}{ 2.05}} & {\textcolor{red}{ 2.18}} & {\textcolor{red}{ 1.80}} & {\textcolor{red}{ 1.58}} \\
                                           synapsis & {\textcolor{red}{ 1.69}} &  1.04 & {\textcolor{red}{ 1.97}} & {\textcolor{red}{ 3.19}} &  0.18 &  0.20 &  0.23 &  0.24 &  0.25 \\
                              synaptic transmission & {\textcolor{red}{ 1.64}} &  0.51 &  0.40 &  0.77 &  0.44 &  0.26 &  0.31 &  0.59 &  0.44 \\
          regulation of neurological system process & {\textcolor{red}{ 1.63}} &  0.90 &  1.06 & {\textcolor{red}{ 1.37}} &  0.53 &  0.45 &  0.78 &  1.06 &  0.56 \\
                             neuron differentiation & {\textcolor{red}{ 1.59}} & {\textcolor{red}{ 1.40}} & {\textcolor{red}{ 2.04}} & {\textcolor{red}{ 1.73}} & {\textcolor{red}{ 2.20}} & {\textcolor{red}{ 1.34}} &  1.21 &  1.25 & {\textcolor{red}{ 1.53}} \\
 positive regulation of neurological system process & {\textcolor{red}{ 1.48}} &  0.61 &  0.67 &  0.46 &  0.11 &  0.13 &  0.17 &  0.19 &  0.20 \\
 \hline
 \end{tabular}}
  \caption{
    Here we list some of the pathways from the {\tt go\_bp} ontology.
    Shown here are only the 32 most significant pathways as determined by $\kappa(\imin,l)$.
    Each pathway is listed alongside approximations to its individual over-representation p-value (estimated using the hypergeometric-distribution).
    The $-\log_{10}(p)$-values are listed for iterations $\imin$-$\imax$ (see top row).
    Those annotations with an individual over-representation p-value smaller than $0.05$ are marked in red.
  }
  \label{tab:genri_neu}
\end{table}

\subsection*{Secondary bicluster:\ }
After discovering and analyzing the primary bicluster within arm-1 (described above), we searched for a secondary bicluster.
We first eliminated the structure associated with the primary bicluster by scrambling the entries of the submatrix $D(\imin)$ (see \cite{Rangan_2018} for details).
We then reran our half-loop algorithm on this scrambled version of arm-1.
While we did find a secondary trace that was indicative of heterogeneity, the overall level of differential-expression was far lower than for the first bicluster (see Fig 25 in S1 Text).
Moreover, the structure associated with this secondary trace did not significantly replicate (see Figs 26 -- 28 in S1 Text).
It is possible that a secondary bicluster exists, but that we could not pinpoint it due to a lack of power in our training-arm.
It is also possible that the scrambled version of arm-1 is heterogeneous, but not in a way that can be described by a bicluster (see \cite{Rangan_2018} for examples along these lines).
In either case, a larger sample size will be required to further probe this residual heterogeneity.

\subsection*{Control biclusters:\ }

Up to this point we have only considered biclusters within the case-population; i.e., subsets of case-subjects which exhibit a genetic-signature that is not shared by the control-subjects.
It is natural to ask if there are also biclusters that exist within the control-population (i.e., whether or not the control-population is homogeneous).
Such `control-biclusters' might be induced by batch effects or issues associated with recruitment; e.g., many of the BD controls may be drawn from another disease study (such as cancer), thus being more likely to share certain genetic features.
It might also be the case that some of the control-biclusters are biologically significant, corresponding to mechanisms which protect against the disease.
In either scenario, a better understanding of the heterogeneity within the control-population can assist in designing homogeneous populations of controls for future studies.

We can easily carry out this analysis simply by reversing the labels within our biclustering algorithm (i.e., swapping $D$ and $X$).
This reversed search will find biclusters that are driven by genetic-signatures which are more prevalent within the controls than within the cases.
As mentioned above, we find that the control-population within arm-1 is quite homogeneous: the trace decays monotonically with no distinguished peaks (see Fig 29 in S1 Text).
This homogeneity can be viewed as a validation of our initial choice of arm-1 as a training- or discovery-arm.

On the other hand, we find strong evidence for heterogeneity within the control-populations of arms 2, 3 and 4 (see Figs 30 -- 32 in S1 Text).
In each case the trace has a significant distinguished maximum involving only a fraction of the control-subjects (i.e,. $13\%$, $28\%$ and $15\%$ of the controls, respectively).

The heterogeneity observed in the control-populations of arms 2, 3 and 4 might be expected; each of these arms comprises multiple smaller studies.
Notably however, the `control-biclusters' within these arms cannot all be easily dismissed as batch-effects.
Indeed, each of the dominant control-biclusters is also quite significant, while also usually well balanced across the ancestry-related covariates and individual cohorts within each arm.
Each of these dominant control-biclusters also replicates across the majority of other arms.

Thus, while a portion of these control-biclusters might be driven by batch-effects or other idiosyncrasies in the control-population, it is possible that that some of these signals have biological relevance, perhaps involving mechanisms which protect against BD (as the control-biclusters were identified specifically because they involved genetic patterns not as prevalent across the cases). 
Consequently, we would recommend considering this heterogeneity when performing other kinds of analysis. 
For example, one should not necessarily assume that the controls are homogeneous, as small subgroups of controls can likely exhibit genetic-signatures that are distinct from the rest. 


\section*{Conclusion}

In this paper we have taken a `genotype-driven' approach to investigating genotypic-heterogeneity.
That is to say, first we used only basic phenotypic classification to divide subjects into cases (BD) and controls (not BD).
We then applied a biclustering analysis to identify genetic subgroups within the case-population.
Analyzing the BDI and BDII cases together as a whole allowed us to identify a genetic subgroup (i.e., the  bicluster described above).
This bicluster involved a genetically homogeneous subset of the BD-cases within the training-arm, which we then used to inform a more robust PRS with better replication across studies.

Our results suggest two hypothesis for future work.
Most directly, our replication- and PRS-analyses indicate that the bicluster we found within the training-arm indeed represents a genetic subgroup of BD which generalizes across data-sets.
More generally, our results provide a proof-of-principle for our overall methodology: a data-driven approach to identifying genetically homogeneous subsets of case-subjects can help construct more robust PRSs, with the potential of improving SNP-replication in BD GWAS and, ultimately, a better understanding of the etiology of Bipolar Disorder.

In some respects our approach can be termed `unsupervised', as we did not use BD-subtype (BDI vs. BDII) or subphenotype information to guide our primary analysis.
This unsupervised approach allows us to circumvent many of the challenges associated with phenotype classification, such as missingness and variation in assessment and collection process (e.g., expert-led vs. self-report).
It also allows us to identify genetic patterns which straddle traditional classifications provided the signature is not present in the control group.
E.g., though our bicluster was enriched for BDI, it was by no means limited to BDI and included many BDII cases.

Along these lines, we believe that a similar unsupervised approach could be used to search for interactions between the signals we have found and other diseases, as well as for cross-psychiatric-disorder signals not present in the control group.
There are many examples of genetic interactions along these lines: the SNPs driving BD have a strong correlation with those driving schizophrenia, and also share overlap with the SNPs driving MDD, OCD, anorexia nervosa, ADHD, ASD and substance-abuse \cite{Kranzler_2019,CrossDisorderGroupofthePsychiatricGenomicsConsortium_2019,Jang_2020}.
Many SNPs have also been associated with other disorders \cite{vanHulzen_2017,OConnell_2018,BipolarDisorderandSchizophreniaWorkingGroupofthePsychiatricGenomicsConsortium_2018,Coleman_2020}.
More generally speaking, BD shows substantial overlap with other disorders; e.g., more than $90\%$ of BD subjects exhibit lifetime comorbidity \cite{Merikangas_2007} with at least one other psychiatric disorder \cite{Frias_2016,Salloum_2017,Eser_2018}, or non-psychiatric disorder \cite{Bortolato_2016,Correll_2017,Vancampfort_2016}.
This high rate of comorbidity implies that BD is one of multiple disorders which perturb several important regulatory systems \cite{RoshanaeiMoghaddam_2009,Kessing_2015}.
Given these relationships, it is possible that the bicluster-score and/or the bicluster-limited PRSs may also correlate with some of the signals of these other disorders.
It is possible that we could discover interesting biclusters which cross psychiatric disorders or are present in the control groups and predict resistance to psychiatric illness more generally; we defer an investigation of these interactions to future work.

The biclustering algorithm we use also offers a `supervised' option which uses additional information (e.g., BD-subtype or other clinical data) to subdivide the case-population while searching for heterogeneity.
Sex might be one important variable to include in such a supervised BD analysis.
For example, while most studies do not indicate large difference in BD prevalence between men and women (indeed, the bicluster we identified was not significantly enriched for sex), there is some evidence of a sex disparity in the prevalence of BDII, rapid-cycling and mixed-episodes \cite{Diflorio_2010,Nivoli_2011}.
Age may also be an important role-player, as an earlier age of onset may be associated with higher severity and a poorer long-term prognosis (possibly due to mis-diagnoses at an early stage) \cite{Zimmerman_2008,Joslyn_2016}.

One limitation of our current study is that it is restricted to common variants (i.e., SNPs with a high minor-allele-frequency).
While it is encouraging that the common variants alone can be used to find replicable and robust signals, it is also likely that the rare variants also play a role in the heterogeneity of BD.
Analyzing the rare variants brings new challenges, as rare variants often require more statistical power to detected and/or validate \cite{Goes_2016,Maaser_2018,Toma_2018,Goes_2019,Forstner_2020,Sul_2020}.

Another more serious limitation is that our training-arm is quite restricted in terms of ancestry.
More generally, almost all the individuals in our data-set are of European descent.
We expect that this lack of diversity will limit our ability to pinpoint the most biologically relevant signals, as many previous GWAS analyses have not generalized well to cohorts of different ancestry \cite{Akinhanmi_2018,Martin_2019,Peterson_2019,Sirugo_2019,Duncan_2019,Mullins_2020}.
An important future direction will be to investigate the interactions between genotypic heterogeneity and ancestry.

We do not expect a full analysis of genetic-heterogeneity to be entirely trivial.
For example, appropriately correcting for ancestry is not always easy, even when searching for homogeneous signals.
When searching for heterogeneity such a correction becomes more complicated and, necessarily, involves more parameters.
Larger (and more diverse) sample sizes will likely be necessary to clarify (i) the disease-specific genetic-subgroups (i.e., biclusters) within BD, as well as (ii) the phenotypic subtypes of BD, and perhaps most importantly: (iii) the interaction between these subgroups and subtypes and other covariates such as ancestry.
We suspect that a careful treatment of the associated statistical issues will pose a significant challenge.
Nevertheless, these advancements will likely further improve our understanding of the etiology of BD.

\section*{Declarations}

\subsection*{Ethics approval and consent to participate}
Not applicable.
\subsection*{Consent for publication}
Not applicable.
\subsection*{Availability of data and materials}
There are no primary data in this paper; all of the code is available at {\tt https://github.com/adirangan/}.
The data which is analyzed can be requested from the Psychiatric Genomics Consortium at: {\tt https://pgc.unc.edu/}.
\subsection*{Competing interests}
The authors declare that they have no competing interests.
\subsection*{Authors' contributions}
CM, AR, AF, JE, JS and JK analyzed the data and contributed to writing the manuscript.
All authors read and approved the final manuscript.
\subsection*{Acknowledgements}
\noindent\makebox[\linewidth]{\rule{5.25in}{0.4pt}}
The chair of the Bipolar Disorder Working Group of the Psychiatric Genomics Consortium is currently Ole A Andreassen (contact email: {\tt o.a.andreassen@medisin.uio.no }).
The contributors to the Bipolar Disorder Working Group are listed below.
\paragraph*{Bipolar Disorder Working Group of the Psychiatric Genomics Consortium:}
{\normalsize Eli A Stahl}\textsuperscript{1,2,3},
{\normalsize Gerome Breen}\textsuperscript{4,5},
{\normalsize Andreas J Forstner}\textsuperscript{6,7,8,9,10},
{\normalsize Andrew McQuillin}\textsuperscript{11},
{\normalsize Stephan Ripke}\textsuperscript{12,13,14},
{\normalsize Vassily Trubetskoy}\textsuperscript{13},
{\normalsize Manuel Mattheisen}\textsuperscript{15,16,17,18,19},
{\normalsize Yunpeng Wang}\textsuperscript{20,21},
{\normalsize Jonathan R I Coleman}\textsuperscript{4,5},
{\normalsize Héléna A Gaspar}\textsuperscript{4,5}
% ...
\noindent\makebox[\linewidth]{\rule{5.25in}{0.4pt}}
\textsuperscript{1} {\small Department of Genetics and Genomic Sciences, Icahn School of Medicine at Mount Sinai, New York, NY, US}.\ \textsuperscript{2} {\small Department of Psychiatry, Icahn School of Medicine at Mount Sinai, New York, NY, US}.\ \textsuperscript{3} {\small Medical and Population Genetics, Broad Institute, Cambridge, MA, US}.\ \textsuperscript{4} {\small MRC Social, Genetic and Developmental Psychiatry Centre, King's College London, London, GB}.\ \textsuperscript{5} {\small NIHR BRC for Mental Health, King's College London, London, GB}.\ \textsuperscript{6} {\small Department of Biomedicine, University of Basel, Basel, CH}.\ \textsuperscript{7} {\small Department of Psychiatry (UPK), University of Basel, Basel, CH}.\ \textsuperscript{8} {\small Institute of Human Genetics, University of Bonn, School of Medicine \& University Hospital Bonn, Bonn, DE}.\ \textsuperscript{9} {\small Centre for Human Genetics, University of Marburg, Marburg, DE}.\ \textsuperscript{10} {\small Institute of Medical Genetics and Pathology, University Hospital Basel, Basel, CH}.\ \textsuperscript{11} {\small Division of Psychiatry, University College London, London, GB}.\ \textsuperscript{12} {\small Stanley Center for Psychiatric Research, Broad Institute, Cambridge, MA, US}.\ \textsuperscript{13} {\small Department of Psychiatry and Psychotherapy, Charité - Universitätsmedizin, Berlin, DE}.\ \textsuperscript{14} {\small Analytic and Translational Genetics Unit, Massachusetts General Hospital, Boston, MA, US}.\ \textsuperscript{15} {\small iSEQ, Center for Integrative Sequencing, Aarhus University, Aarhus, DK}.\ \textsuperscript{16} {\small Department of Biomedicine - Human Genetics, Aarhus University, Aarhus, DK}.\ \textsuperscript{17} {\small Department of Clinical Neuroscience, Centre for Psychiatry Research, Karolinska Institutet, Stockholm, SE}.\ \textsuperscript{18} {\small Department of Psychiatry, Psychosomatics and Psychotherapy, Center of Mental Health, University Hospital Würzburg, Würzburg, DE}.\ \textsuperscript{19} {\small iPSYCH, The Lundbeck Foundation Initiative for Integrative Psychiatric Research, DK}.\ \textsuperscript{20} {\small Institute of Biological Psychiatry, Mental Health Centre Sct. Hans, Copenhagen, DK}.\ \textsuperscript{21} {\small Institute of Clinical Medicine, University of Oslo, Oslo, NO}.\ \textsuperscript{22} {\small Department of Complex Trait Genetics, Center for Neurogenomics and Cognitive Research, Amsterdam Neuroscience, Vrije Universiteit Amsterdam, Amsterdam, NL}.\ \textsuperscript{23} {\small deCODE Genetics / Amgen, Reykjavik, IS}.\ \textsuperscript{24} {\small Queensland Brain Institute, The University of Queensland, Brisbane, QLD, AU}.\ \textsuperscript{25} {\small Institute for Molecular Bioscience, The University of Queensland, Brisbane, QLD, AU}.\ \textsuperscript{26} {\small Division of Endocrinology and Center for Basic and Translational Obesity Research, Boston Children’s Hospital, Boston, MA, US}.\ \textsuperscript{27} {\small Medical Research Council Centre for Neuropsychiatric Genetics and Genomics, Division of Psychological Medicine and Clinical Neurosciences, Cardiff University, Cardiff, GB}.\ \textsuperscript{28} {\small National Centre for Register-Based Research, Aarhus University, Aarhus, DK}.\ \textsuperscript{29} {\small Centre for Integrated Register-based Research, Aarhus University, Aarhus, DK}.\ \textsuperscript{30} {\small Molecular \& Behavioral Neuroscience Institute, University of Michigan, Ann Arbor, MI, US}.\ \textsuperscript{31} {\small Department of Neuroscience, IRCCS - Istituto Di Ricerche Farmacologiche Mario Negri, Milan, IT}.\ \textsuperscript{32} {\small Department of Psychiatry and Behavioral Neuroscience, University of Chicago, Chicago, IL, US}.\ \textsuperscript{33} {\small Psychiatry, Berkshire Healthcare NHS Foundation Trust, Bracknell, GB}.\ \textsuperscript{34} {\small Psychiatry, Rush University Medical Center, Chicago, IL, US}.\ \textsuperscript{35} {\small Center for Neonatal Screening, Department for Congenital Disorders, Statens Serum Institut, Copenhagen, DK}.\ \textsuperscript{36} {\small Department of Psychiatry, Weill Cornell Medical College, New York, NY, US}.\ \textsuperscript{37} {\small Department of Psychiatry and Psychotherapy, University Hospital Carl Gustav Carus, Technische Universität Dresden, Dresden, DE}.\ \textsuperscript{38} {\small Department of Medical Epidemiology and Biostatistics, Karolinska Institutet, Stockholm, SE}.\ \textsuperscript{39} {\small Department of Psychiatric Research, Diakonhjemmet Hospital, Oslo, NO}.\ \textsuperscript{40} {\small Psychiatry, UMC Utrecht Brain Center Rudolf Magnus, Utrecht, NL}.\ \textsuperscript{41} {\small Human Genetics, University of California Los Angeles, Los Angeles, CA, US}.\ \textsuperscript{42} {\small Institute of Psychiatric Phenomics and Genomics (IPPG), University Hospital, LMU Munich, Munich, DE}.\ \textsuperscript{43} {\small Department of Psychiatry and Human Behavior, University of California, Irvine, Irvine, CA, US}.\ \textsuperscript{44} {\small Molecular \& Behavioral Neuroscience Institute and Department of Computational Medicine \& Bioinformatics, University of Michigan, Ann Arbor, MI, US}.\ \textsuperscript{45} {\small Psychiatry, University of California San Francisco, San Francisco, CA, US}.\ \textsuperscript{46} {\small Instituto de Salud Carlos III, Biomedical Network Research Centre on Mental Health (CIBERSAM), Madrid, ES}.\ \textsuperscript{47} {\small Department of Psychiatry, Hospital Universitari Vall d´Hebron, Barcelona, ES}.\ \textsuperscript{48} {\small Department of Psychiatry and Forensic Medicine, Universitat Autònoma de Barcelona, Barcelona, ES}.\ \textsuperscript{49} {\small Psychiatric Genetics Unit, Group of Psychiatry Mental Health and Addictions, Vall d´Hebron Research Institut (VHIR), Universitat Autònoma de Barcelona, Barcelona, ES}.\ \textsuperscript{50} {\small Department of Psychiatry, Mood Disorders Program, McGill University Health Center, Montreal, QC, CA}.\ \textsuperscript{51} {\small Division of Psychiatry, University of Edinburgh, Edinburgh, GB}.\ \textsuperscript{52} {\small University of Iowa Hospitals and Clinics, Iowa City, IA, US}.\ \textsuperscript{53} {\small Translational Genomics, USC, Phoenix, AZ, US}.\ \textsuperscript{54} {\small Department of Translational Research in Psychiatry, Max Planck Institute of Psychiatry, Munich, DE}.\ \textsuperscript{55} {\small Centre for Psychiatry, Queen Mary University of London, London, GB}.\ \textsuperscript{56} {\small UCL Genetics Institute, University College London, London, GB}.\ \textsuperscript{57} {\small Department of Psychiatry, Laboratory of Psychiatric Genetics, Poznan University of Medical Sciences, Poznan, PL}.\ \textsuperscript{58} {\small Department of Neurosciences, University of California San Diego, La Jolla, CA, US}.\ \textsuperscript{59} {\small Department of Radiology, University of California San Diego, La Jolla, CA, US}.\ \textsuperscript{60} {\small Department of Psychiatry, University of California San Diego, La Jolla, CA, US}.\ \textsuperscript{61} {\small Department of Cognitive Science, University of California San Diego, La Jolla, CA, US}.\ \textsuperscript{62} {\small Applied Molecular Genomics Unit, VIB Department of Molecular Genetics, University of Antwerp, Antwerp, Belgium}.\ \textsuperscript{63} {\small Department of Psychiatry and Behavioral Sciences, Johns Hopkins University School of Medicine, Baltimore, MD, US}.\ \textsuperscript{64} {\small Department of Medical Genetics, Oslo University Hospital Ullevål, Oslo, NO}.\ \textsuperscript{65} {\small NORMENT, KG Jebsen Centre for Psychosis Research, Department of Clinical Science, University of Bergen, Bergen, NO}.\ \textsuperscript{66} {\small Department of Neurology, Oslo University Hospital, Oslo, NO}.\ \textsuperscript{67} {\small NORMENT, KG Jebsen Centre for Psychosis Research, Oslo University Hospital, Oslo, NO}.\ \textsuperscript{68} {\small Center for Statistical Genetics and Department of Biostatistics, University of Michigan, Ann Arbor, MI, US}.\ \textsuperscript{69} {\small Department of Medical \& Molecular Genetics, Indiana University, Indianapolis, IN, US}.\ \textsuperscript{70} {\small Department of Genetic Epidemiology in Psychiatry, Central Institute of Mental Health, Medical Faculty Mannheim, Heidelberg University, Mannheim, DE}.\ \textsuperscript{71} {\small Center for Neurobehavioral Genetics, University of California Los Angeles, Los Angeles, CA, US}.\ \textsuperscript{72} {\small Department of Molecular Medicine and Surgery, Karolinska Institutet and Center for Molecular Medicine, Karolinska University Hospital, Stockholm, SE}.\ \textsuperscript{73} {\small Department of Clinical Neuroscience, Karolinska Institutet and Center for Molecular Medicine, Karolinska University Hospital, Stockholm, SE}.\ \textsuperscript{74} {\small Child and Adolescent Psychiatry Research Center, Stockholm, SE}.\ \textsuperscript{75} {\small Department of Psychiatry and Psychotherapy, University Medical Center Göttingen, Göttingen, DE}.\ \textsuperscript{76} {\small Department of Psychiatry, Dalhousie University, Halifax, NS, CA}.\ \textsuperscript{77} {\small Genetics and Computational Biology, QIMR Berghofer Medical Research Institute, Brisbane, QLD, AU}.\ \textsuperscript{78} {\small Department of Psychological Medicine, University of Worcester, Worcester, GB}.\ \textsuperscript{79} {\small School of Biomedical Sciences, Plymouth University Peninsula Schools of Medicine and Dentistry, University of Plymouth, Plymouth, GB}.\ \textsuperscript{80} {\small School of Psychiatry, University of New South Wales, Sydney, NSW, AU}.\ \textsuperscript{81} {\small Bioinformatics Research Centre, Aarhus University, Aarhus, DK}.\ \textsuperscript{82} {\small Biostatistics, University of Minnesota System, Minneapolis, MN, US}.\ \textsuperscript{83} {\small Mental Health Department, University Regional Hospital, Biomedicine Institute (IBIMA), Málaga, ES}.\ \textsuperscript{84} {\small Department of Psychology, Eberhard Karls Universität Tübingen, Tubingen, DE}.\ \textsuperscript{85} {\small Department of Psychiatry and Behavioral Sciences, Howard University Hospital, Washington, DC, US}.\ \textsuperscript{86} {\small Center for Multimodal Imaging and Genetics, University of California San Diego, La Jolla, CA, US}.\ \textsuperscript{87} {\small Psychiatrie Translationnelle, Inserm U955, Créteil, FR}.\ \textsuperscript{88} {\small Faculté de Médecine, Université Paris Est, Créteil, FR}.\ \textsuperscript{89} {\small Campbell Family Mental Health Research Institute, Centre for Addiction and Mental Health, Toronto, ON, CA}.\ \textsuperscript{90} {\small Neurogenetics Section, Centre for Addiction and Mental Health, Toronto, ON, CA}.\ \textsuperscript{91} {\small Department of Psychiatry, University of Toronto, Toronto, ON, CA}.\ \textsuperscript{92} {\small Institute of Medical Sciences, University of Toronto, Toronto, ON, CA}.\ \textsuperscript{93} {\small Department of Psychiatry, Psychosomatic Medicine and Psychotherapy, University Hospital Frankfurt, Frankfurt am Main, DE}.\ \textsuperscript{94} {\small Cell Biology, SUNY Downstate Medical Center College of Medicine, Brooklyn, NY, US}.\ \textsuperscript{95} {\small Institute for Genomic Health, SUNY Downstate Medical Center College of Medicine, Brooklyn, NY, US}.\ \textsuperscript{96} {\small ISGlobal, Barcelona, ES}.\ \textsuperscript{97} {\small Psychiatry, Altrecht, Utrecht, NL}.\ \textsuperscript{98} {\small Psychiatry, GGZ inGeest, Amsterdam, NL}.\ \textsuperscript{99} {\small Psychiatry, VU medisch centrum, Amsterdam, NL}.\ \textsuperscript{100} {\small Psychiatry, North East London NHS Foundation Trust, Ilford, GB}.\ \textsuperscript{101} {\small Clinic for Psychiatry and Psychotherapy, University Hospital Cologne, Cologne, DE}.\ \textsuperscript{102} {\small Psychiatric and Neurodevelopmental Genetics Unit, Massachusetts General Hospital, Boston, MA, US}.\ \textsuperscript{103} {\small HudsonAlpha Institute for Biotechnology, Huntsville, AL, US}.\ \textsuperscript{104} {\small Department of Human Genetics, University of Michigan, Ann Arbor, MI, US}.\ \textsuperscript{105} {\small Psychiatry, University of Illinois at Chicago College of Medicine, Chicago, IL, US}.\ \textsuperscript{106} {\small Max Planck Institute of Psychiatry, Munich, DE}.\ \textsuperscript{107} {\small Mental Health, NHS 24, Glasgow, GB}.\ \textsuperscript{108} {\small Division of Psychiatry, Centre for Clinical Brain Sciences, University of Edinburgh, Edinburgh, GB}.\ \textsuperscript{109} {\small Psychiatry, Brigham and Women's Hospital, Boston, MA, US}.\ \textsuperscript{110} {\small Department of Psychiatry and Psychotherapy, University of Bonn, Bonn, DE}.\ \textsuperscript{111} {\small Department of Genetics, Harvard Medical School, Boston, MA, US}.\ \textsuperscript{112} {\small Department of Psychiatry, University of Michigan, Ann Arbor, MI, US}.\ \textsuperscript{113} {\small Genetic Cancer Susceptibility Group, International Agency for Research on Cancer, Lyon, FR}.\ \textsuperscript{114} {\small Estonian Genome Center, University of Tartu, Tartu, EE}.\ \textsuperscript{115} {\small Discipline of Biochemistry, Neuroimaging and Cognitive Genomics (NICOG) Centre, National University of Ireland, Galway, Galway, IE}.\ \textsuperscript{116} {\small Neuropsychiatric Genetics Research Group, Dept of Psychiatry and Trinity Translational Medicine Institute, Trinity College Dublin, Dublin, IE}.\ \textsuperscript{117} {\small Institute of Neuroscience and Medicine (INM-1), Research Centre Jülich, Jülich, DE}.\ \textsuperscript{118} {\small Research/Psychiatry, Veterans Affairs San Diego Healthcare System, San Diego, CA, US}.\ \textsuperscript{119} {\small Department of Clinical Sciences, Psychiatry, Umeå University Medical Faculty, Umeå, SE}.\ \textsuperscript{120} {\small Department of Clinical Psychiatry, Psychiatry Clinic, Clinical Center University of Sarajevo, Sarajevo, BA}.\ \textsuperscript{121} {\small Department of Neurobiology, Care sciences, and Society, Karolinska Institutet and Center for Molecular Medicine, Karolinska University Hospital, Stockholm, SE}.\ \textsuperscript{122} {\small Psychiatry, Harvard Medical School, Boston, MA, US}.\ \textsuperscript{123} {\small Division of Clinical Research, Massachusetts General Hospital, Boston, MA, US}.\ \textsuperscript{124} {\small Outpatient Clinic for Bipolar Disorder, Altrecht, Utrecht, NL}.\ \textsuperscript{125} {\small Department of Psychiatry, Washington University in Saint Louis, Saint Louis, MO, US}.\ \textsuperscript{126} {\small Department of Biochemistry and Molecular Biology II, Institute of Neurosciences, Center for Biomedical Research, University of Granada, Granada, ES}.\ \textsuperscript{127} {\small Department of Neuroscience, Icahn School of Medicine at Mount Sinai, New York, NY, US}.\ \textsuperscript{128} {\small Medicine, Psychiatry, Biomedical Informatics, Vanderbilt University Medical Center, Nashville, TN, US}.\ \textsuperscript{129} {\small Department of Health Sciences Research, Mayo Clinic, Rochester, MN, US}.\ \textsuperscript{130} {\small Psychiatry and Behavioral Sciences, Stanford University School of Medicine, Stanford, CA, US}.\ \textsuperscript{131} {\small Rush University Medical Center, Chicago, IL, US}.\ \textsuperscript{132} {\small Scripps Translational Science Institute, La Jolla, CA, US}.\ \textsuperscript{133} {\small Neuroscience Research Australia, Sydney, NSW, AU}.\ \textsuperscript{134} {\small Faculty of Medicine, Department of Psychiatry, School of Health Sciences, University of Iceland, Reykjavik, IS}.\ \textsuperscript{135} {\small Division of Mental Health and Addiction, Oslo University Hospital, Oslo, NO}.\ \textsuperscript{136} {\small NORMENT, University of Oslo, Oslo, NO}.\ \textsuperscript{137} {\small Psychiatry and the Behavioral Sciences, University of Southern California, Los Angeles, CA, US}.\ \textsuperscript{138} {\small Mood Disorders, PsyQ, Rotterdam, NL}.\ \textsuperscript{139} {\small Institute for Medical Sciences, University of Aberdeen, Aberdeen, UK}.\ \textsuperscript{140} {\small Research Division, Federal Institute for Drugs and Medical Devices (BfArM), Bonn, DE}.\ \textsuperscript{141} {\small Centre for Addiction and Mental Health, Toronto, ON, CA}.\ \textsuperscript{142} {\small Neurogenomics, TGen, Los Angeles, AZ, US}.\ \textsuperscript{143} {\small Psychiatry, Psychiatrisches Zentrum Nordbaden, Wiesloch, DE}.\ \textsuperscript{144} {\small Computational Sciences Center of Emphasis, Pfizer Global Research and Development, Cambridge, MA, US}.\ \textsuperscript{145} {\small Department of Biostatistics, Princess Margaret Cancer Centre, Toronto, ON, CA}.\ \textsuperscript{146} {\small Dalla Lana School of Public Health, University of Toronto, Toronto, ON, CA}.\ \textsuperscript{147} {\small Psychological Medicine, Institute of Psychiatry, Psychology \& Neuroscience, King's College London, London, GB}.\ \textsuperscript{148} {\small Department of Mental Health, Johns Hopkins University Bloomberg School of Public Health, Baltimore, MD, US}.\ \textsuperscript{149} {\small Institute of Genetic Medicine, Johns Hopkins University School of Medicine, Baltimore, MD, US}.\ \textsuperscript{150} {\small NORMENT, KG Jebsen Centre for Psychosis Research, Division of Mental Health and Addiction, Institute of Clinical Medicine and Diakonhjemmet Hospital, University of Oslo, Oslo, NO}.\ \textsuperscript{151} {\small National Institute of Mental Health, Klecany, CZ}.\ \textsuperscript{152} {\small Department of Psychiatry, University of Melbourne, Melbourne, Victoria, AU}.\ \textsuperscript{153} {\small Department of Psychiatry and Addiction Medicine, Assistance Publique - Hôpitaux de Paris, Paris, FR}.\ \textsuperscript{154} {\small Paris Bipolar and TRD Expert Centres, FondaMental Foundation, Paris, FR}.\ \textsuperscript{155} {\small UMR-S1144 Team 1: Biomarkers of relapse and therapeutic response in addiction and mood disorders, INSERM, Paris, FR}.\ \textsuperscript{156} {\small Psychiatry, Université Paris Diderot, Paris, FR}.\ \textsuperscript{157} {\small Psychiatry, University of Pennsylvania, Philadelphia, PA, US}.\ \textsuperscript{158} {\small Department of Psychiatry, University of Münster, Münster, DE}.\ \textsuperscript{159} {\small Division of Endocrinology, Children's Hospital Boston, Boston, MA, US}.\ \textsuperscript{160} {\small Centre for Affective Disorders, Institute of Psychiatry, Psychology and Neuroscience, London, GB}.\ \textsuperscript{161} {\small Department of Psychiatry \& Psychology, Mayo Clinic, Rochester, MN, US}.\ \textsuperscript{162} {\small School of Medical Sciences, University of New South Wales, Sydney, NSW, AU}.\ \textsuperscript{163} {\small Department of Human Genetics, University of Chicago, Chicago, IL, US}.\ \textsuperscript{164} {\small Biometric Psychiatric Genetics Research Unit, Alexandru Obregia Clinical Psychiatric Hospital, Bucharest, RO}.\ \textsuperscript{165} {\small Institute of Neuroscience and Physiology, University of Gothenburg, Gothenburg, SE}.\ \textsuperscript{166} {\small INSERM, Paris, FR}.\ \textsuperscript{167} {\small Department of Medical \& Molecular Genetics, King's College London, London, GB}.\ \textsuperscript{168} {\small Neuroscience Therapeutic Area, Janssen Research and Development, LLC, Titusville, NJ, US}.\ \textsuperscript{169} {\small Cancer Epidemiology and Prevention, M. Sklodowska-Curie Cancer Center and Institute of Oncology, Warsaw, PL}.\ \textsuperscript{170} {\small School of Psychology, The University of Queensland, Brisbane, QLD, AU}.\ \textsuperscript{171} {\small Research Institute, Lindner Center of HOPE, Mason, OH, US}.\ \textsuperscript{172} {\small Centre for Cognitive Ageing and Cognitive Epidemiology, University of Edinburgh, Edinburgh, GB}.\ \textsuperscript{173} {\small Human Genetics Branch, Intramural Research Program, National Institute of Mental Health, Bethesda, MD, US}.\ \textsuperscript{174} {\small Division of Mental Health and Addiction, Oslo University Hospital, Oslo, NO}.\ \textsuperscript{175} {\small Division of Mental Health and Addiction, University of Oslo, Institute of Clinical Medicine, Oslo, NO}.\ \textsuperscript{176} {\small Institute of Molecular and Cell Biology, University of Tartu, Tartu, EE}.\ \textsuperscript{177} {\small Mental Health, Faculty of Medicine and Health Sciences, Norwegian University of Science and Technology - NTNU, Trondheim, NO}.\ \textsuperscript{178} {\small Psychiatry, St Olavs University Hospital, Trondheim, NO}.\ \textsuperscript{179} {\small Psychosis Research Unit, Aarhus University Hospital, Risskov, DK}.\ \textsuperscript{180} {\small Munich Cluster for Systems Neurology (SyNergy), Munich, DE}.\ \textsuperscript{181} {\small University of Liverpool, Liverpool, GB}.\ \textsuperscript{182} {\small Psychiatry and Human Genetics, University of Pittsburgh, Pittsburgh, PA, US}.\ \textsuperscript{183} {\small Mental Health Services in the Capital Region of Denmark, Mental Health Center Copenhagen, University of Copenhagen, Copenhagen, DK}.\ \textsuperscript{184} {\small Division of Psychiatry, Haukeland Universitetssjukehus, Bergen, NO}.\ \textsuperscript{185} {\small Faculty of Medicine and Dentistry, University of Bergen, Bergen, NO}.\ \textsuperscript{186} {\small Human Genetics and Computational Biomedicine, Pfizer Global Research and Development, Groton, CT, US}.\ \textsuperscript{187} {\small College of Medicine Institute for Genomic Health, SUNY Downstate Medical Center College of Medicine, Brooklyn, NY, US}.\ \textsuperscript{188} {\small Department of Clinical Genetics, Amsterdam Neuroscience, Vrije Universiteit Medical Center, Amsterdam, NL}.\ \textsuperscript{189} {\small Department of Neurology and Neurosurgery, McGill University, Faculty of Medicine, Montreal, QC, CA}.\ \textsuperscript{190} {\small Montreal Neurological Institute and Hospital, Montreal, QC, CA}.\ \textsuperscript{191} {\small Department of Biomedical and NeuroMotor Sciences, University of Bologna, Bologna, IT}.\ \textsuperscript{192} {\small Department of Psychiatry, Massachusetts General Hospital, Boston, MA, US}.\ \textsuperscript{193} {\small Psychiatric and Neurodevelopmental Genetics Unit (PNGU), Massachusetts General Hospital, Boston, MA, US}.\ \textsuperscript{194} {\small Faculty of Medicine, University of Iceland, Reykjavik, IS}.\ \textsuperscript{195} {\small Department of Psychiatry, Hospital Namsos, Namsos, NO}.\ \textsuperscript{196} {\small Department of Neuroscience, Norges Teknisk Naturvitenskapelige Universitet Fakultet for naturvitenskap og teknologi, Trondheim, NO}.\ \textsuperscript{197} {\small Department of Genetics, University of North Carolina at Chapel Hill, Chapel Hill, NC, US}.\ \textsuperscript{198} {\small Department of Psychiatry, University of North Carolina at Chapel Hill, Chapel Hill, NC, US}.\ \textsuperscript{199} {\small Department of Psychiatry, McGill University, Montreal, QC, CA}.\ \textsuperscript{200} {\small Dept of Psychiatry, Sankt Olavs Hospital Universitetssykehuset i Trondheim, Trondheim, NO}.\ \textsuperscript{201} {\small Clinical Institute of Neuroscience, Hospital Clinic, University of Barcelona, IDIBAPS, CIBERSAM, Barcelona, ES}.\ \textsuperscript{202} {\small Institute of Biological Psychiatry, MHC Sct. Hans, Mental Health Services Copenhagen, Roskilde, DK}.\ \textsuperscript{203} {\small Department of Clinical Medicine, University of Copenhagen, Copenhagen, DK}.\ \textsuperscript{204} {\small Psychiatry, Indiana University School of Medicine, Indianapolis, IN, US}.\ \textsuperscript{205} {\small Biochemistry and Molecular Biology, Indiana University School of Medicine, Indianapolis, IN, US}.\ \textsuperscript{206} {\small Department of Pathology and Laboratory Medicine, University of California Los Angeles, Los Angeles, CA, US}.\ \textsuperscript{\dag} deceased.\ 
\bigskip

\noindent\makebox[\linewidth]{\rule{5.25in}{0.4pt}}



\setcounter{figure}{12}

\part*{Supplementary Information (S1 Text)}

This part contains a detailed description of our methods, including an outline of the steps involved and the considerations we made along the way. Also contains the supporting figures referenced in the main text.

\section{Outline of analysis pipeline}
In this section we briefly outline the major steps of our analysis pipeline. This includes (1) our initial biclustering analysis of the training arm-1, (2) our subsequent replication study on replication arms 2-4, and (3) the following PRS analysis on replication arms 2-4.
\begin{enumerate}
\setcounter{enumi}{-1}
\item As a preliminary step, we requested individual level BD data from the PGC\footnote{Currently, no permissions are required beyond a PGC BD group approval. We initially requested access in November, 2013. See \text{\tt https://pgcdataaccess.formstack.com/forms/pgc\_data\_access\_bip}}.
The data-set we requested includes the 27 cohorts listed in Fig \ref{fig:arms}.
This dataset includes the raw genotyped and $1000$-genomes European reference panel imputed data for more than $18K$ case-subjects and $29K$ control-subjects, as well as the associated genome-wide principal-components associated with each subject and used as a proxy for genetic ancestry.
This data-set also includes metadata such as subtype information, which was not used to guide our analysis below.
\item In our primary analysis, we focused on the individual level raw genotyped data (i.e., we did not consider imputed data for the primary analysis).
\begin{enumerate}
\item First, we compared the available genotyped SNP lists from each cohort to divide the data into training and replication arms (see Fig \ref{fig:arms} in the main text).
\item We selected BDRN for our training arm-1. This was the largest single study available to us, containing $2524/4106$ cases/controls, and allowed for a large replication arm-2 with $5781/8289$ cases/controls sharing $\sim 85\%$ of the genotyped SNPs.
\item We used an F-test to determine which of the genome-wide principal-components to include as proxies for ancestry in our primary analysis of arm-1.
\item We then ran our biclustering algorithm on the raw genotyped data for the 2524 BD cases and 4106 controls in arm-1. We discuss this step in more detail within the next section. As a brief summary:
\begin{enumerate}
\item We included only the raw genotyped data with minor-allele-frequency (maf) $>25\%$, and we corrected for case-control status as well as the genome-wide principal-components identified by the F-test above.
\item We ran this algorithm with an `elimination-fraction' of $\gamma$ $=$ $0.5^{8}$ $\sim$ $0.004$, which is sufficient to ensure convergence (see supplementary Fig 32 in supplementary section 7.3 of \cite{Rangan_2018}).
\item We used a permutation test to determine the overall significance-level of the heterogeneity we observed. We found an overall significance level of $p\lesssim 1/64$, with the most prominent individual iterations in the range $i\in[\imin,\imax]$. 
\end{enumerate}
\end{enumerate}
\item After running our biclustering-algorithm on arm-1, we then used the structure of the bicluster to perform a replication-study in arms 2, 3 and 4.
\begin{enumerate}
\item For each replication-arm, we determined which SNPs from the bicluster at iteration $i$ were also present in that replication-arm. 
\item We then generated the dominant principal-component for the bicluster from arm-1 at iteration $i$, restricted to the SNPs in the intersection between arm-1 and the replication-arm. This principal-component is referred to as $v(i)$ in the main text. \item We then generated a `bicluster-score' for each case- and control-subject by projecting the arm-1 and replication-arm data onto this dominant principal-component $v(i)$. In the main text we refer to these bicluster-scores in arm-1 as $u_{j_{D}}(i)$ and $u_{j_{X}}(i)$, for the cases and controls respectively. For the replication-arm we use the notation $u'_{j_{D}'}(i)$ and $u'_{j_{X}'}(i)$.
\item We compute the covariate-corrected $\auc$ for arm-1, denoted as $A(i)$, by comparing the values of $u_{j_{D}}(i)$ to the values of $u_{j_{X}}(i)$. Similarly, we calculate the covariate-corrected $\auc$ for the replication-arm, denoted as $A'(i)$, by comparing the values of $u'_{j_{D}'}$ to those of $u'_{j_{X}'}$. These $\auc$-values are corrected for the seven covariates used in \cite{Stahl_2019} (i.e., $\{U_{1},\ldots,U_{6}\}$ as well as $U_{19}$).
\item To assign a p-value to each covariate-corrected $\auc$-value we once again use a permutation-test involving random permutations of the case-control labels.
\item After calculating these p-values, we concluded that the arm-1 bicluster-score indeed replicated across arms 2-4.
\end{enumerate}
\item Given the results of the replication-study above, we switched from analyzing the raw genotyped data to analyzing the imputed data available for the subjects (see \cite{Stahl_2019})
\begin{enumerate}
\item We start by calculating, for each p-value threshold $\tp$, the values $\prsorig(j',\tp)$ and $\prsbicl(j',i,\tp)$ for each subject $j'$, and bicluster iteration $i$ where we see strong evidence of replication (i.e., for $i\in[\imin,\imax]$).
\begin{enumerate}
\item To account for linkage-disequilibrium (LD), we use Plink’s ‘clump’ function on the available imputed data for arm-1. We perform this clumping step using the same parameters as in \cite{Stahl_2019} (i.e., info-score threshold of $0.9$, R2-threshold of $0.1$, genomic window of $500$Kb, and minor-allele-frequency threshold of $0.05$). We use the HRC EUR panel as our LD reference \cite{HRCERP_2016}.
\item For each p-value threshold $\tp$, we calculate each PRS-value for each individual as the sum of the risk-allele counts multiplied by the natural-log of the risk-allele odds-ratio. In this calculation we restrict the SNP-weight-vector to include only those SNPs with individual GWAS $p$-values that are more significant than the threshold $\tp$ (when forming the PRS).
\begin{enumerate}
\item For the $\prsorig(j',\tp)$ values we use weights drawn from the summary statistics in \cite{Stahl_2019} for arm-1 (`BDRN'). These summary statistics were generated by correcting for the genome-wide principal-components mentioned above (i.e., $\{U_{1},\ldots,U_{6}\}$ and $U_{19}$).
\item For the $\prsbicl(j',i,\tp)$ values we generate bicluster- and iteration-specific weights by running a covariate-corrected GWAS. For this calculation we restrict the case-subjects to the subset $\cJ(i)$ delineated by iteration $i$ of the bicluster. However, we do not restrict the controls, and for this calculation we use all the controls from arm-1. Once again, to remain consistent with \cite{Stahl_2019}, we correct for the same genome-wide principal-components $\{U_{1},\ldots,U_{6}\}$ and $U_{19}$.
\end{enumerate}
\end{enumerate}
\item Once we have calculated $\prsorig(j',\tp)$ on arms 2-4, we can consider the values of $\prsorig(j',\tp)$ for a particular arm, comparing the results between cases and controls. We quantify this comparison by calculating a covariate-corrected $\auc$ (correcting again for $\{U_{1},\ldots,U_{6}\}$ and $U_{19}$) within this arm, denoted by $\aucorig(\tp)$ in the main text.
\item In a similar fashion we compare the case- and control-values of $\prsbicl(j',i,\tp)$ by calculating the covariate-corrected $\auc$ $\aucbicl(i,\tp)$. 
\item By examining the values of $\aucorig(\tp)$ and $\aucbicl(i,\tp)$, as well as their dependence on $\tp$, we conclude that the bicluster-informed PRS-scores do a better job of highlighting SNPs that are useful for linear prediction (see Fig \ref{fig:prs_comparison_trn4_tsty_nixxx} in the main text).
\item We then calculate $\aucorigBDI(\tp)$ and $\aucbiclBDI(i,\tp)$, restricting ourselves to a comparison between the case-subjects diagnosed with BDI and the control-subjects. This analysis indicates that the signal we saw across the replication arms in Fig \ref{fig:prs_comparison_trn4_tsty_nixxx} is likely carried by the BDI subjects (see Fig \ref{fig:prs_BD1_comparison_trn4_tsty_nixxx} in the main text).
\item We then calculate $\aucorigBDII(\tp)$ and $\aucbiclBDII(i,\tp)$, restricting ourselves to a comparison between the BDII case-subjects and the control-subjects. This analysis indicates that the signal carried by the BDI subjects is not as strongly carried by the BDII subjects (c.f. Figs \ref{fig:prs_BD1_comparison_trn4_tsty_nixxx} and \ref{fig:prs_BD2_comparison_trn4_tsty_nixxx} in the main text).
\item Finally, for each value of $\tp$, we pool the values of $\prsorig(j',\tp)$ across all replication-arms, and recalculate the associated values of $\aucorig(\tp)$, $\aucorigBDI(\tp)$ and $\aucorigBDII(\tp)$. Similarly, for each value of $\tp$ and $i$, we pool the values of $\prsbicl(j',i,\tp)$ across all replication-arms and recalculate the associated values of $\aucbicl(i,\tp)$, $\aucbiclBDI(i,\tp)$ and $\aucbiclBDII(i,\tp)$. We convert these $\auc$-values to $R^{2}$-values on a liability-scale (see \cite{LGWV_2012}), producing $\cRorig(\tp)$, $\cRorigBDI(\tp)$ and $\cRorigBDII(\tp)$, as well as $\cRbicl(\tp)$, $\cRbiclBDI(\tp)$ and $\cRbiclBDII(\tp)$. We show the results in Fig \ref{fig:prs_BDX_liability_trn4_tsty_nixxx}. This analysis corroborates our conclusions above: the BDI subjects seem to be driving the bicluster-informed PRS-associated signal we see across the replication-arms.
\end{enumerate}
\end{enumerate}

\section{Details regarding the biclustering method}

\subsection*{Selection of minor-allele-frequency (maf) threshold for primary analysis:}
In our primary analysis we restricted ourselves to maf$\geq25\%$ for the discovery arm.
The main reason we did this was because we did not want our initial training to be too dependent on allele-combinations that were too rare.
More specifically, the cohorts within the replication arms were of varying sizes, and several were relatively small (i.e., containing only a few hundred cases and controls).
With an maf of only $05\%$, we would expect allele-combinations (e.g., homozygous recessive) with a prevalence of only $(1/20)^2 = 1/400$, and these allele-combinations might not even show up in the smaller cohorts.
Thus, to be conservative and focus on allele-combinations that we would reasonably expect to see in a smaller study later on, we limited ourselves to maf$\geq25\%$.
With this larger maf of $25\%$, the rarest allele-combinations would have a prevalence of $(1/4)^2 = 1/16$, meaning that we should still expect to see a few dozen or so such combinations in each of the smaller cohorts.

Secondarily, we wanted to conduct our primary analysis (i.e., discovery in arm-1) with an eye towards the PRS-analysis we followed up with after our replication-study.
This PRS-analysis involves imputed data, and there is evidence that imputation accuracy diminishes with maf (see \cite{SGK_2021}).
Thus, restricting ourselves to SNPs with a sufficiently high maf also helps ensure that the SNPS we use to determine our signal will be accurately imputed in other data-sets.

With this being said, it is certainly reasonable to ask what kind of bicluster we would have found if we lowered the maf-threshold to maf$\geq05\%$ in arm-1.
This lower maf-threshold produces a bicluster with many more SNPs (i.e., including many with maf ranging from $05\%$ to $25\%$) but a strongly overlapping set of subjects.
Intriguingly, the overlap in subjects (relative to chance) peaks within the iteration-interval $[\imin-\imax]$ and maintains highly significant enrichment (p in the range $10^{-35}$ to $10^{-45}$ throughout).
We view this as corroboration that our selected range of iterations is a roughly accurate delineation of the true signal (see Fig \ref{fig:maf25_vs_maf05_rdrop}).

In terms of next steps, it is quite natural to combine these (and similar) results with our primary biclustering analysis.
By collecting the results of several biclustering analyses (each with different parameters), we could redefine the `core' of the bicluster to be those case-subjects that lie in the intersection of the majority of trials (e.g., the case-subjects that lie in the intersection of the bicluster found using maf$\geq25\%$ and the bicluster found using  maf$\geq05\%$).
One might also consider further restricting this core to include only those case-subjects that persist over the sensitivity-tests described below.
We fully intend to pursue this strategy in future work; we have refrained from this approach in the current paper because our goal here is to describe a robust recipe that could be applied from scratch (without, e.g., knowing ahead of time which biclusters would and would not replicate).

\subsection*{Correcting for ancestry:}

Before we run the biclustering method on a data-set, we need to choose how to correct for ancestry.
Following the research of \cite{Stahl_2019}, we use genome-wide principal-components as a proxy for genetic ancestry.
In \cite{Stahl_2019}, the genome-wide principal-components $U_{1},\ldots,U_{6}$ and $U_{19}$ were found to be significantly associated with case-control status across the studies considered.
In our case we have fewer subjects in arm-1, and so we determined which genome-wide principal-components were significant for this arm by running an $F$-test applied to nested logistic regression on the sequence $U_{1}$, $U_{2}$, $U_{3}$ and so forth.
This $F$-test selected only $U_{1}$ and $U_{2}$ as significant for arm-1.
Thus, when conducting our initial biclustering analysis in arm-1 we only correct for $U_{1}$ and $U_{2}$.
However, when conducting our replication-study in arms 2, 3 and 4 we stay consistent with \cite{Stahl_2019} and correct for components $U_{1},\ldots,U_{6}$ and $U_{19}$.
We also correct for $U_{1},\ldots,U_{6}$ and $U_{19}$ when conducting our PRS analysis later on.

\subsection*{Iterative biclustering-algorithm (recapitulated from the main text):}
As mentioned in the main text, we use the half-loop method of \cite{Rangan_2018}.
We start by re-coding each SNP via its three allele-combinations: heterozygous and homozygous dominant and recessive.
After recoding we'll assume the data-set contains $M_{D}$ case-subjects and $M_{X}$ control-subjects, each measured across $N$ allele-combinations.
We denote the array of case-subjects by $D$, with $D(j_{D},k)$ referring to allele-combination-$k$ in case-subject-$j_{D}$.
Similarly, we denote the array of control-subjects by $X$, with $X(j_{X},k)$ referring to allele-combination-$k$ in control-subject-$j_{X}$.
We use the generic subject-index $j$ to refer to both the $j_{D}$ and the $j_{X}$.

In its most basic form, the half-loop algorithm then proceeds as follows:
\begin{description}
\item[Step-0] First we load/initialize the data-arrays $D$ and $X$.
\item[Step-1] For each case $j_{D}$ and allele-combination $k$, we measure the fraction of other cases in $D$ which share that allele-combination, denoted by $[D\leftarrow D](j_{D},k)$. Similarly, we measure the fraction of controls in $X$ which share that allele-combination, denoted by $[D\leftarrow X](j_{D},k)$. The difference between these two values, denoted by $Q(j_{D},k) = [D\leftarrow D](j_{D},k) - [D\leftarrow X](j_{D},k)$ is a measure of the differential-expression of allele-combination $k$ when contrasting that particular case-subject (at case-index $j_{D}$) with the control-population.
\item[Step-2] After calculating $Q(j_{D},k)$, we form the `row-scores' $Q^{\row}(j_{D}) = \sum_{k} Q(j_{D},k)$. This row-score is a measure of the total differential-expression between case-index $j_{D}$ and the control-population, accumulated across all allele-combinations. Similarly, we calculate the `column-scores' $Q^{\col}(k)$ $=$ $\sum_{j_{D}} Q(j_{D},k)$. This column-score is a measure of the total differential-expression associated with allele-combination $k$, as measured between the case-population and the control-population. Finally, we calculate the overall `trace' $\bar{Q}$ $=$ $\sum_{j_{D},k} Q(j_{D},k)$, which is a measure of the overall differential-expression exhibited between the case-patients in $D$ and the controls in $X$, accumulated across all the allele-combinations in the data-array.
\item[Step-3] At this point we remove a small fraction of those case-subjects and allele-combinations from $D$ with the lowest row- and column-scores. For this analysis, the fraction we choose is $\gamma$ $=$ $0.5^{8}\sim0.004$.
\item[Step-4] We return to Step-1, iterating until there are no more case-subjects within $D$.
\end{description}

The biclustering-algorithm above proceeds iteratively; at each iteration $i$ we remove a small fraction $\gamma$ of the remaining case-subjects and allele-combinations. As described in supplementary section 7.3 in \cite{Rangan_2018}, the overall speed and accuracy of the biclustering-algorithm depends on $\gamma$. For large values of $\gamma$ the algorithm is fast, but not particularly accurate. For smaller values of $\gamma$ the algorithm becomes somewhat slower, but more accurate. In practice, the algorithm's detection-thresholds converge for values of $\gamma\lesssim 0.5^{5}\sim 0.03$ (see Fig 32 in supplementary section 7.3 in \cite{Rangan_2018}). Thus, in this analysis we make a conservative choice of $\gamma=0.5^{8}\sim0.004$, which is roughly an order of magnitude smaller than the value required for convergence. 

After each iteration $i$, we are left with a subset $\cJ(i)$ comprising $M(i)$ case-subjects and a subset $\cK(i)$ comprising $N(i)$ allele-combinations. Together, these row- and column-subsets form an $M(i)\times N(i)$ sub-array $D(i)$ of the original $D$.
If the case-array $D$ were to contain a bicluster with a sufficiently strong signal, then the rows and columns of that bicluster would be retained until the end, with the other rows and columns eliminated earlier (see supplementary sections 2 and 15.2 of \cite{Rangan_2018} for statistical guarantees).

This half-loop method has detection-thresholds similar to spectral-clustering and message-passing \cite{Alon_1998,Deshpande_2015}, but has several additional useful features.
First, the half-loop method allows us to search for disease-specific heterogeneity by directly correcting for control-subjects.
This case-control correction also motivates the null-hypothesis H0 described below; the permutation-test allows us to avoid spurious structures that are unrelated to the disease-label.
Second, the half-loop scores in Step-1 allow us to (implicitly) correct for linkage-disequilibrium (LD).
More specifically, subsets of SNPs which are in equally strong LD in both the case- and control-populations will be excluded as the algorithm proceeds, unless some of those SNPs are involved in a pattern of differential-expression specific to the remaining case-subjects, in which case they will be retained (as desired).
Third, the method also allows us to correct for continuous covariates.
This covariate-correction is described in detail in supplementary section 10 of \cite{Rangan_2018}, but essentially amounts to a reweighting of the $Q(j,k)$ in Step-1 to reduce the overall level of differential-expression contributed by structures which are not evenly distributed in covariate-space.
Finally, the method itself is rather straightforward and does not require the fine-tuning of parameters; the accuracy of the algorithm converges as $\gamma\rightarrow 0$, and as long as $\gamma$ is sufficiently small the detection thresholds for the algorithm will not depend on $\gamma$.

%

\subsection*{Identifying biclusters:}

As described above, our algorithm for biclustering proceeds as follows:

First, we run the biclustering algorithm on the original data-set.
As the algorithm proceeds through its iterations, it produces a nested sequence of subsets $\cJ(i)$ and $\cK(i)$ (with $i$ referring to the iteration-index).
Along the way we record the trace $\bar{Q}(i)$, which is a covariate-corrected measure of the overall level of differential expression between the remaining case-subjects (i.e., $\cJ(i)$) and the controls, restricted to the allele-combinations within $\cK(i)$.
This trace is shown as the red-curve in Fig \ref{fig:trace}, and does not by itself pick out any particular bicluster, nor any particular iteration.

We determine if there is sufficient statistical evidence for a bicluster by performing a permutation-test:
We randomly permute the case-control labels of that data-set to generate random trials from a null-hypothesis.
This null hypothesis (referred to as $H0$ in the main text) corresponds to the hypothesis that any heterogeneity in the data-set is not linked to case-control status.
As a technical detail, we preserve the covariate-structure determined by the genome-wide principal-components within this permuted data by permuting case-control labels while respecting proximity in the covariate-space (see supplementary section 10 in \cite{Rangan_2018}).
We then rerun the biclustering algorithm on all these label-shuffled data-sets, producing a trace for each label-shuffled trial drawn from $H0$ (see black curves in Fig \ref{fig:trace} in the main text).
We then compare the original trace (red) to the label-shuffled traces (black), and measure the overall level of significance (see supplementary section 14.2 in \cite{Rangan_2018}).

For a typical `noisy’ data-set, this first biclustering run would not return a significant overall p-value, and we would stop, concluding that there is not sufficient evidence for heterogeneity in the form of a bicluster.
However, if (and only if) the original trace is significant, we can conclude that there is sufficient evidence for heterogeneity in the form of a bicluster.
This is the case for arm-1 (with the overall p-value $\lesssim 1/64$).

The actual ‘boundary’ of this bicluster is then determined by comparing the original trace with the label-shuffled traces (e.g., looking for iterations where the difference is significant).
In some cases the red-trace will have a distinguished maximum, clearly indicating an iteration $i$ which can be used to cleanly delineate the bicluster (i.e., comprising $\cJ(i)$ and $\cK(i)$).
However, in many real situations (such as this analysis), the scenario is less straightforward.
While the overall difference between the red- and black-traces was significant, there was no clear maximum towards the middle of the red-trace. 
Rather, there was a `plateau’ extending across a range of iterations $i\in[\imin,\imax]$. 
Thus, in order to be conservative, we thought of the first bicluster as a `fuzzy’ bicluster defined by a subset of case-subjects somewhere in-between $\cJ(\imax)$ and $\cJ(\imin)$ and a subset of allele-combinations somewhere in-between $\cK(\imax)$ and $\cK(\imin)$.
Recall that these are nested in one another, with $\cJ(\imax)\subset\cJ(\imin)$ and $\cK(\imax)\subset\cK(\imin)$. 

\subsection*{Searching for multiple biclusters:}

Only once we believe that there is sufficient statistical evidence for a first bicluster do we search for a second.
To perform this second search, we scramble the entries in the bicluster identified from the original trace and then repeat our biclustering-algorithm on this scrambled data-set, attempting to search for a second bicluster.
In this manner, the search for the second bicluster requires us to delineate the first bicluster (otherwise the search for a second bicluster would return exactly the same results as the first search).
In our case, just to be conservative, we scrambled the entries in the bicluster delineated by $\cJ(163)$ and $\cK(163)$, which includes all the case-subjects and allele-combinations that could have fallen into the fuzzy range of our first-bicluster.

If this second bicluster were also significant, we would have searched for a third, and so forth.
As illustrated in Figs \ref{fig:trace} and \ref{fig:trace_secondary}, when we used this strategy to search arm-1 for case-biclusters we only found a single significant bicluster: the second case-bicluster in arm-1 was only marginally significant (i.e., an overall p-value of $\sim 0.05$), and so we stopped our search there. In principle we could have searched for a third bicluster, but because this second bicluster failed to replicate in arms 2, 3 and 4 (see Figs \ref{fig:AUC_trn4_tst1_secondary}-\ref{fig:AUC_trn4_tst3_secondary}), we did not continue.

When searching for the control-biclusters in arm-1 our first trace was not significant (Fig \ref{fig:trace_cl4_X}), and so we stopped our search (i.e., we never searched for a second bicluster within the control-subjects in arm-1).
However, when searching for control biclusters in arms 2, 3 and 4, we actually found several significant control-biclusters.
Many of these control-biclusters then replicated in the other arms (see, e.g., Figs \ref{fig:trace_cl1_X}-\ref{fig:trace_cl3_X} for the first control-bicluster found in each of these other arms).
However, we elected to defer study of these control-biclusters to a later paper, as we already had enough to report on from the case-bicluster in arm-1.

Incidentally, we also searched arms 2, 3 and 4 for case-biclusters (i.e., treating each as a discovery-arm).
In each case there was some evidence for significant case-biclusters, with particularly strong evidence in arms 3 and 4.
The traces for these biclusters are illustrated in Figs \ref{fig:trace_cl1_maf01_dex_p25_D_m2r2_g004}-\ref{fig:trace_cl3_maf01_dex_p25_D_m2r2_g004}.
For this paper we elected not to pursue these biclusters, mainly because of the evident heterogeneity in the control-populations for these arms.
To elaborate, the reason we were wary is that our biclustering algorithm is designed to work most cleanly in an environment where the control-subjects are a well-sampled representation of the true control-population (as this is where we can guarantee the statistical bounds described in sections 18-22 of the supplementary information in \cite{Rangan_2018}).
Because the controls in arms 2, 3 and 4 are not homogeneous, it is possible that there is batch-specific structure to the control-heterogeneity that is not directly related to disease-label.
For example, one of the control cohorts (bmpo) was recruited from a cancer study.
Thus, even though we corrected our search for the case-population and the resulting control-biclusters are significantly related to control-case label, it might be the case that this association is driven by recruitment artifacts or covariates that we don’t have access to.
This potential for an unknown (but structured) source of heterogeneity for the controls in arms 2, 3 and 4 made us hesitant to pursue the case-biclusters found in these arms in this paper.
However, we intend to investigate these structures in future work.

\subsection*{Sensitivity:}
Our biclustering method (taken from \cite{Rangan_2018}) is more-or-less automatic, and once the experimental design is determined (i.e., the arrangement of cases, controls and covariates) and a sufficiently small elimination-fraction $\gamma$ is chosen, there aren't any additional parameters that can be tweaked.
Nevertheless, we can still perform a kind of sensitivity analysis by deliberately forcing the algorithm to produce a different output by changing which subjects and SNPs are excluded at each iteration.
One strategy for such a sensitivity analysis is to force the loop-counting method to exclude subjects and SNPs one at a time -- recalculating the loop-scores after each individual exclusion.
This dramatically slows down the overall algorithm, but can force the subjects and SNPs to be excluded in a different order (particularly during the first few iterations at the beginning).

When we perform this sensitivity test we find that, indeed, the subjects and SNPs are excluded in a different order.
However, largely the same subset of subjects were retained during the iterations corresponding to the interval highlighted in Figs \ref{fig:AUC_trn4_tst1}-\ref{fig:AUC_trn4_tst3}.
More specifically, the range we highlight in Figs \ref{fig:AUC_trn4_tst1}-\ref{fig:AUC_trn4_tst3} corresponds to retaining $47\%$ -- $21\%$ of the original case-subjects.
When we perform our sensitivity test and look at the iterations for which the number of case-subjects fall into this same range, we find an overlap of $\sim 75\%$ with our original subset, a very significant enrichment ($p\leq 10^{-80}$, see Fig \ref{fig:g004_vs_g001_rdrop}).

As described above in the discussion regarding minor-allele-frequency, we believe it's an excellent idea to use these kinds of sensitivity-tests to whittle down the bicluster its `core'.
For example, we could retain only those case-subjects that persist across a variety of sensitivity-tests.
As mentioned above, we refrained from doing this for the present paper because we wanted our analysis pipeline to be simple to understand, and we didn't want to add extra hyperparameters (such as how many and what kinds of sensitivity-tests to run, how to intersect the various biclusters found across these sensitivity tests, and so forth).

\subsection*{Overfitting:}

Overfitting is a very real problem, particularly when attempting to characterize heterogeneity.
In our case we have tried our best to avoid overfitting by adopting the following safeguards.

First, we use a permutation-test to characterize the significance of the original biclustering-algorithm (within the discovery-process in arm-1), as described above and illustrated in Fig \ref{fig:trace}.
Briefly, this permutation-test involves running the same biclustering-algorithm (with the same parameters, principal-components, etc.) but on a randomly permuted copy of arm-1 for which the case-control labels were shuffled.
Thus, much of the structure and correlations in this label-shuffled data matches that of the original data; the main difference being that the disease-label no longer correlates with the disease.
By comparing the trace $\bar{Q}(i)$ for each iteration of the original biclustering run (red curve in Fig \ref{fig:trace}) against the distribution of traces obtained from the shuffled-data (grey curves in Fig \ref{fig:trace}), we can determine an overall p-value for the entire biclustering process.
If this overall p-value were not significant, we would have stopped and not pursued any further analysis.
However, in the case of our first biclustering-run, this overall p-value was $\lesssim 1/64$, giving us confidence that there was indeed a significant signal linking the bicluster to the disease-labels in arm-1.

Now, while we believe that our first bicluster from arm-1 is a significant signal, we can't be sure that this bicluster is disease-related, as there could be some problem with the labels themselves (e.g., a batch-effect we were unaware of).
Thus, as a second safeguard against overfitting, we check for replication in three independent data-sets.
That is to say, we used the bicluster to define bicluster-scores, and then checked to see if the bicluster-scores retained information related to the disease-label in the replication-arms (i.e., Figs \ref{fig:AUC_trn4_tst1}-\ref{fig:AUC_trn4_tst3}).

This second safeguard helped us reassure ourselves that the bicluster we found was indeed linked to both the disease-label as well as the SNPs selected (and unlikely to be entirely driven by an unaccounted-for covariate).
However, the bicluster could still be driven by some idiosyncrasy within the genotyping process, and might not generalize to other genotyping platforms.
For example, the replication in Figs \ref{fig:AUC_trn4_tst2} and \ref{fig:AUC_trn4_tst3} is not as strong as that in Fig \ref{fig:AUC_trn4_tst1}; we currently believe that this phenomenon is due to the fact that the number of genotyped SNPs shared between arm-1 and arm-2 is greater than the number of genotyped SNPs shared between arm-1 and arms-3 and 4.
However, it is also possible that this phenomenon occurs because our bicluster is platform-specific, and does not extend to data genotyped on other platforms.

Thus, as a third safeguard, we performed the PRS-analysis illustrated in Figs \ref{fig:prs_comparison_trn4_tst1_nixxx} and \ref{fig:prs_comparison_trn4_tsty_nixxx}.
This analysis is not as strongly tethered to exactly which SNPs were genotyped.
Indeed, since we are able to use imputed data to conduct the PRS-analysis, we conducted this analysis using all SNPs available with minor-allele-frequence $\geq 05\%$.
Thus, the number of accessible SNPs for this analysis is comparable across arms 2, 3 and 4.
Put another way, the PRS-analysis is not directly hampered by a reduction in overlap-fraction, and arms 3 and 4 seem to have the potential to replicate just as well as arm-2 (see, e.g., the $i=225$ subplot within Fig \ref{fig:prs_comparison_trn4_tsty_nixxx}).
We believe that this PRS-analysis helps strengthen our claim that this bicluster is not an artefact of the genotyping platform for arm-1, and is actually selecting a subset of patients that are correlated in a disease-relevant manner.

\subsection*{Comparison to other methods:}

As far as we are aware, our biclustering method is the only available open-source software we know of which can tackle large arrays of categorical data (i.e., hundreds of thousands to millions of genotyped SNPs) while correcting for cases and controls (i.e., disease status), and also correcting for continuous covariates (i.e. genome-wide principal-components).
Thus, we were not able to apply another biclustering method to the genotyped bipolar data.
However, we have certainly used other data-sets to compare our biclustering method to other methods in the literature.

Originally, we used synthetic data-sets to compare the performance of our biclustering algorithm to several other algorithms (see supplementary section 7.4 in \cite{Rangan_2018}).
As demonstrated in this paper, our method performs quite favorably, with higher accuracy than the classical methods tested.

More recently, we have applied our biclustering methods to snRNA-seq data, and we have compared the performance of our method against two popular methods in the literature (UMap and Louvain-clustering).
These comparisons (involving both simulated and real data) are described in the appendix S1 of \cite{ZLLetal_2024}.
In summary, we find that both UMap and Louvain-clustering have a difficult time determining which clusters/biclusters are statistically significant, and which are spurious.
By contrast, our approach uses the permutation-test described above to estimate the overall p-value of any potential bicluster.
Moreover, as described above, we are quite conservative in our approach, searching for biclusters one at a time, and only proceeding if the previous searches return statistically significant results.
As a result of these safeguards our biclustering method is much more accurate than UMap and Louvain-clustering at recovering biologically relevant structure.

\subsection*{Possible limitations:}

One limitation of our methodology is that, so far, we have only used the raw genotyped data as input to detemine biclusters in our training arm-1.
We did so to avoid concerns of spurious correlations that might manifest within imputed data \cite{McGrouther_2012}.
However, this choice limited the range of SNPs available for our replication-study in Figs \ref{fig:AUC_trn4_tst1}-\ref{fig:AUC_trn4_tst3}.
Going forward, we intend to extend our approach to allow for imputated data within the training-arm.
In the best case scenario this imputed data will be fully accurate and will not introduce any spurious correlations.
More generally, we will need to characterize -- and correct for -- any spurious structure introduced by the imputation.
Doing so will open the door for higher powered analyses.
Additionally, if we are able to deduce genetic subgroup membership directly from high-quality imputed data, then we will have even more options available for replication and validation.

Another shortcoming of our method is that it is likely {\em too} conservative when it comes to identifying heterogeneity.
As an example, when studying Alzheimer's disease we encountered scenarios where there was clearly heterogeneity present, but the metric we used to compare the original-trace to the label-shuffled traces was not sensitive enough to pick out the nature of this heterogeneity (see, e.g., the section entitled "Search for disease-specific biclusters" and Figure S4 in \cite{ESR_2024}).
In this study we suspect that the heterogeneity in the data is more complicated than can be characterized by a simple bicluster.
That is, the heterogeneity may involve structures that are more nuanced than a distinguished set of subjects and SNPs (see section 7.3 and Fig 35 in \cite{Rangan_2018}), and we would need a more sophisticated strategy to identify it.
We believe that something similar is happening with the second case-bicluster found in arm-1 (see Fig \ref{fig:trace_secondary}), and we intend to pursue this line of inquiry in future work.

\section{Additional supporting information and supporting figures}

\subsection*{Additional replication analyses in arm-2}

\begin{figure}
  \centering
  \includegraphics[width=6.5in]{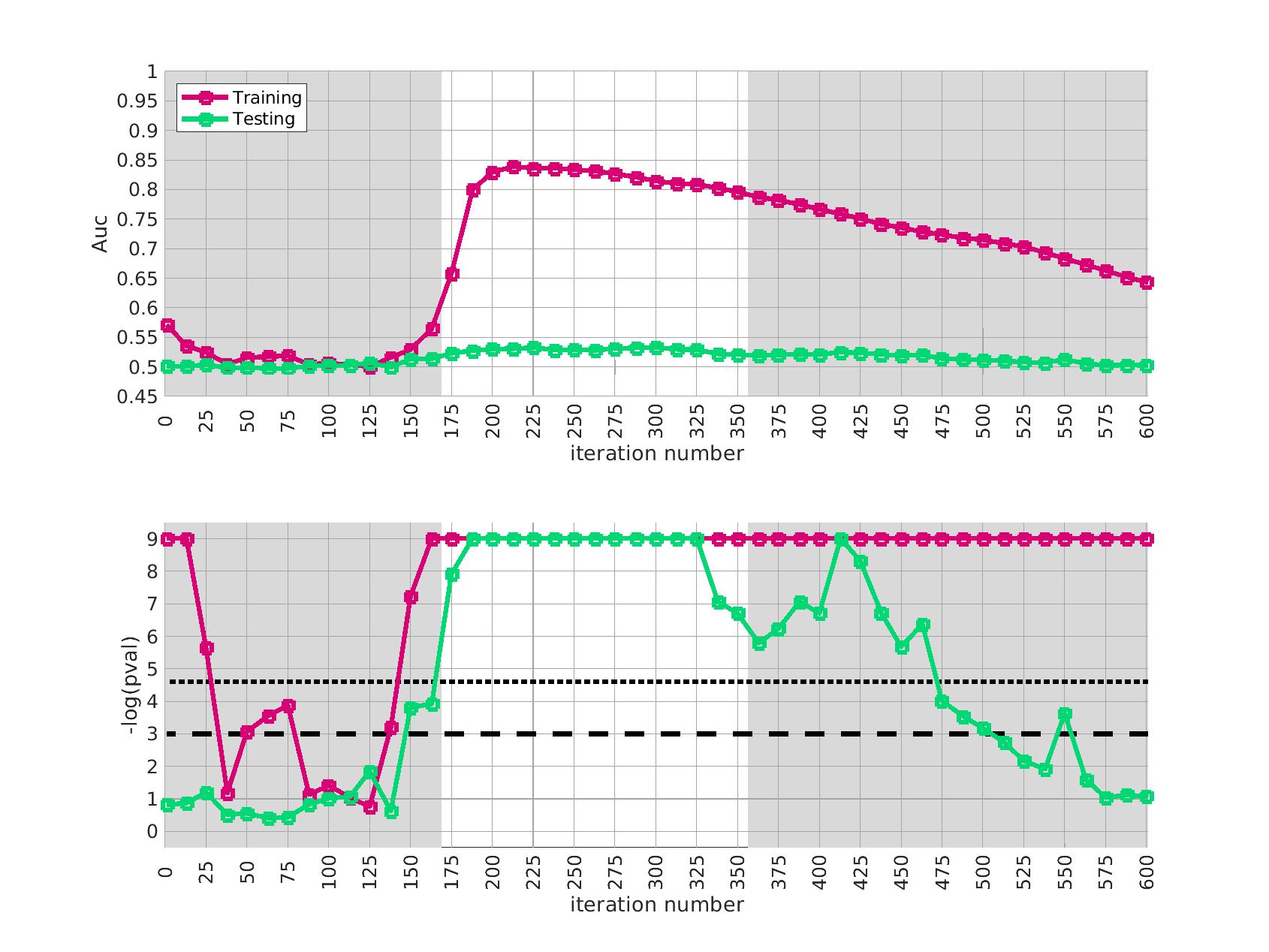}
  \caption{
    This figure is similar to Fig \ref{fig:AUC_trn4_tst1}, except that we randomly eliminate SNPs from arm-2 until the SNP-overlap between arm-2 and the training-arm is equal to the SNP-overlap between arm-3 and the training-arm.
  }
  \label{fig:AUC_trn4_tst1cap2}
\end{figure}

\begin{figure}
  \centering
  \includegraphics[width=6.5in]{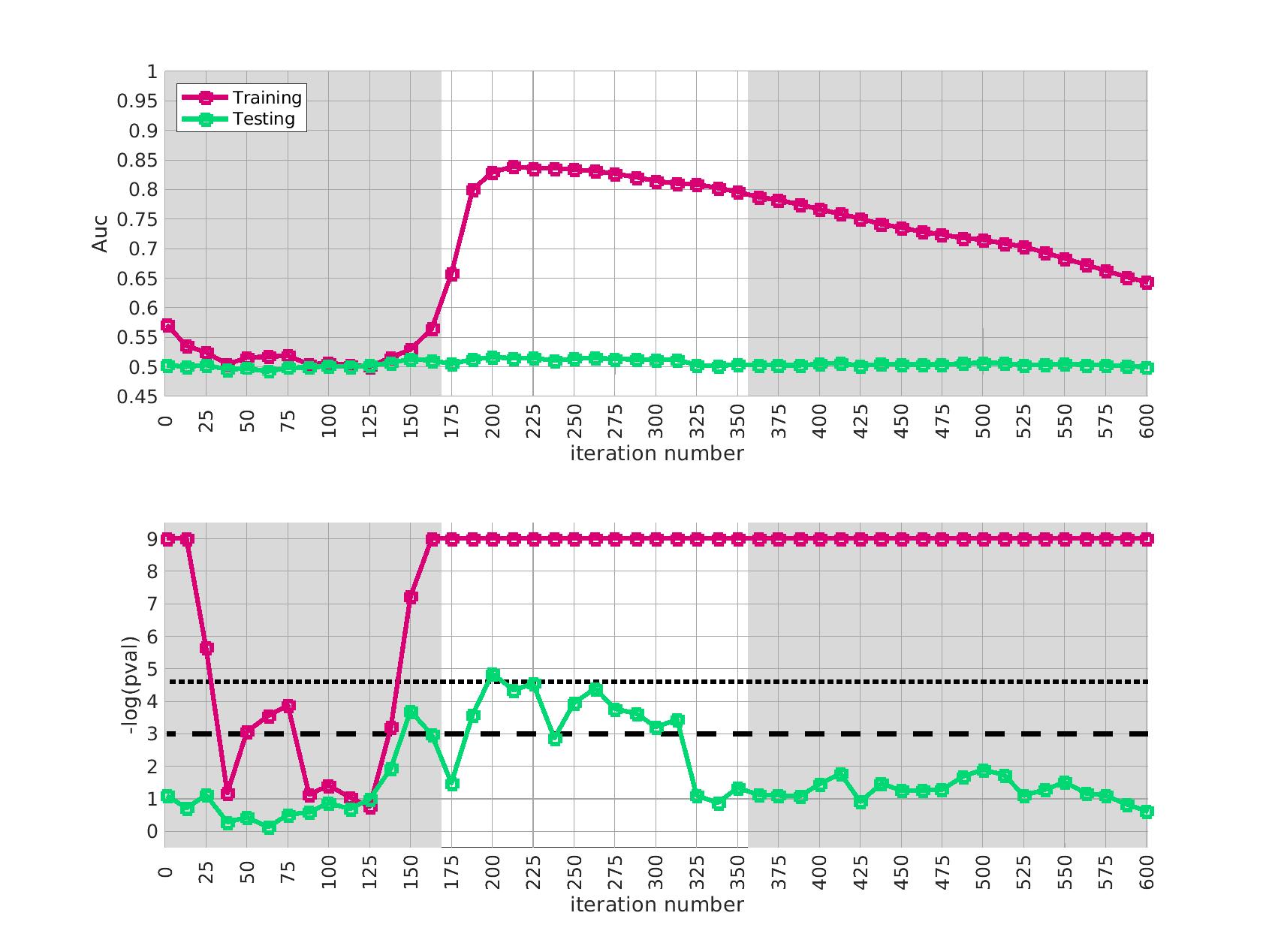}
  \caption{
    This figure is similar to Fig \ref{fig:AUC_trn4_tst1}, except that we randomly eliminate SNPs from arm-2 until the SNP-overlap between arm-2 and the training-arm is equal to the SNP-overlap between arm-4 and the training-arm.
  }
  \label{fig:AUC_trn4_tst1cap3}
\end{figure}

Figs \ref{fig:AUC_trn4_tst1cap2} and \ref{fig:AUC_trn4_tst1cap3} illustrate the replication observed in arm-2 when the SNP-overlap between arm-1 and arm-2 is artificially decreased.

\subsection*{Influence of covariates on the bicluster}

\begin{figure}
  \centering
  \includegraphics[width=6.5in]{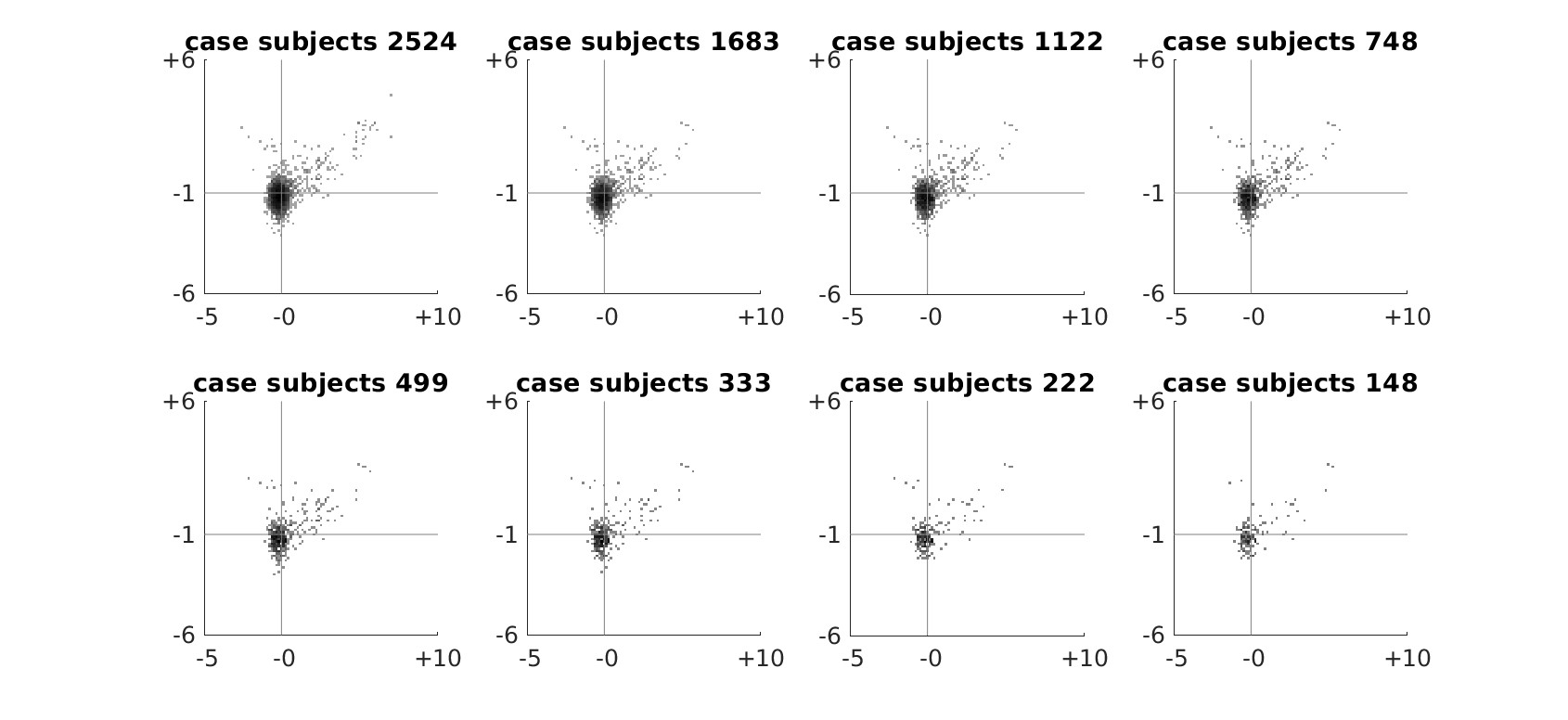}
  \caption{
    This figure illustrates the distribution of ancestry-associated principal-components $U^{1}_{1}$ and $U^{1}_{2}$ amongst the remaining case-subjects as the half-loop algorithm proceeds (specifically, at iterations $i\in\{ 96, 189, 276, 359, 442, 511, 585 \}$). Each subplot displays a scatterplot of remaining case-subjects plotted with respect to $U^{1}_{1}$ and $U^{1}_{2}$. The number of remaining case-subjects is shown above each subplot. The horizontal and vertical lines indicate the median values for the original distribution. Note that the overall shape of the distribution does not change much as the algorithm proceeds.
  }
  \label{fig:mds_distribution_A}
\end{figure}

\begin{figure}
  \centering
  \includegraphics[width=6.5in]{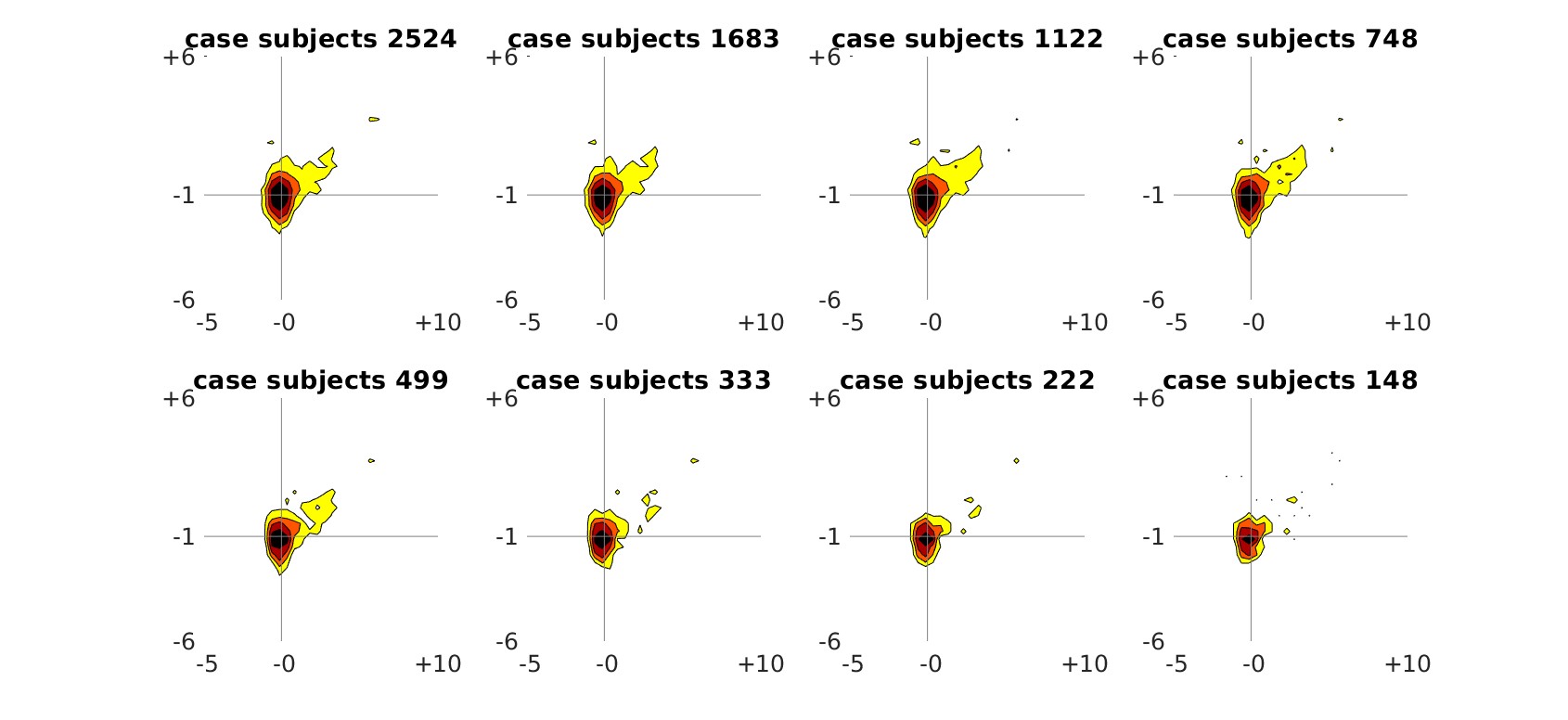}
  \caption{
    This figure is similar to Fig \ref{fig:mds_distribution_A}, except that a heat-map of the distribution is shown (rather than a scatterplot).
    The colors in the heat-map correspond to the logarithm of the density in the underlying distribution.
    Four different contours are shown, ranging from yellow to maroon, corresponding to the $20\%$, $40\%$, $60\%$ and $80\%$ percentiles of the log-density.
  }
  \label{fig:mds_distribution_B}
\end{figure}

\begin{figure}
  \centering
  \includegraphics[width=5.5in]{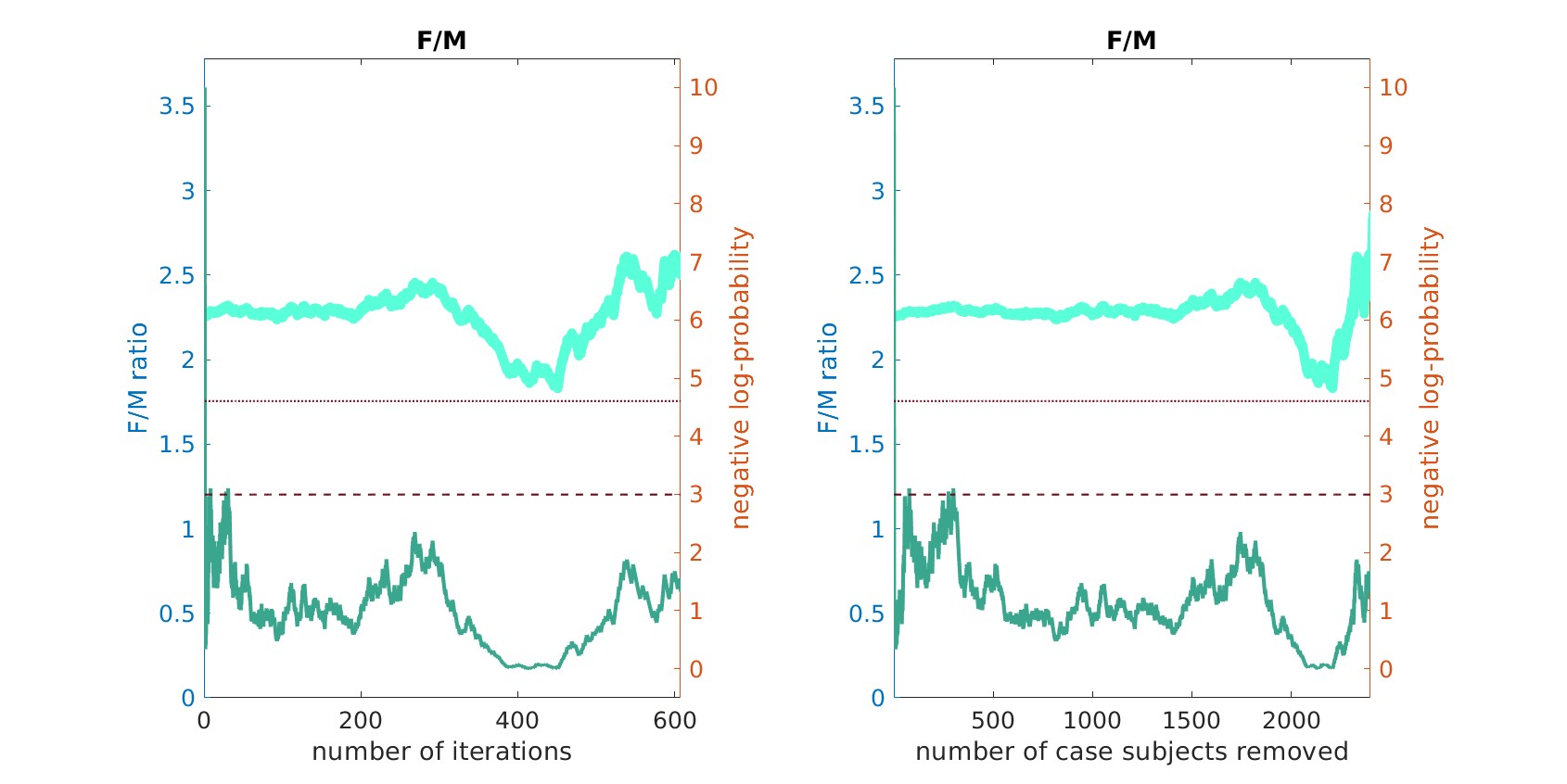}
  \caption{
    This figure plots the ratio of female to male subjects within $\cJ(i)$ (light-teal, left y-axis) as a function of the iteration $i$ (left) and the number of remaining case-subjects (right). The dark-teal line corresponds to the negative-log-probability (right y-axis) of observing a ratio at least as large by chance.
    The dashed and dotted horizontal lines indicate $0.05$ and $0.01$ significance values, respectively.
    Note that the female population is not over-represented across the range of iterations including $i\in[\imin,\imax]$, implying that the bicluster we observe is not significantly enriched for female subjects.
  }
  \label{fig:sex_stack}
\end{figure}

Figs \ref{fig:mds_distribution_A} and \ref{fig:mds_distribution_B} illustrate the association between the bicluster found in arm-1 and the ancestry-related covariates. Fig \ref{fig:sex_stack} illustrates the association with sex.

\subsection*{Interaction between bicluster-score and population-wide PRS}
\begin{figure}
   \centering
   \includegraphics[width=5.5in]{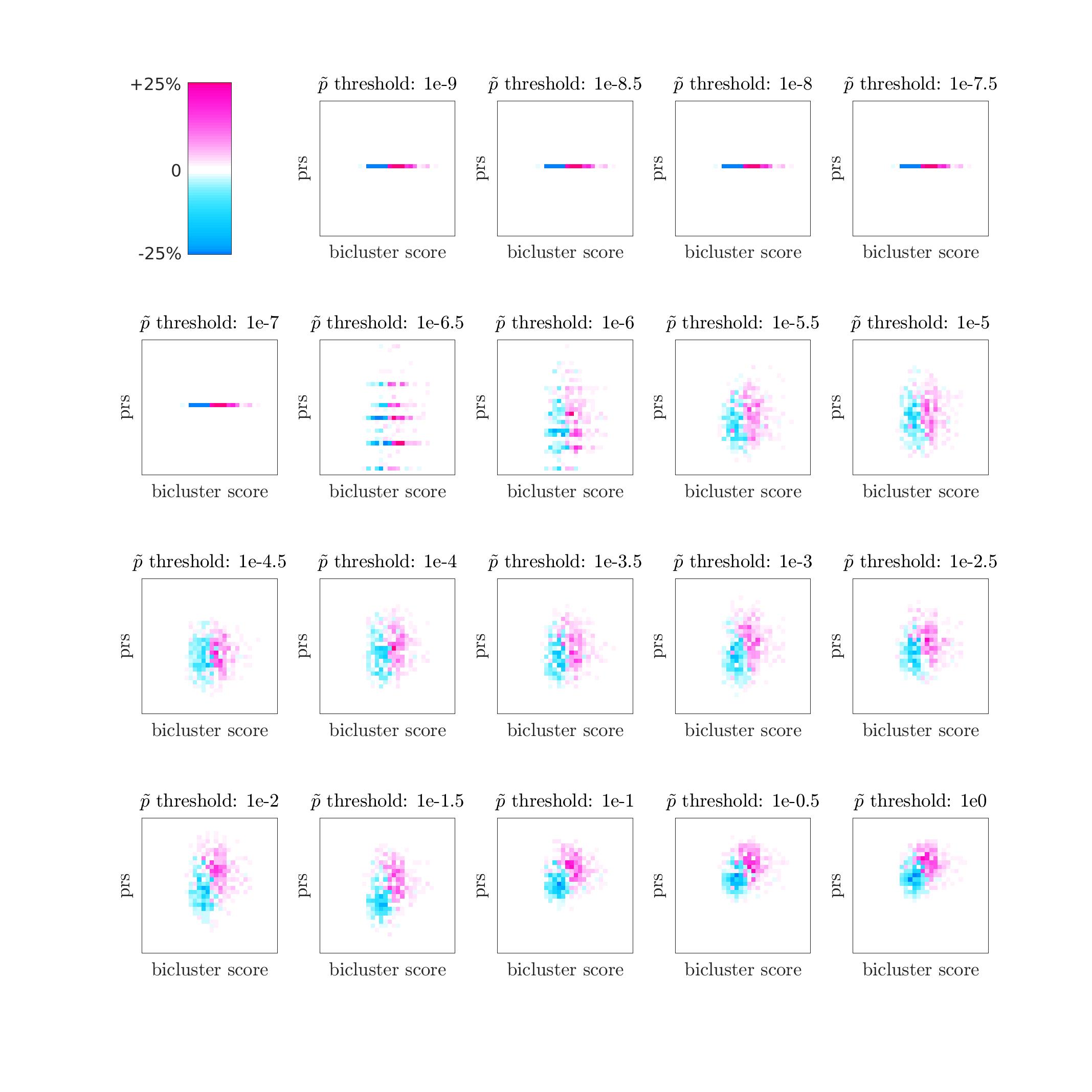}
   \vspace{-2em}
   \caption{
     In this figure we illustrate the correspondence between the population-wide $\prs(j';\tp)$ and bicluster-score $u'_{j'}(i)$ for arm-2 at $i=\imin$.
     Each subplot visualizes the distribution of subjects in arm-2 as a function of bicluster-score $u'_{j'}(i)$ (horizontal) and $\prsorig(j';\tp)$ (vertical), with the SNP-$p$-value threshold $\tp$ varying across the subplots.
     In each subplot a heatmap is shown, representing the difference between the density of cases and controls.
     The color pink corresponds to areas with a higher case-density than control-density, while blue corresponds to areas with a higher control-density than case-density.
     The colorbar (upper-left) ranges across $\pm 25\%$ of the maximum density (taken across both the case- and control-distributions).
     Note that, while the bicluster-score is correlated with the population-wide PRS when $\tp$ is sufficiently high, the correlation is far from perfect.
     Note that, when $\tp$ is high (i.e., $\tp\sim 1$ and all the SNPs are used to generate the PRS), there is a marked correlation between case-control status and high-values of $\prsorig(j';\tp)$ and bicluster-score $u'_{j'}(i)$.
     However, for lower values of $\tp$ this structure shifts, and $\prsorig(j';\tp)$ is no longer a useful indicator of case-control status while the bicluster-score $u'_{j'}(i)$ is still a useful indicator. 
     This trend persists for other iterations, as illustrated in Figs \ref{fig:prs_nagelkerke_scatterplot_trn4_tst1_ni225} and \ref{fig:prs_nagelkerke_scatterplot_trn4_tst1_ni350}.
   }
   \label{fig:prs_nagelkerke_scatterplot_trn4_tst1_ni175}
 \end{figure}

\begin{figure}
  \centering
  \includegraphics[width=5.5in]{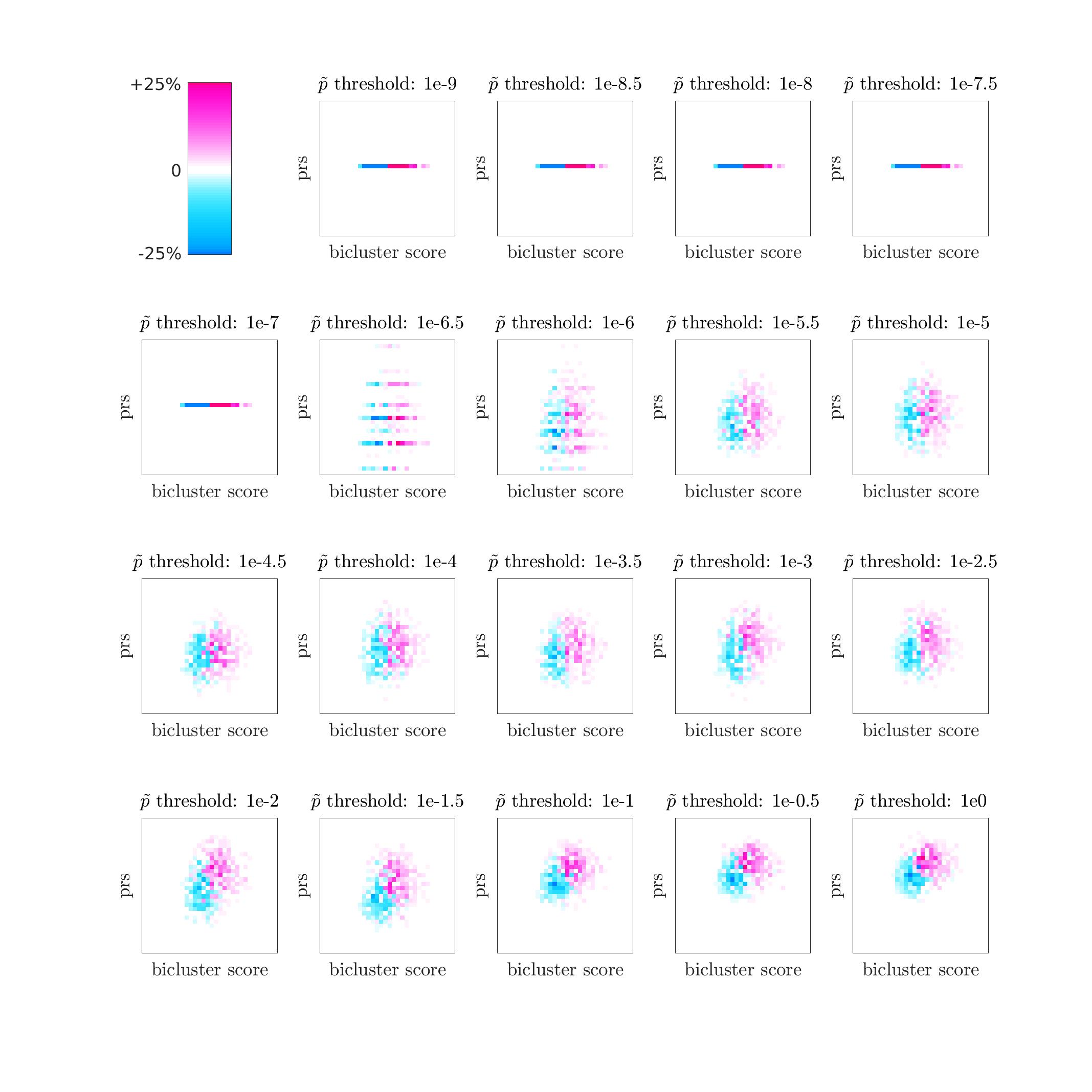}
  \caption{
    This figure is analogous to Fig \ref{fig:prs_nagelkerke_scatterplot_trn4_tst1_ni175}, except for $i=225$.
  }
  \label{fig:prs_nagelkerke_scatterplot_trn4_tst1_ni225}
\end{figure}

\begin{figure}
  \centering
  \includegraphics[width=5.5in]{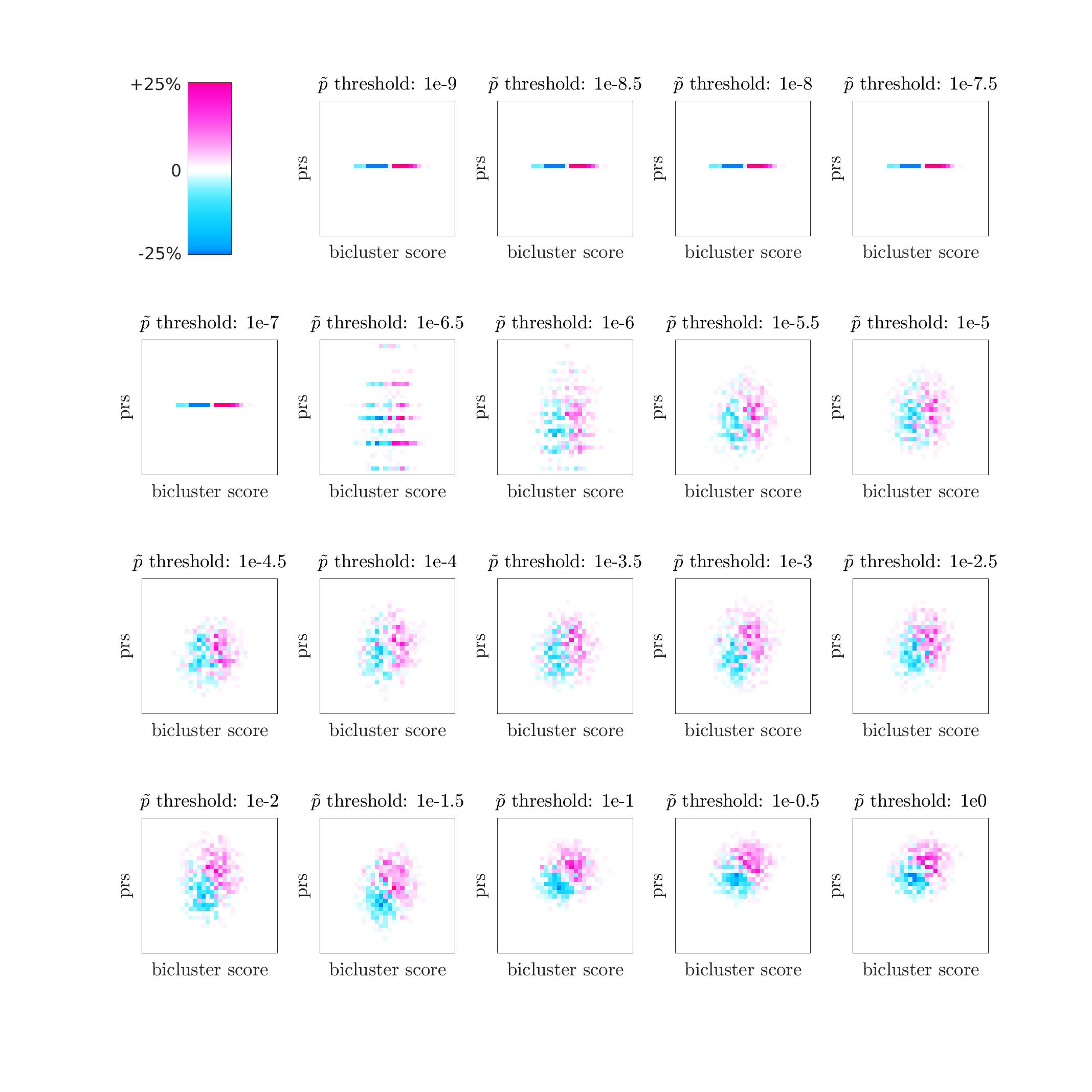}
  \caption{
    This figure is analogous to Fig \ref{fig:prs_nagelkerke_scatterplot_trn4_tst1_ni175}, except for $i=350$.
  }
  \label{fig:prs_nagelkerke_scatterplot_trn4_tst1_ni350}
\end{figure}

As described in the Methods section, we calculated the population-wide $\prsorig(j';\tp)$ and the bicluster-informed $\prsbicl(j';i,\tp)$ across a variety of iterations $i$ and $\tp$-thresholds.
In Fig \ref{fig:prs_nagelkerke_scatterplot_trn4_tst1_ni175} we illustrate the correspondence between the population-wide $\prsorig(j';\tp)$ and bicluster-score $u'_{j'}(i)$ for arm-2 at $i=\imin$.
This trend persists for other iterations, as illustrated in Figs \ref{fig:prs_nagelkerke_scatterplot_trn4_tst1_ni225} and \ref{fig:prs_nagelkerke_scatterplot_trn4_tst1_ni350}.

\subsection*{Nagelkerke $R^{2}$ for a series of linear-models}
To complement the observations of Figs \ref{fig:prs_nagelkerke_scatterplot_trn4_tst1_ni175}, \ref{fig:prs_nagelkerke_scatterplot_trn4_tst1_ni225} and \ref{fig:prs_nagelkerke_scatterplot_trn4_tst1_ni350}, we investigate how the bicluster-informed PRSs and the bicluster-scores themselves contribute to risk-prediction.
For a given $i$ and $\tp$ we build a series of linear-models to predict case-control status.
These linear-models use: (i) the ancestry-related covariates, (ii) the population-wide $\prsorig(j';\tp)$, (iii) the bicluster-informed $\prsbicl(j';i,\tp)$, and (iv) the bicluster-score $u'_{j'}(i)$.

The series of linear-models we consider involve successively more and more of these terms.
Thus, the null linear-model attempts to predict case-control status using no terms at all (i.e., using only the average prevalence of cases and controls).
The first linear-model attempts to predict case-control status using only term (i), that is, only the ancestry-related covariates.
The second linear-model uses both term (i) and (ii), i.e., both the ancestry-related covariates and the population-wide PRS.
The third linear-model uses terms (i), (ii) and (iii).
And finally, the fourth linear-model uses terms (i)-(iv).
For each linear-model we measure the Nagelkerke pseudo-$r^{2}$ value between that model and the null model
The difference in pseudo-$r^{2}$ values gives an estimate of the additional explanatory power provided by each term in succession.

\begin{figure}
  \centering
  \begin{tabular}{cc}
    \includegraphics[width=2.75in]{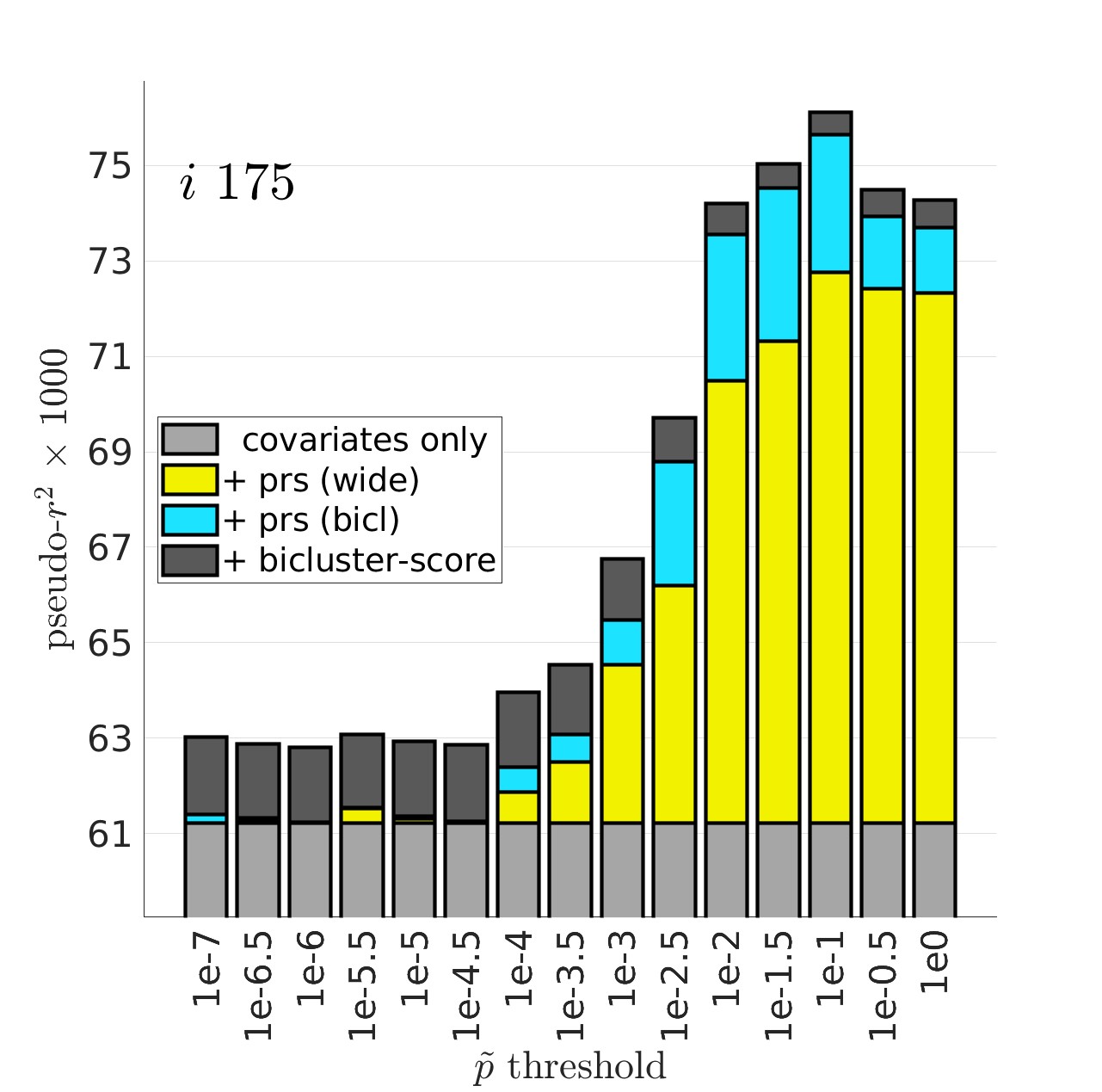}
    &
    \includegraphics[width=2.75in]{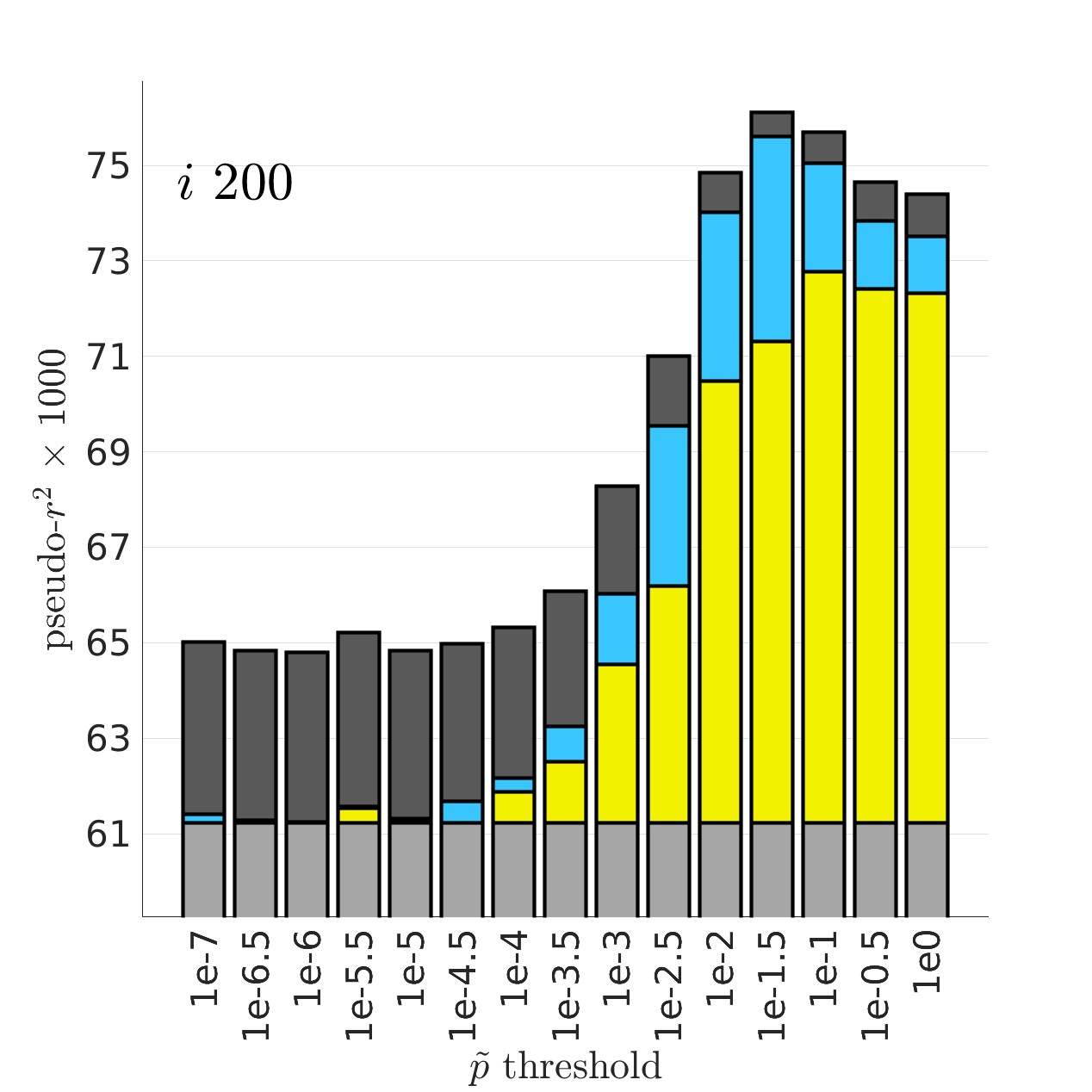}
    \\
    \includegraphics[width=2.75in]{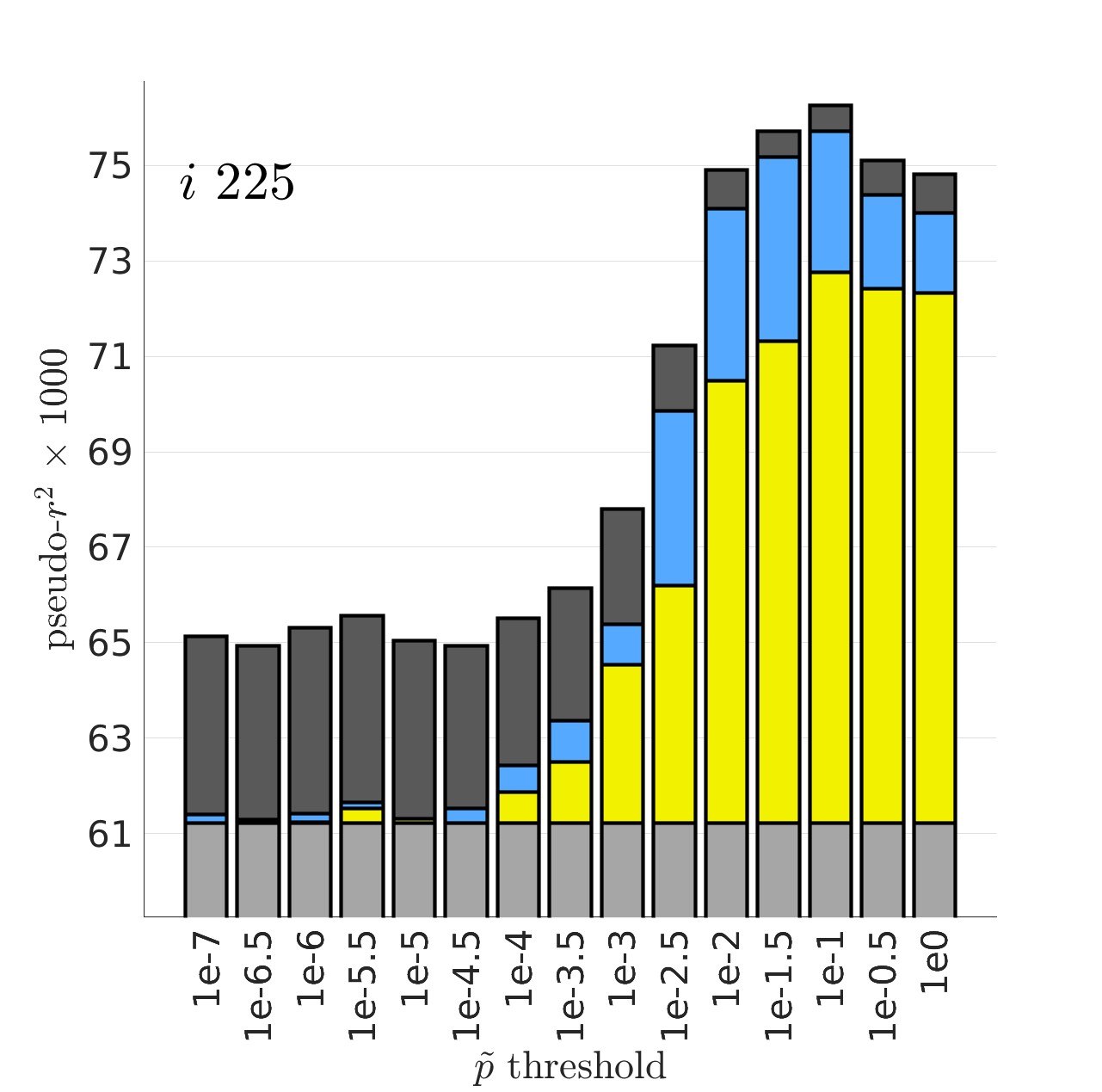}
    &
    \includegraphics[width=2.75in]{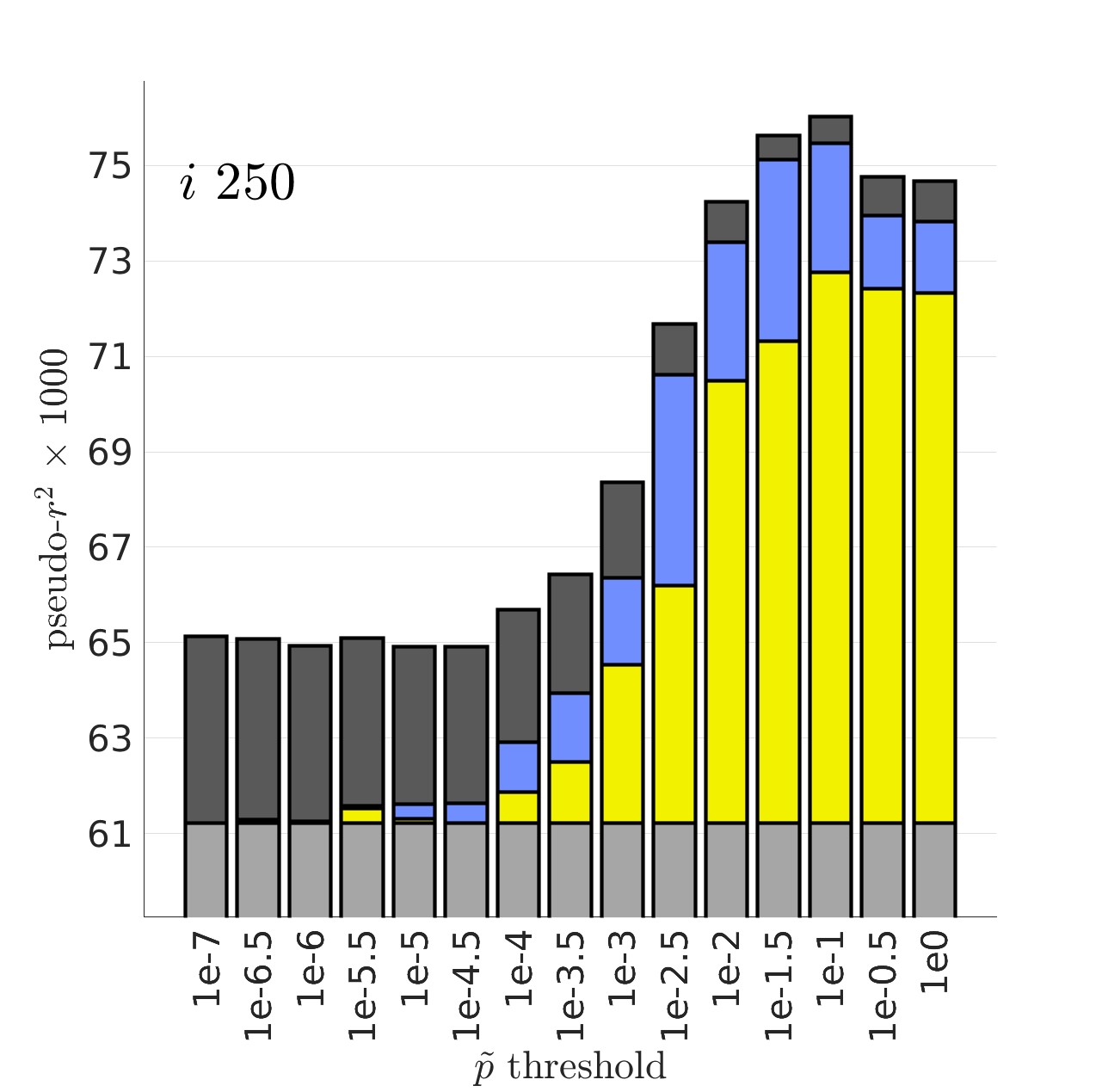}
  \end{tabular}
  \caption{
    In this figure we illustrate the explanatory power of the bicluster-informed $\prsbicl(j';i,\tp)$ in combination with the bicluster-score $u'_{j'}(i)$ for arm-2, with the iteration number $i$ varying across the subplots.
    Within each subplot the $\tp$-threshold is shown along the horizontal-axis.
    We measure the incremental Nagelkerke pseudo-$r^{2}$ value associated with the following terms: (i) the ancestry-related covariates, (ii) the population-wide $\prsorig(j';\tp)$, (iii) the bicluster-informed $\prsbicl(j';i,\tp)$, and (iv) the bicluster-score $u'_{j'}(i)$.
    For each $i$ and $\tp$ we show a vertical bar divided into segments illustrating the contribution of each term (i)-(iv), with colors light-grey, yellow, blue-pink and dark-grey (respectively).
    The colors used for the third term (ranging from blue to pink) correspond to the colors used in Fig \ref{fig:prs_comparison_trn4_tst1_nixxx} below. 
  }
  \label{fig:prs_nagelkerke_stack_trn4_tst1_nixxx}
\end{figure}

\begin{figure}
  \centering
  \begin{tabular}{cc}
    \includegraphics[width=2.75in]{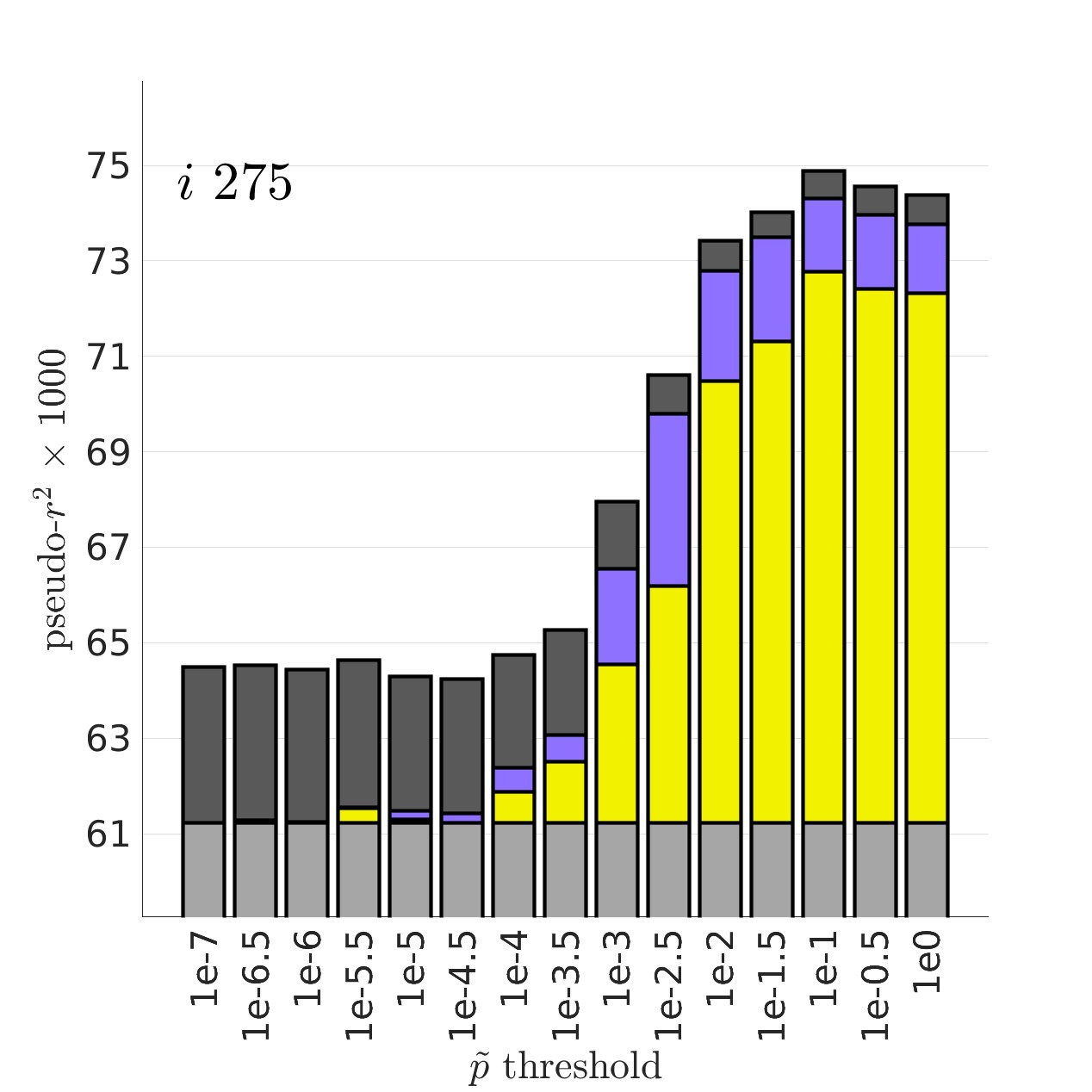}
    &
    \includegraphics[width=2.75in]{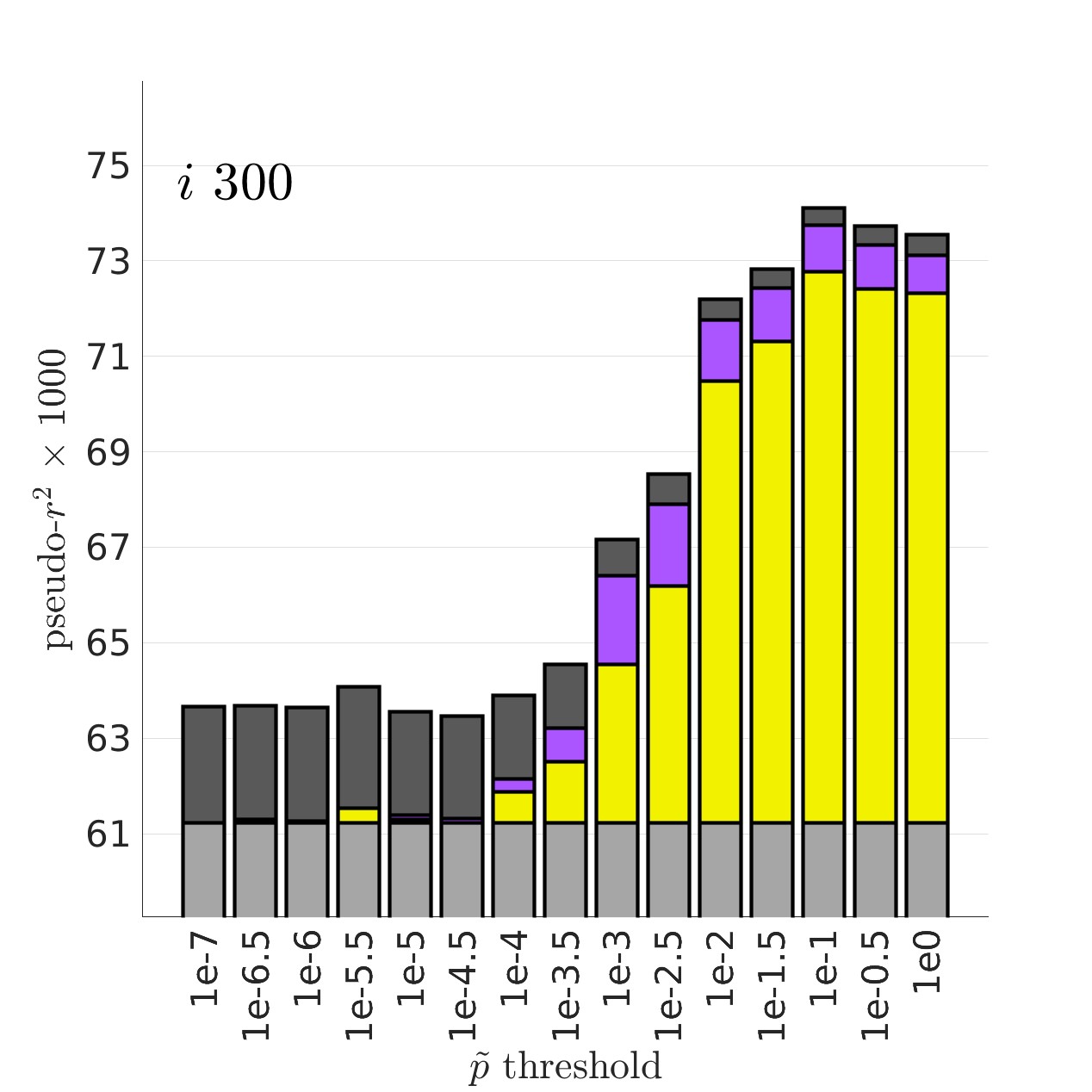}
    \\
    \includegraphics[width=2.75in]{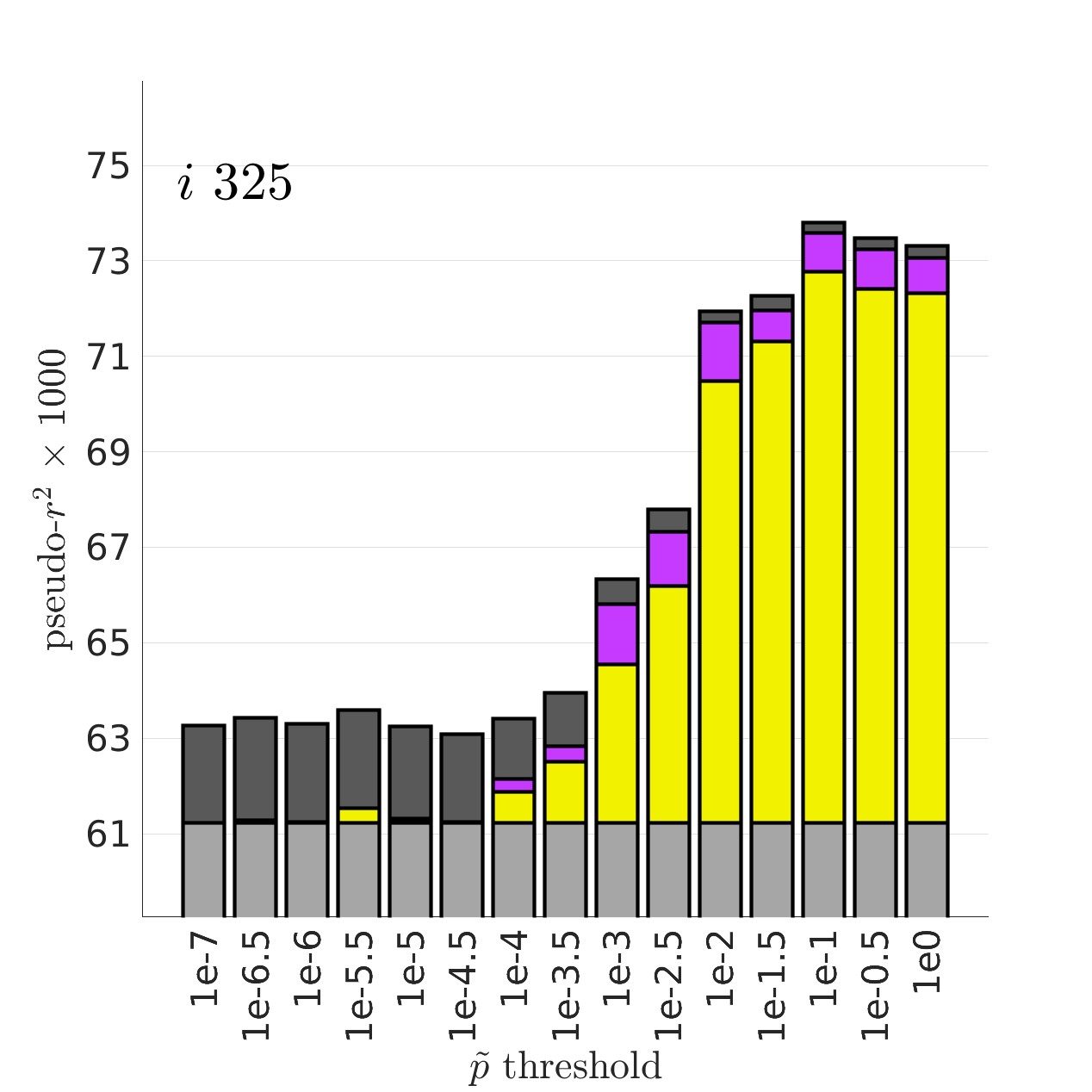}
    &
    \includegraphics[width=2.75in]{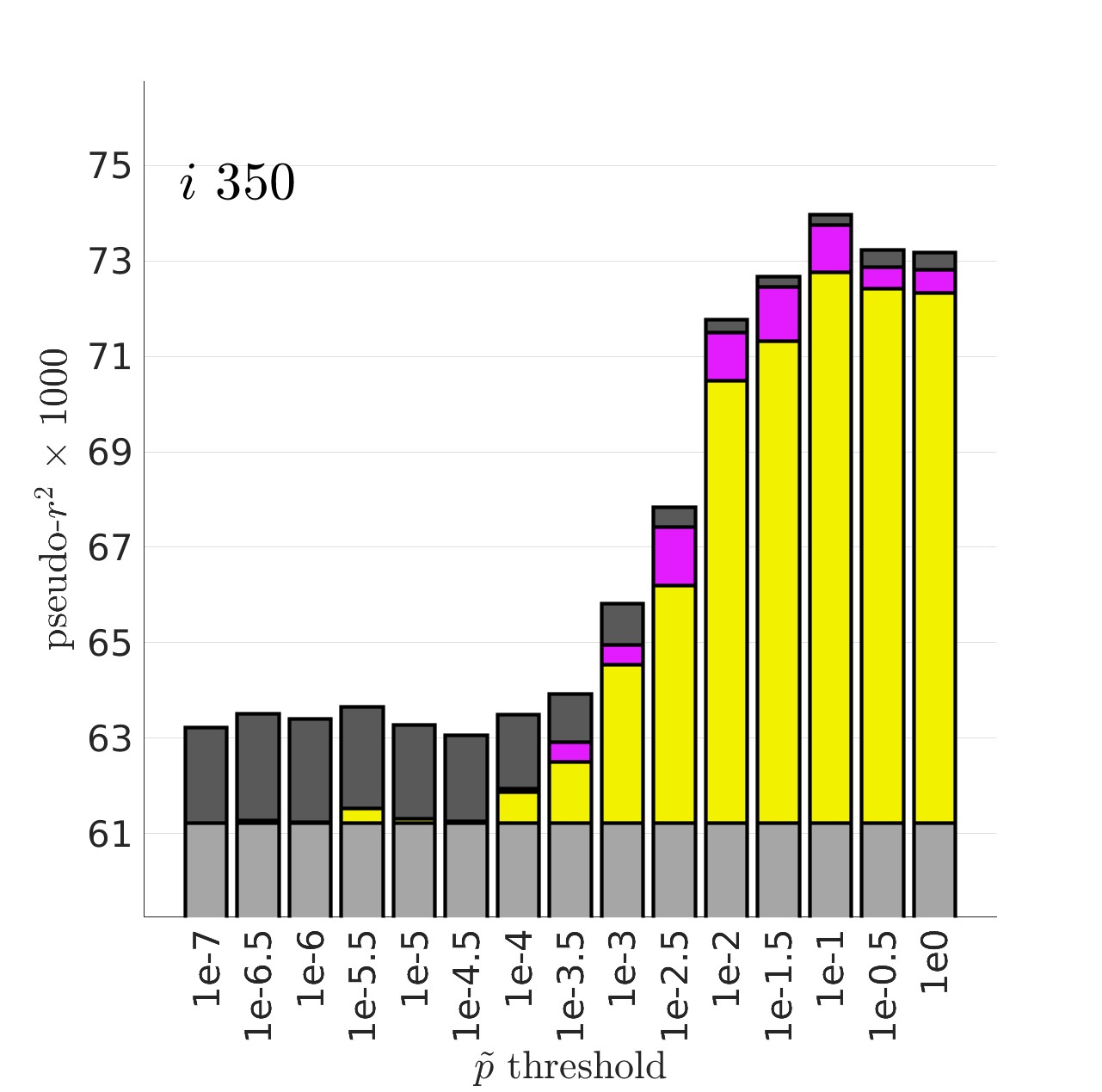}
  \end{tabular}
  \caption{
    This is analogous to Fig \ref{fig:prs_nagelkerke_stack_trn4_tst1_nixxx}, for a different set of iterations.
  }
  \label{fig:prs_nagelkerke_stack_trn4_tst1_niyyy}
\end{figure}

Results for various $i$ and $\tp$ are shown in Figs \ref{fig:prs_nagelkerke_stack_trn4_tst1_nixxx} and \ref{fig:prs_nagelkerke_stack_trn4_tst1_niyyy}.
Each vertical bar in this figure corresponds to a particular $i$ and $\tp$.
Each vertical bar is further divided into segments illustrating the incremental Nagelkerke pseudo-$r^{2}$ value associated with each term.
Note that the bicluster-informed PRS, as well as the bicluster-score $u'_{j'}(i)$ add explanatory power to the underlying linear model.
This phenomena is most pronounced when $i$ is in the middle of the range $i\in[\imin,\imax]$ and the threshold $\tp$ is small.

The observations in Figs \ref{fig:prs_nagelkerke_scatterplot_trn4_tst1_ni175}-\ref{fig:prs_nagelkerke_stack_trn4_tst1_niyyy} suggest that the bicluster might contain information useful for improving risk-prediction. Moroever, the peak in overall $R^{2}$ when $\tp\sim 1e-2$ suggests that the signal in our bicluster-score involves subsets of SNPs which do not individually achieve genome-wide significance.

\subsection*{Replication results for arm-3 and arm-4}

\begin{figure}
  \centering
  \includegraphics[width=7.5in]{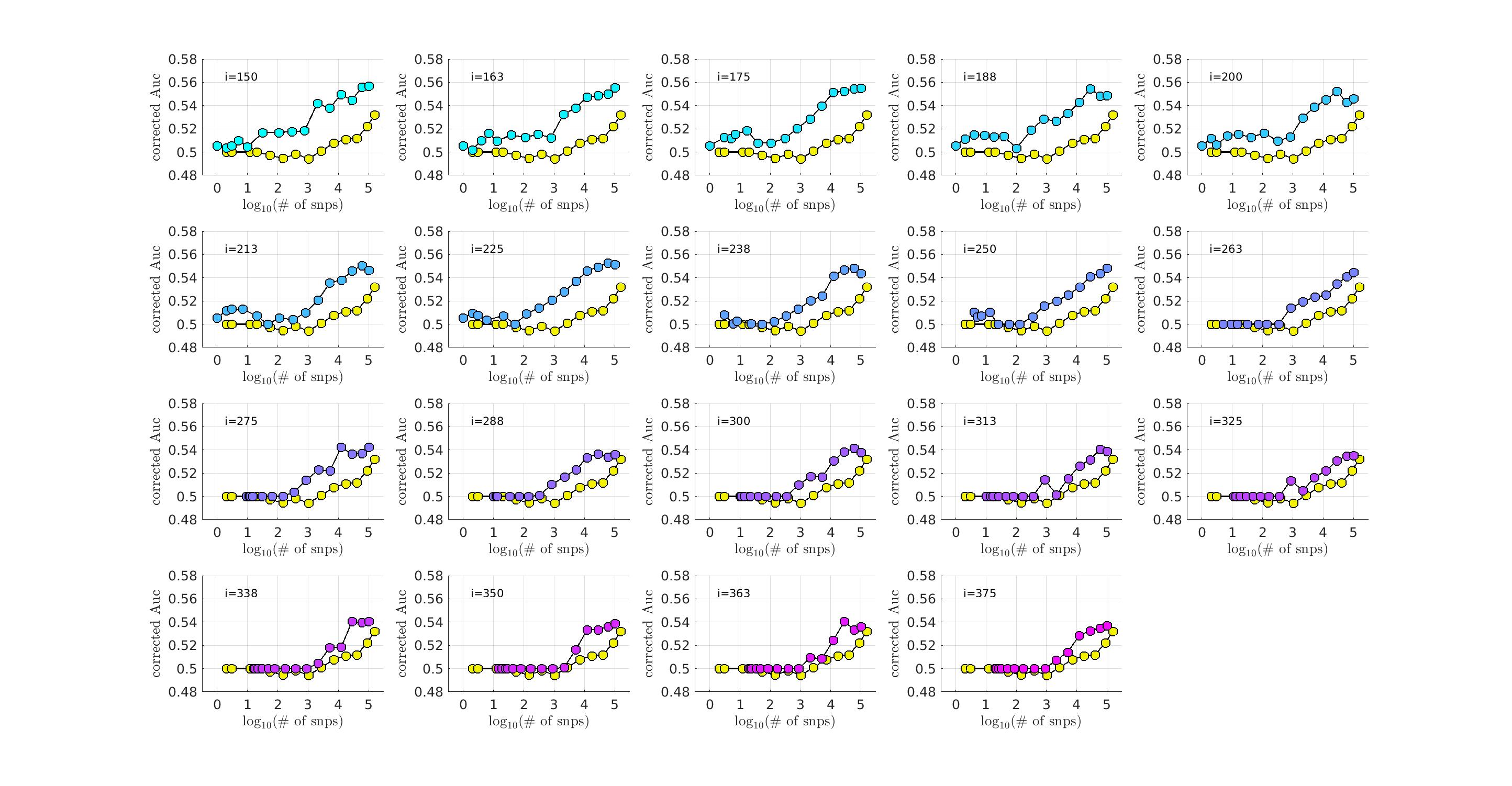}
  \caption{
    This figure is analogous to Fig \ref{fig:prs_comparison_trn4_tst1_nixxx}, except that we test on arm-3 rather than arm-2.
  }
  \label{fig:prs_comparison_trn4_tst2_nixxx}
\end{figure}

\begin{figure}
  \centering
  \includegraphics[width=7.5in]{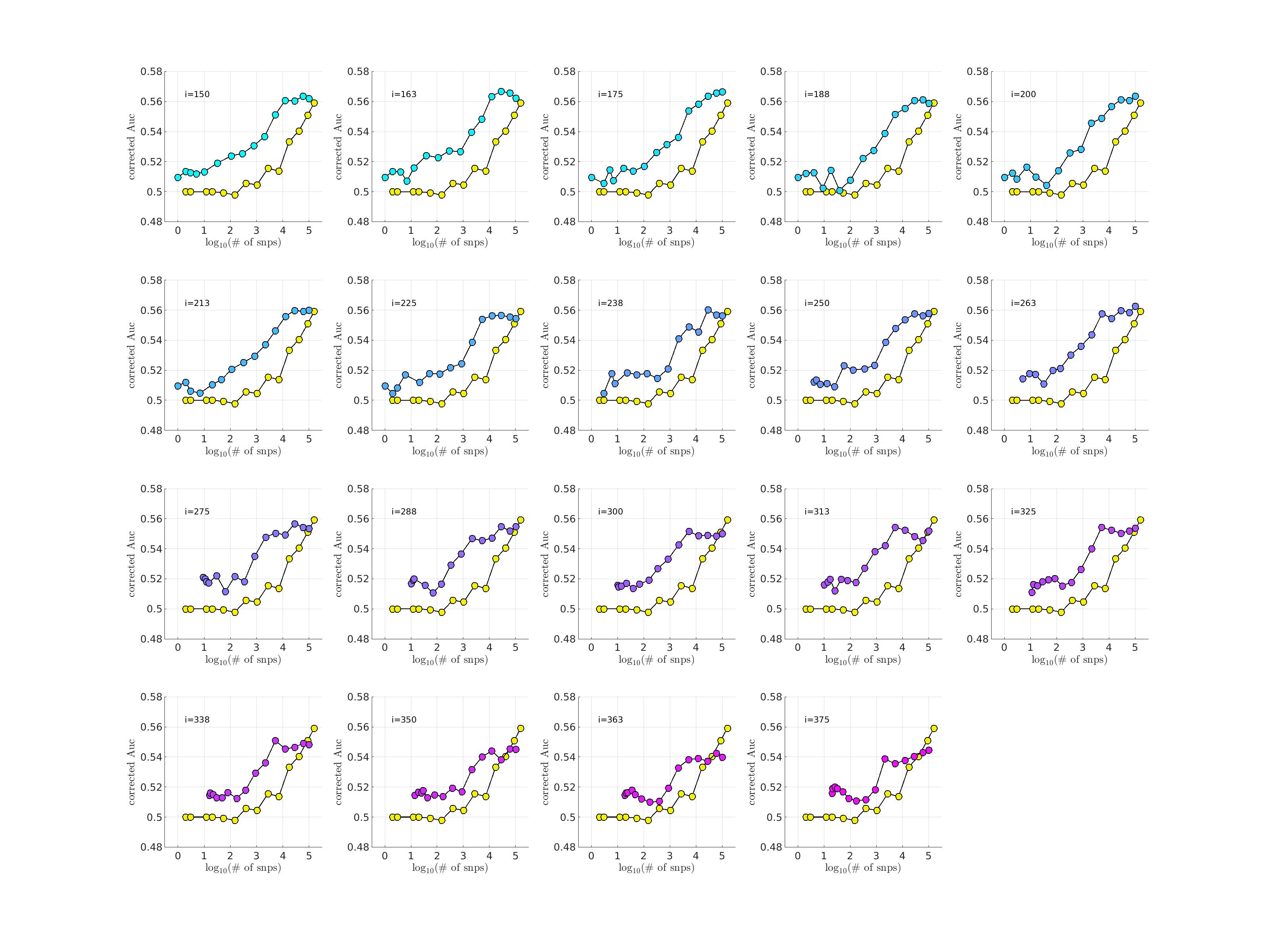}
  \caption{
    This figure is analogous to Fig \ref{fig:prs_comparison_trn4_tst1_nixxx}, except that we test on arm-4 rather than arm-2.
  }
  \label{fig:prs_comparison_trn4_tst3_nixxx}
\end{figure}

Figs \ref{fig:prs_comparison_trn4_tst2_nixxx} and \ref{fig:prs_comparison_trn4_tst3_nixxx} illustrate the replication analyses on arms 3 and 4.

\subsection*{Significance of secondary bicluster in arm-1}

\begin{figure}
  \centering
  \includegraphics[width=5.5in]{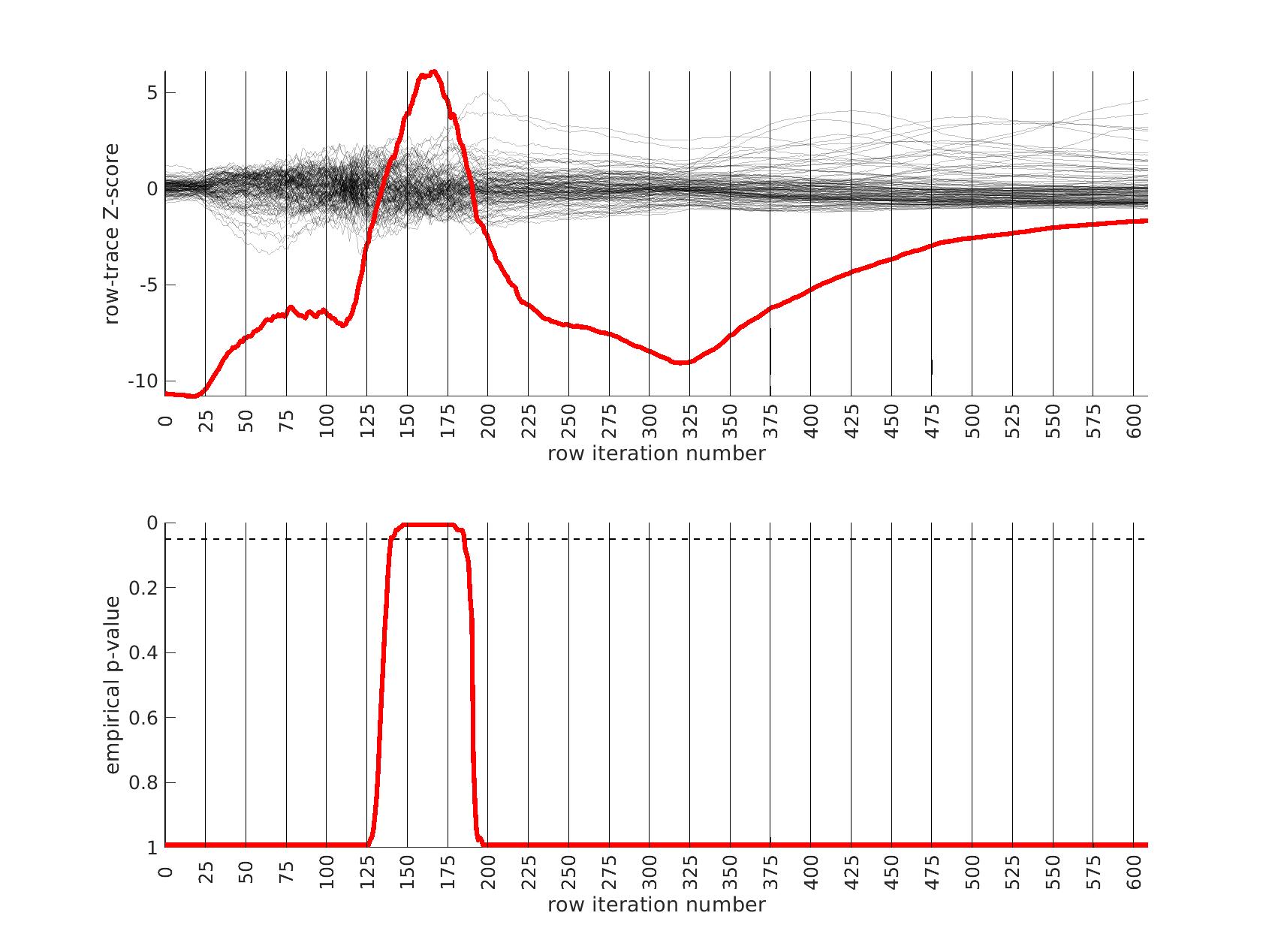}
  \caption{
    This figure is analogous to Fig \ref{fig:trace}, except that the red trace corresponds to a search for a secondary bicluster.
    Note that the red trace is only somewhat significant over a small range of iterations, including $i\in\left[150,175\right]$.
  }
  \label{fig:trace_secondary}
\end{figure}

\begin{figure}
  \centering
  \includegraphics[width=6.5in]{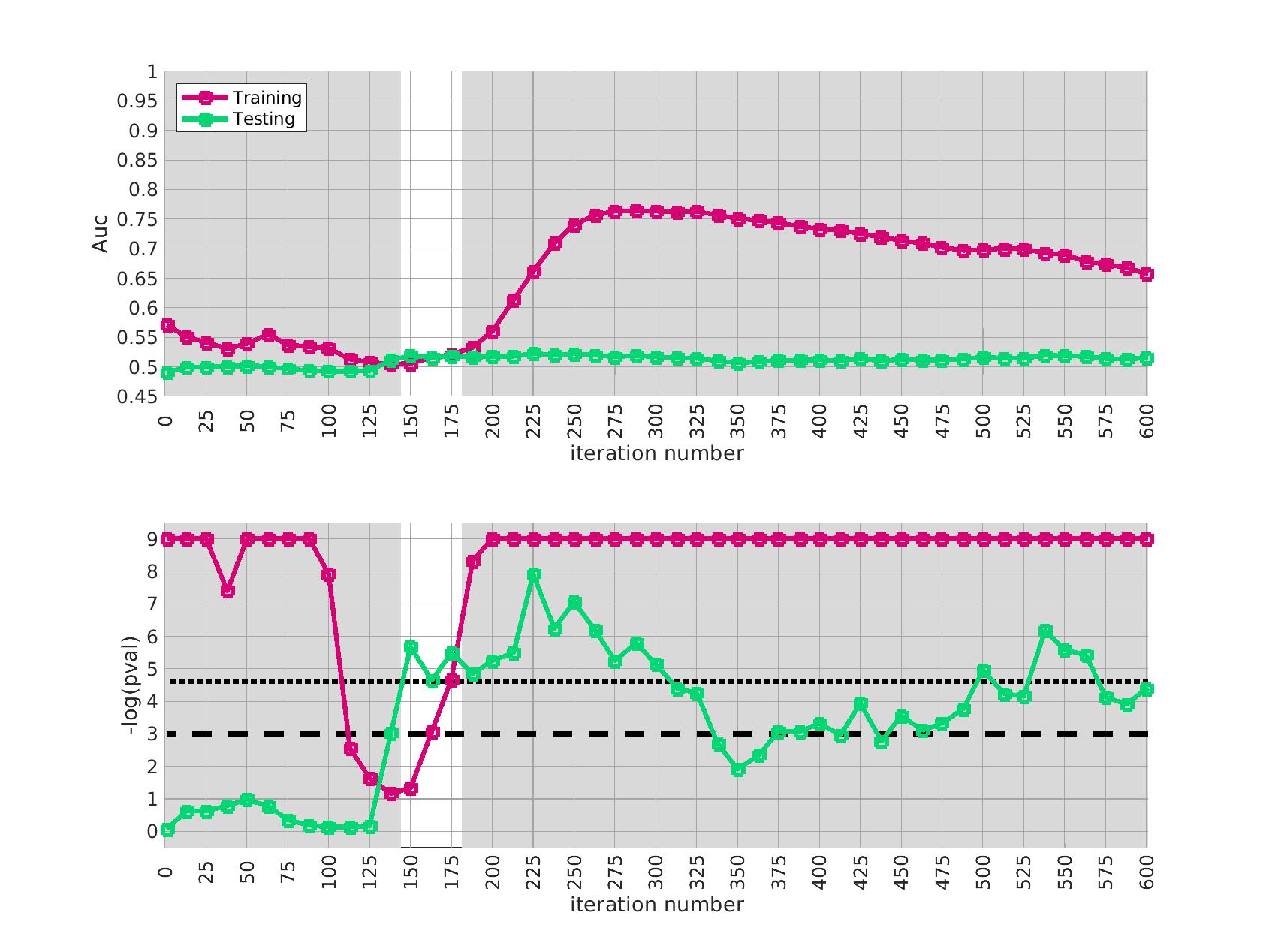}
  \caption{
    In this figure we illustrate the replication of the secondary bicluster in arm-2. The overall replication p-value is $p=0.86$ (i.e., not significant).
  }
  \label{fig:AUC_trn4_tst1_secondary}
\end{figure}

\begin{figure}
  \centering
  \includegraphics[width=6.5in]{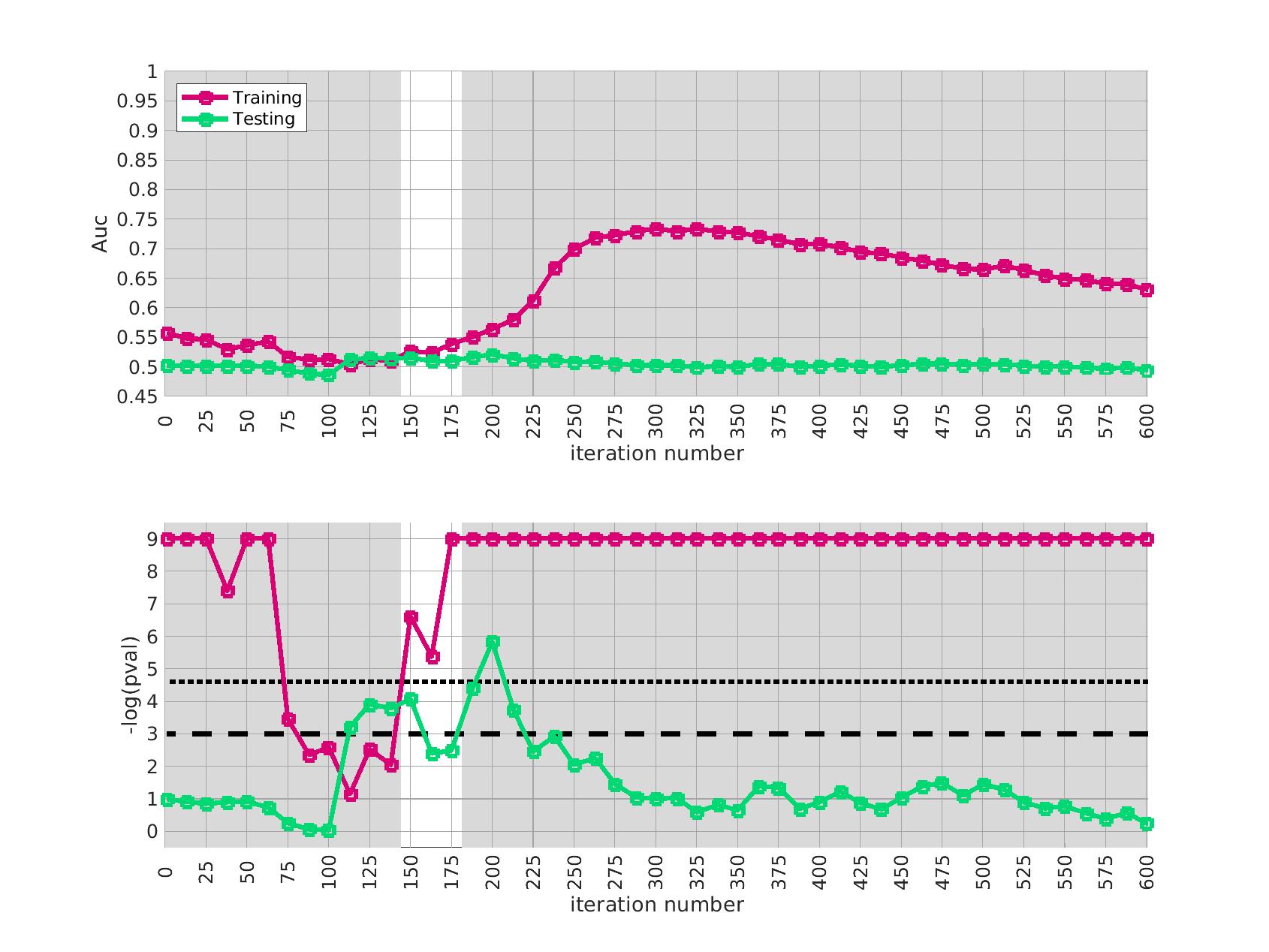}
  \caption{
    In this figure we illustrate the replication of the secondary bicluster in arm-3. The overall replication p-value is $p=0.41$ (i.e., not significant).
  }
  \label{fig:AUC_trn4_tst2_secondary}
\end{figure}

\begin{figure}
  \centering
  \includegraphics[width=6.5in]{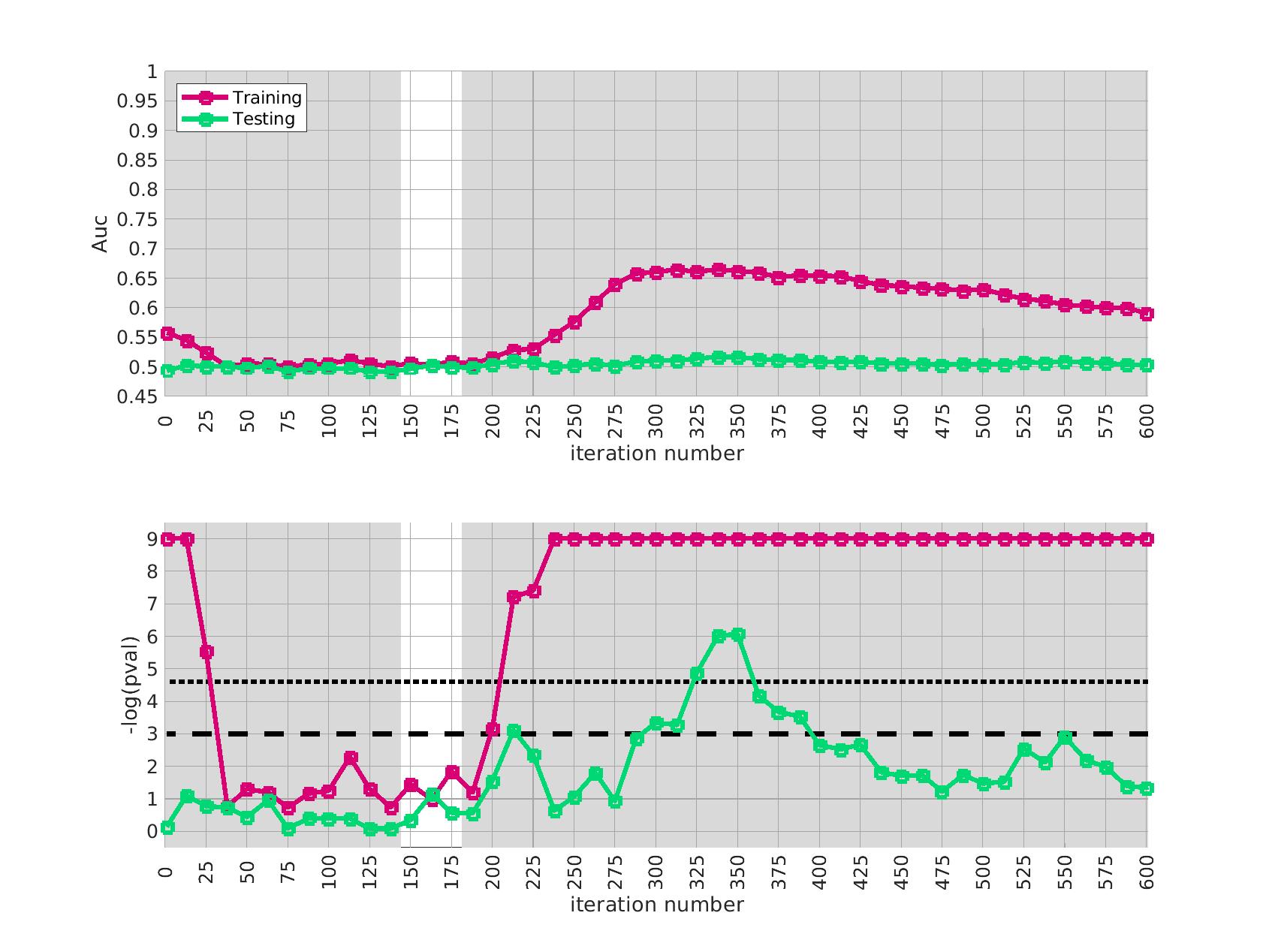}
  \caption{
      In this figure we illustrate the replication of the secondary bicluster in arm-4. The overall replication p-value is $p=0.85$ (i.e., not significant).
  }
  \label{fig:AUC_trn4_tst3_secondary}
\end{figure}

Figs \ref{fig:trace_secondary} \ref{fig:AUC_trn4_tst1_secondary} \ref{fig:AUC_trn4_tst2_secondary} and \ref{fig:AUC_trn4_tst3_secondary} illustrate the significance of the secondary bicluster found in arm-1. 

\subsection*{Control-specific biclusters}

\begin{figure}
  \centering
  \includegraphics[width=5.5in]{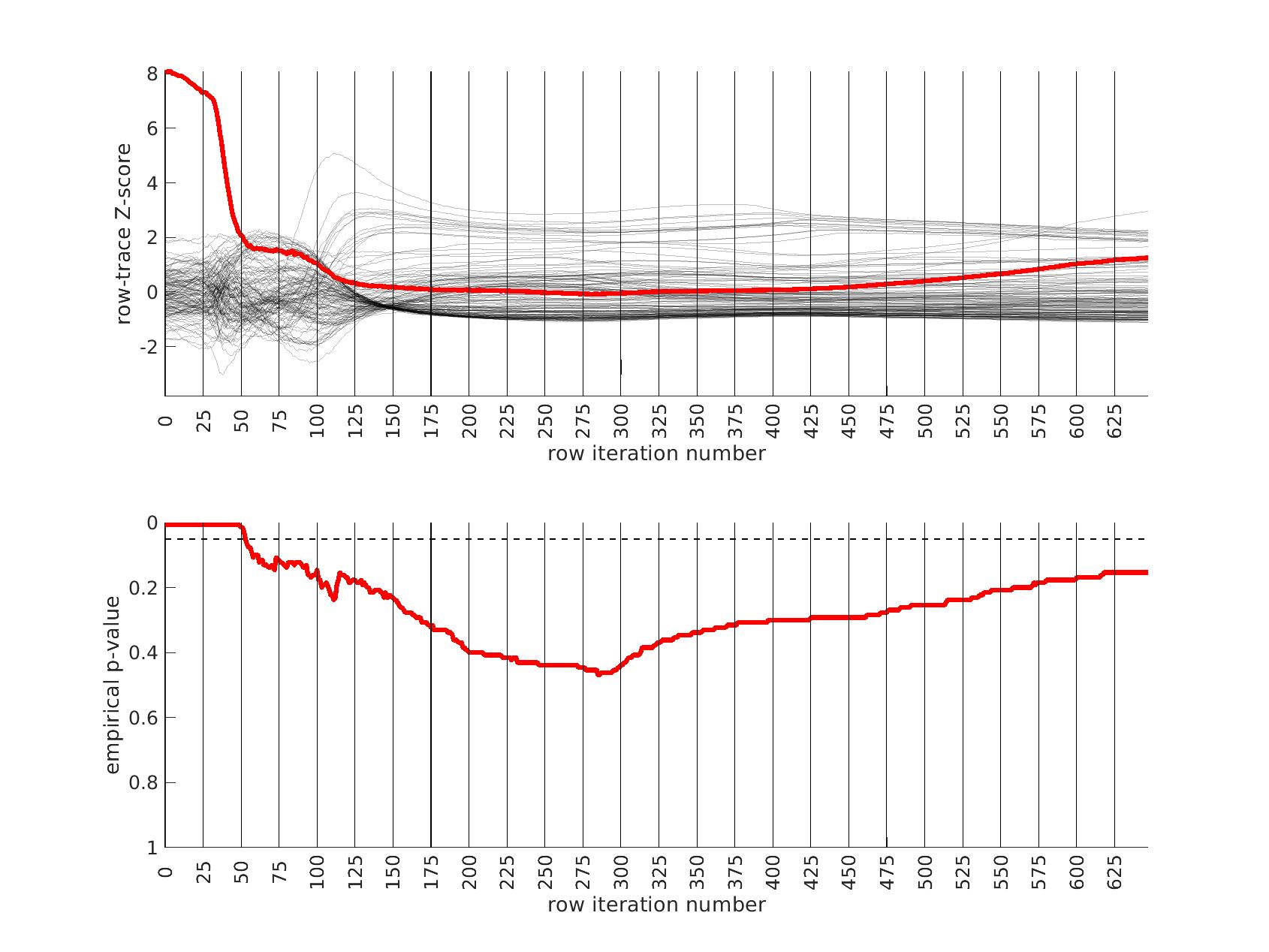}
  \caption{
    This figure is analogous to Fig \ref{fig:trace}, except that the red trace corresponds to a search for a bicluster within the control-population of arm-1, rather than the case-population.
    Note that the red trace decays monotonically, with no distinguished peaks as the algorithm proceeds.
  }
  \label{fig:trace_cl4_X}
\end{figure}

\begin{figure}
  \centering
  \includegraphics[width=5.5in]{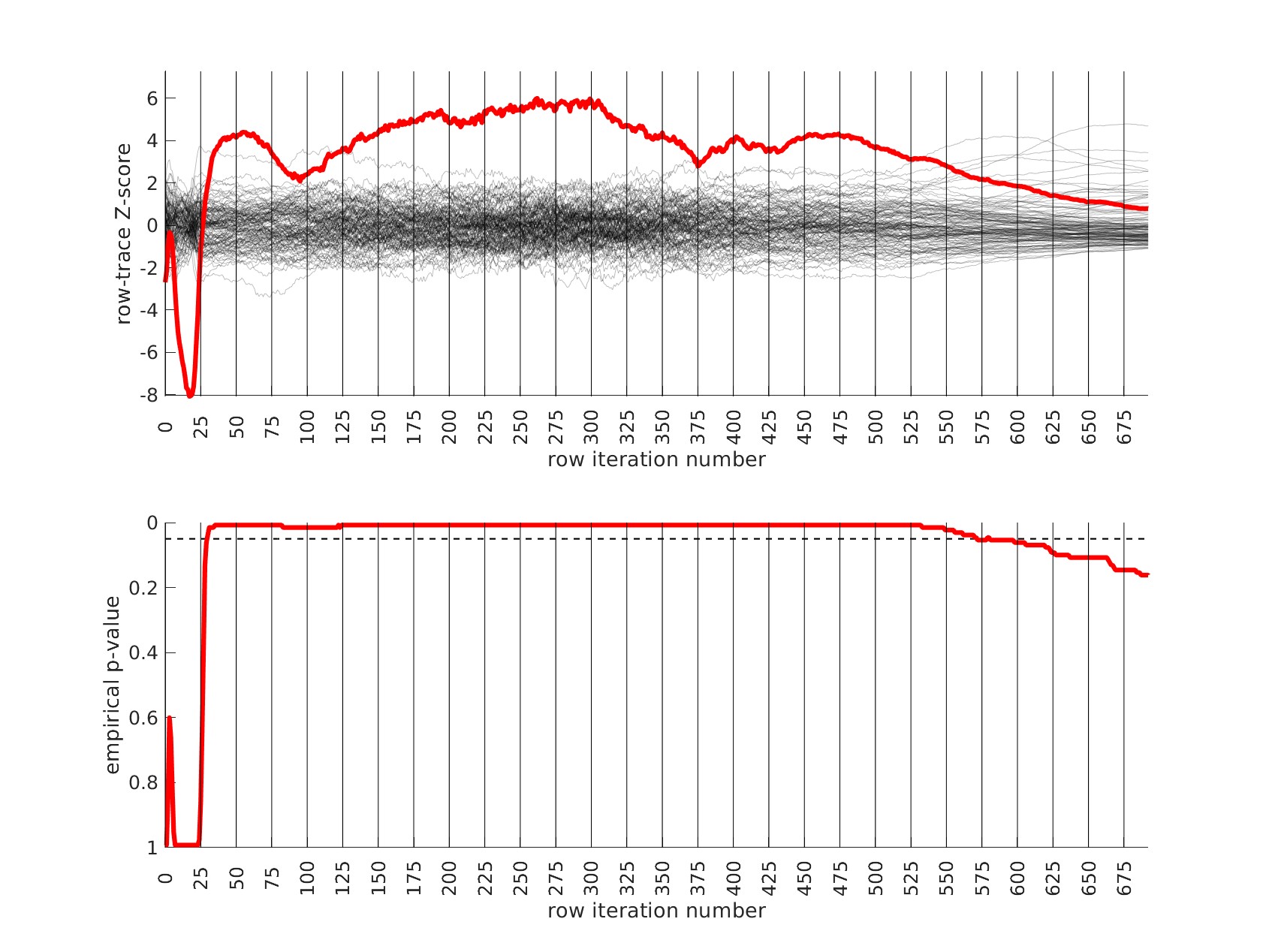}
  \caption{
    This figure is analogous to Fig \ref{fig:trace_cl4_X}, except that we consider arm-2 as a training-arm (rather than arm-1).
    Note that there are multiple distinguished peaks to the red trace, indicating (at least) one bicluster.
    The overall replication p-value of the dominant bicluster indicated by this trace is $p=0.030$ in arm-3, $p=0.17$ in arm-4 and $p=0.00015$ in arm-1. 
  }
  \label{fig:trace_cl1_X}
\end{figure}

\begin{figure}
  \centering
  \includegraphics[width=5.5in]{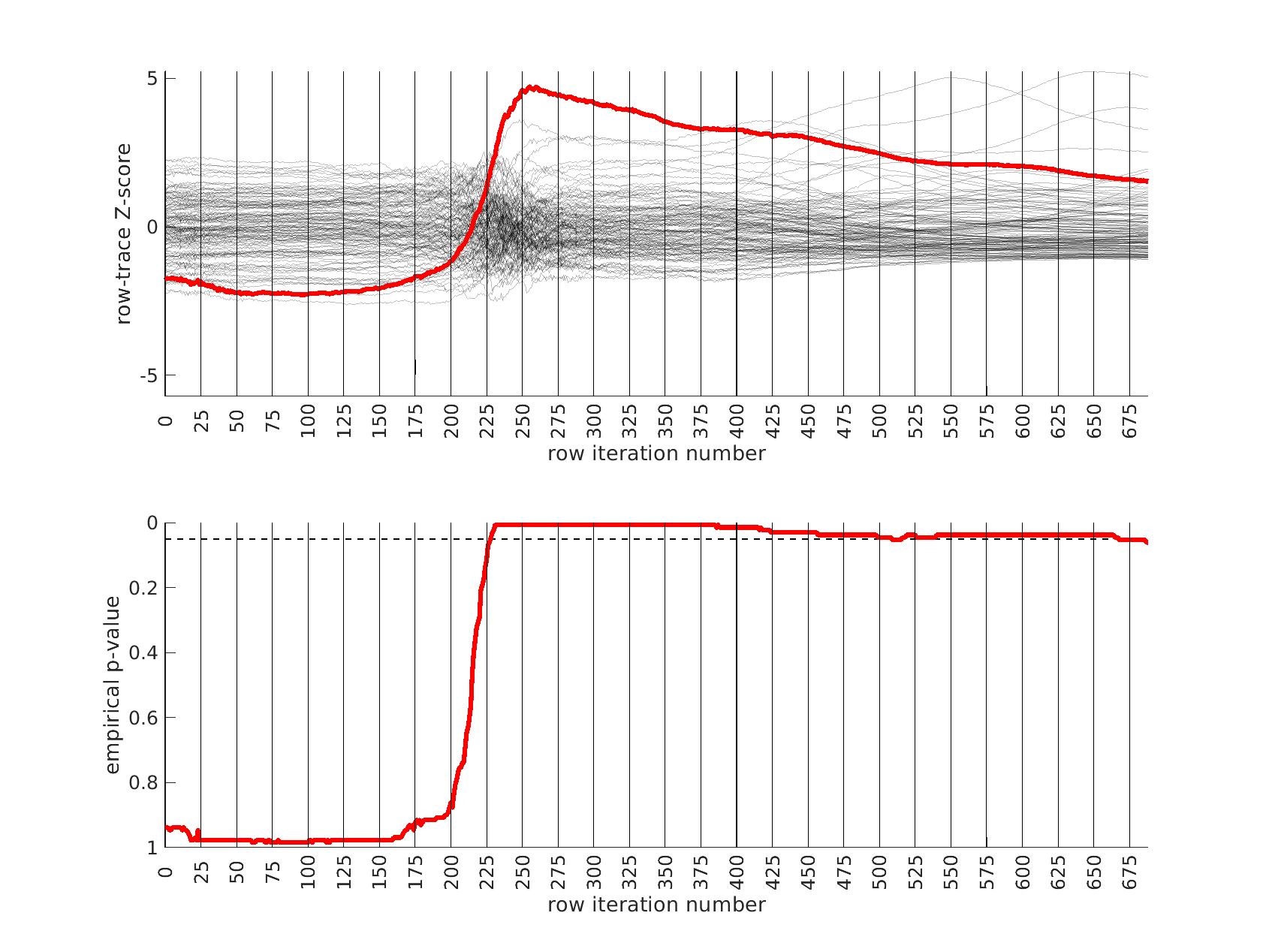}
  \caption{
    This figure is analogous to Fig \ref{fig:trace_cl4_X}, except that we consider arm-3 as a training-arm (rather than arm-1).
    Note that there is a distinguished peak to the red trace, indicating one bicluster.
    The overall replication p-value of this dominant bicluster is $p=0.00010$ in arm-2, $p=0.0025$ in arm-4, and $p=0.0063$ in arm-1.
  }
  \label{fig:trace_cl2_X}
\end{figure}

\begin{figure}
  \centering
  \includegraphics[width=5.5in]{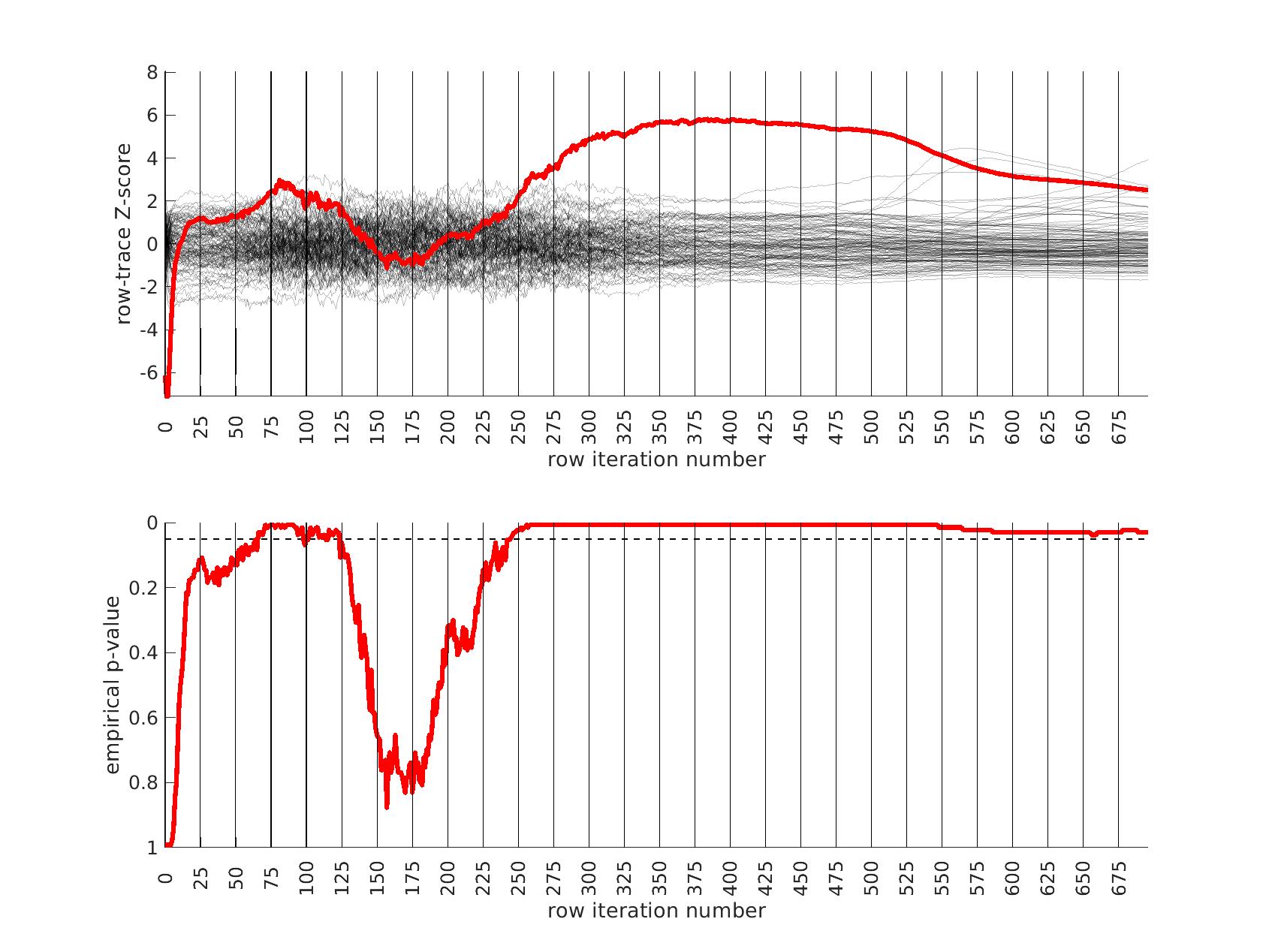}
  \caption{
    This figure is analogous to Fig \ref{fig:trace_cl4_X}, except that we consider arm-4 as a training-arm (rather than arm-1).
    Note that there are multiple distinguished peaks to the red trace, indicating (at least) one bicluster.
    The overall replication p-value of the dominant bicluster indicated by this trace is $p=0.000025$ in arm-2, $p=0.00013$ in arm-3, and $p=0.040$ in arm-1.
  }
  \label{fig:trace_cl3_X}
\end{figure}

Figs \ref{fig:trace_cl4_X}, \ref{fig:trace_cl1_X}, \ref{fig:trace_cl2_X} and \ref{fig:trace_cl3_X} illustrate the replication analyses for the control-biclusters found in arms 2, 3 and 4. 

\subsection*{Dependence of biclustering on minor-allele-frequency}

\begin{figure}
  \centering
  \includegraphics[trim=350 50 200 050,clip,width=6.5in]{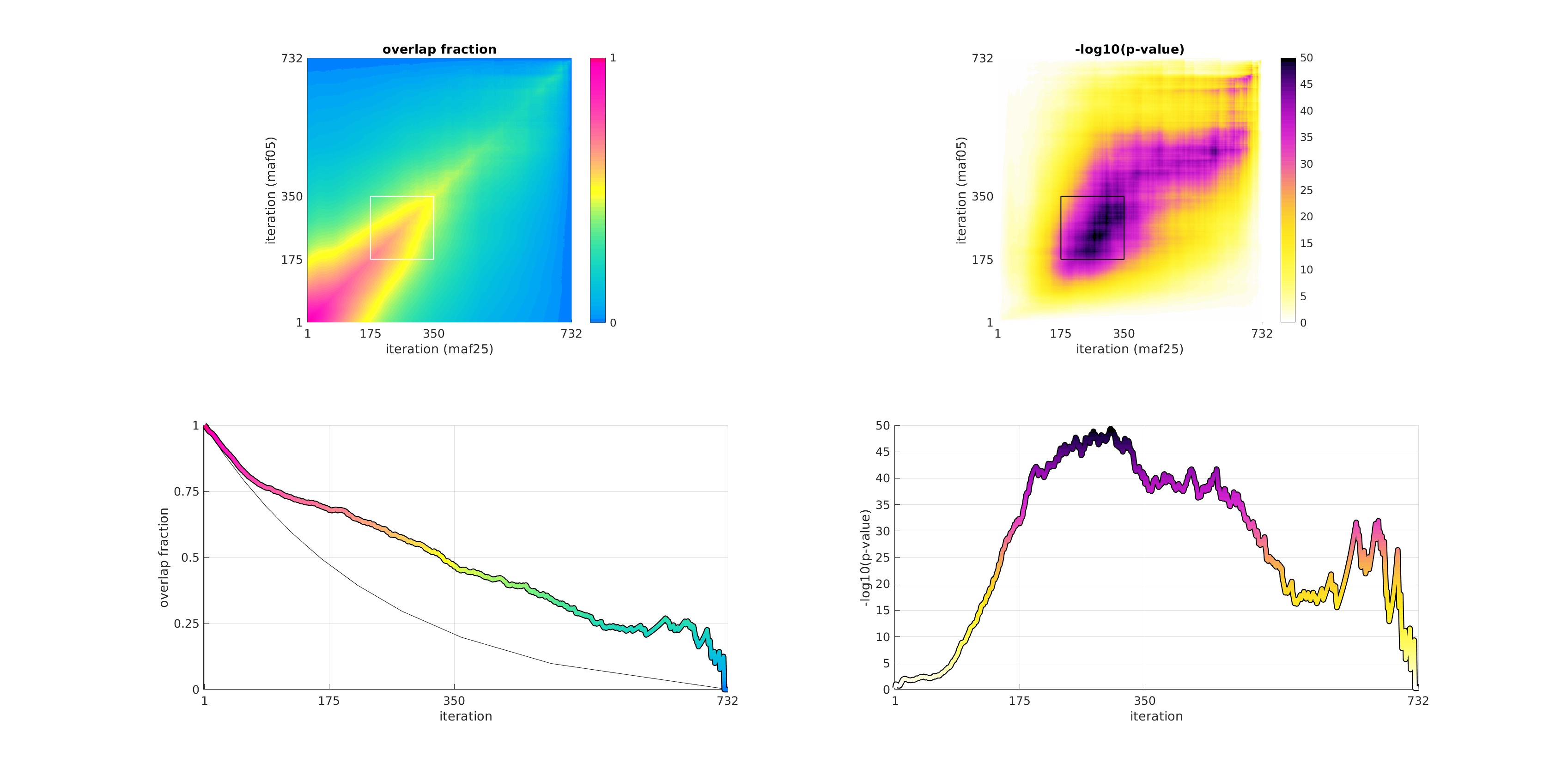}
  \caption{
    In the main text we used a minor-allele-frequency (i.e., maf) threshold of $0.25$ when searching for the bicluster described in Fig \ref{fig:trace}.
    If, instead, we use an maf-threshold of $0.05$, we also find a significant bicluster, albeit one that includes more allele-combinations and a slightly different selection of case-subjects.
    Illustrated in this figure is the overlap between these two biclusters.
    The first panel shows the overlap in case-subjects as a function of iteration. The iterations for maf-threshold $0.25$ (i.e., the bicluster described in the main text) are shown along the horizontal, while the vertical axis corresponds to iterations for maf-threshold $0.05$.
    In the lower-left subplot we examine comparable iterations (i.e., involving the same number of case-subjects).
    Here, we see that the overlap in case-subjects (colored using the colormap in the upper-left) is much higher than chance-level (black).
    The corresponding p-value for this case-subject overlap is shown in the upper-right.
    In the lower-right subplot we illustrate the p-value for comparable iterations.
    Note that the p-value peaks for iterations in the range $[175-350]$, which is the range where the original bicluster was most significant.
    We view this as corroboration that our discovery process (within arm-1) is robust.
    }
  \label{fig:maf25_vs_maf05_rdrop}
\end{figure}

\begin{figure}
  \centering
  \includegraphics[trim=350 50 200 050,clip,width=6.5in]{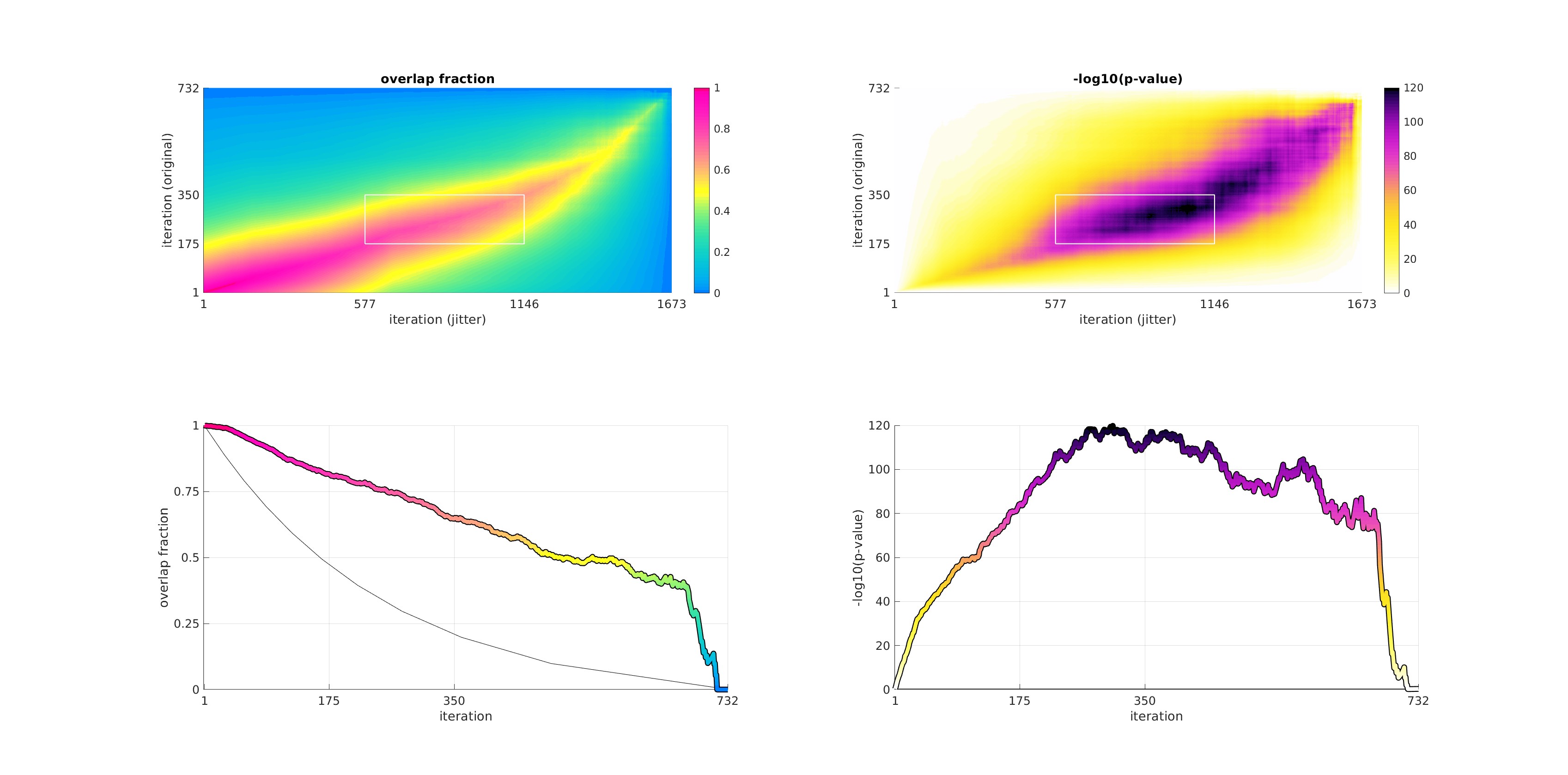}
  \caption{
    This figure is similar to Fig \ref{fig:maf25_vs_maf05_rdrop}, except that this time we compare the biclustering results from the original biclustering run with a biclustering run where we recalculate the scores after each individual subject and allele-combination is removed (adding a small amount of `jitter' to the original algorithm). In the upper-left we show the overlap in case-subjects between the jittered algorithm (horizontal) and original algorithm (vertical). The iterations $[175,350]$ for the original algorithm correspond to iterations $[577,1146]$ for the jittered algorithm (see highlighted box in upper subplots).
    The overlap-fraction and p-values for comparable iterations are shown in the bottom subplots.
    Again, the overlap in case-subjects is very significant, with p-values below $10^{-80}$ for the range of iterations considered in the main text.
    }
  \label{fig:g004_vs_g001_rdrop}
\end{figure}

Fig \ref{fig:maf25_vs_maf05_rdrop} illustrate the relationship between the trace computed using maf$\geq 05\%$, and the trace in the main-text, which was computed using maf$\geq 25\%$.
A similar comparision is shown in Fig \ref{fig:g004_vs_g001_rdrop} between a trace computed with a small amount of jitter (in this case eliminating only a single subject or allele-combination per iteration) and the trace in the main-text (which uses an elimination-fraction of $\gamma=0.5^{8}\sim0.004$).

\subsection*{Illustration of case- and control-subject distribution}

\begin{figure}
  \centering
  \includegraphics[trim=20 0 0 0,clip,width=5.5in]{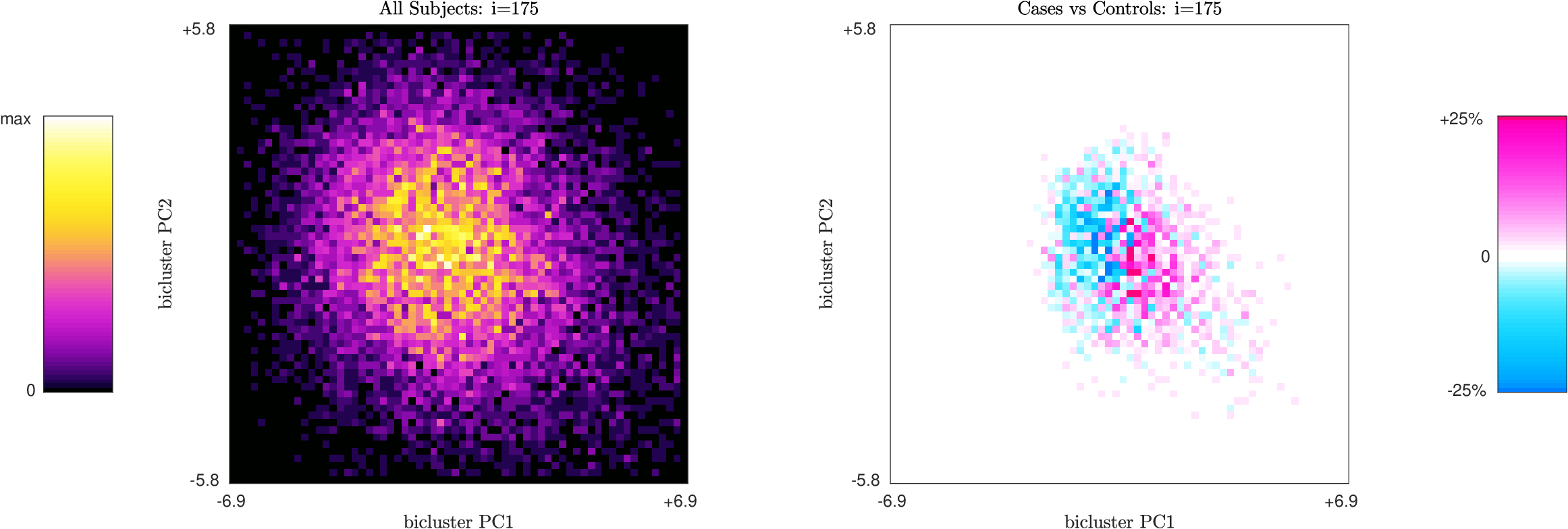}
  \caption{
    This figure illustrates a `heatmap' displaying the subjects in arm-2 projected onto the first- and second-principal-components of the bicluster discovered in arm-1 and defined using iteration $i=175$. The first principal-component (horizontal) is the bicluster-score we use in our replication-study. The second principal-component (vertical) is not used in our analysis.
    The left subplot shows a heatmap illustrating the distribution of all the subjects in arm-2 (cases and controls).
    The right subplot shows the difference between the density of cases and controls.
    The color pink corresponds to areas with a higher case-density than control-density, while blue corresponds to areas with a higher control-density than case-density. The colorbar (far right) ranges across $\pm25\%$ of the maximum density (taken across both the case- and control-distributions).
    }
  \label{fig:pca_heatmap_ver4_trn4_tst1_ni175}
\end{figure}

\begin{figure}
  \centering
  \includegraphics[trim=20 0 0 0,clip,width=5.5in]{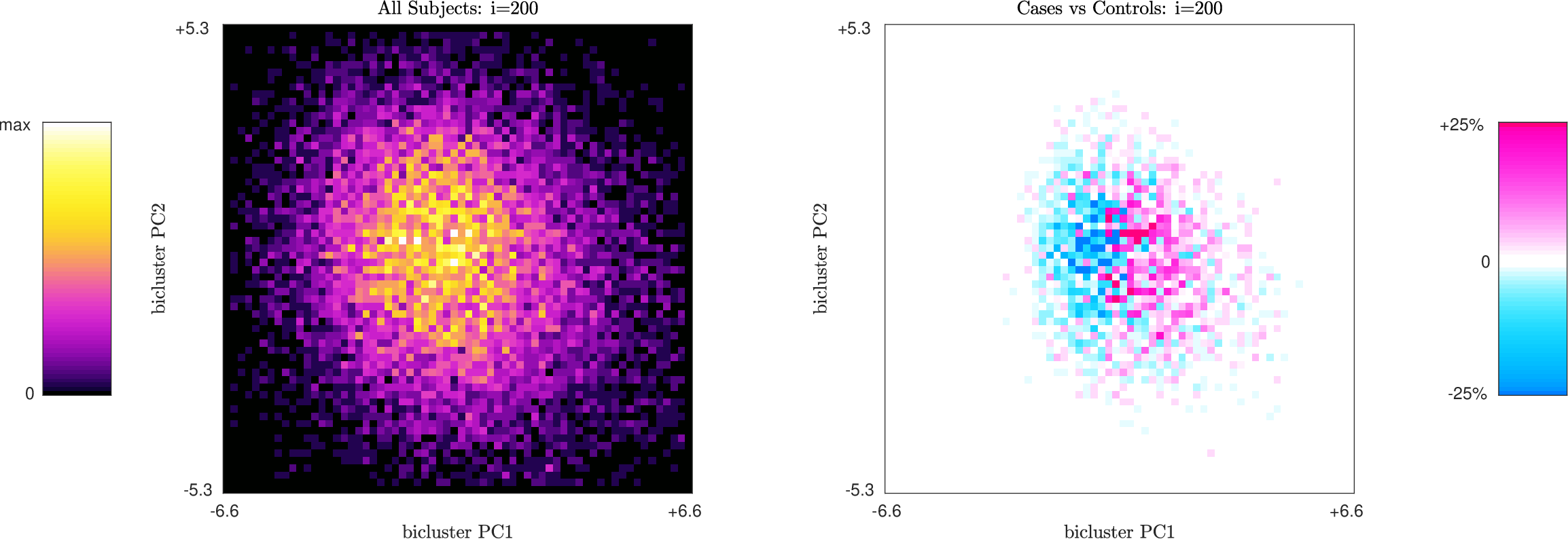}
  \caption{
    This figure is analogous to Fig \ref{fig:pca_heatmap_ver4_trn4_tst1_ni175}, except that we use iteration $i=200$.
    }
  \label{fig:pca_heatmap_ver4_trn4_tst1_ni200}
\end{figure}

\begin{figure}
  \centering
  \includegraphics[trim=20 0 0 0,clip,width=5.5in]{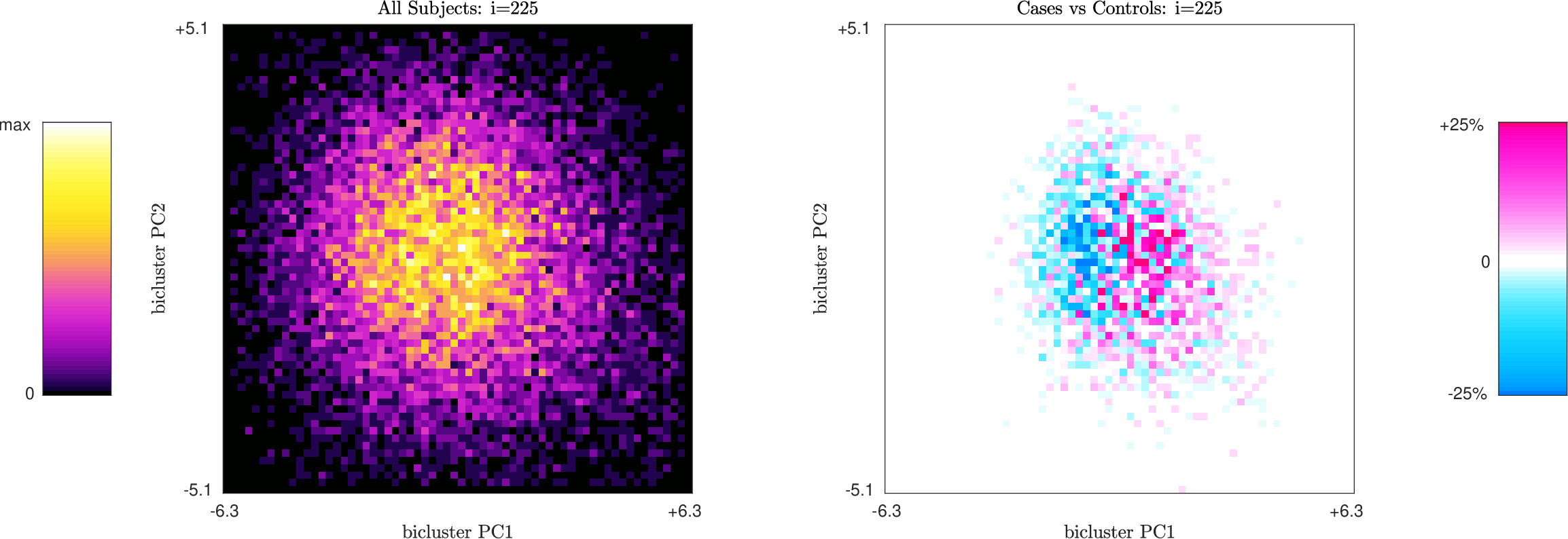}
  \caption{
    This figure is analogous to Fig \ref{fig:pca_heatmap_ver4_trn4_tst1_ni175}, except that we use iteration $i=225$.
    }
  \label{fig:pca_heatmap_ver4_trn4_tst1_ni225}
\end{figure}

\begin{figure}
  \centering
  \includegraphics[trim=20 0 0 0,clip,width=5.5in]{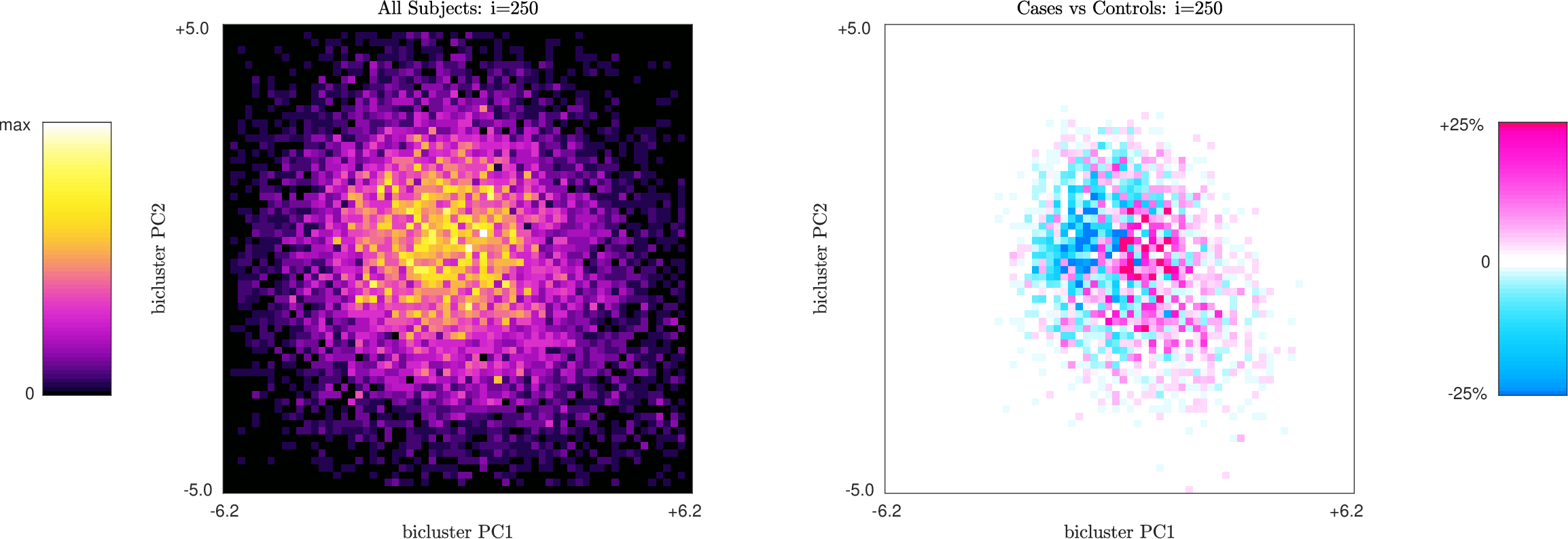}
  \caption{
    This figure is analogous to Fig \ref{fig:pca_heatmap_ver4_trn4_tst1_ni175}, except that we use iteration $i=250$.
    }
  \label{fig:pca_heatmap_ver4_trn4_tst1_ni250}
\end{figure}

\begin{figure}
  \centering
  \includegraphics[trim=20 0 0 0,clip,width=5.5in]{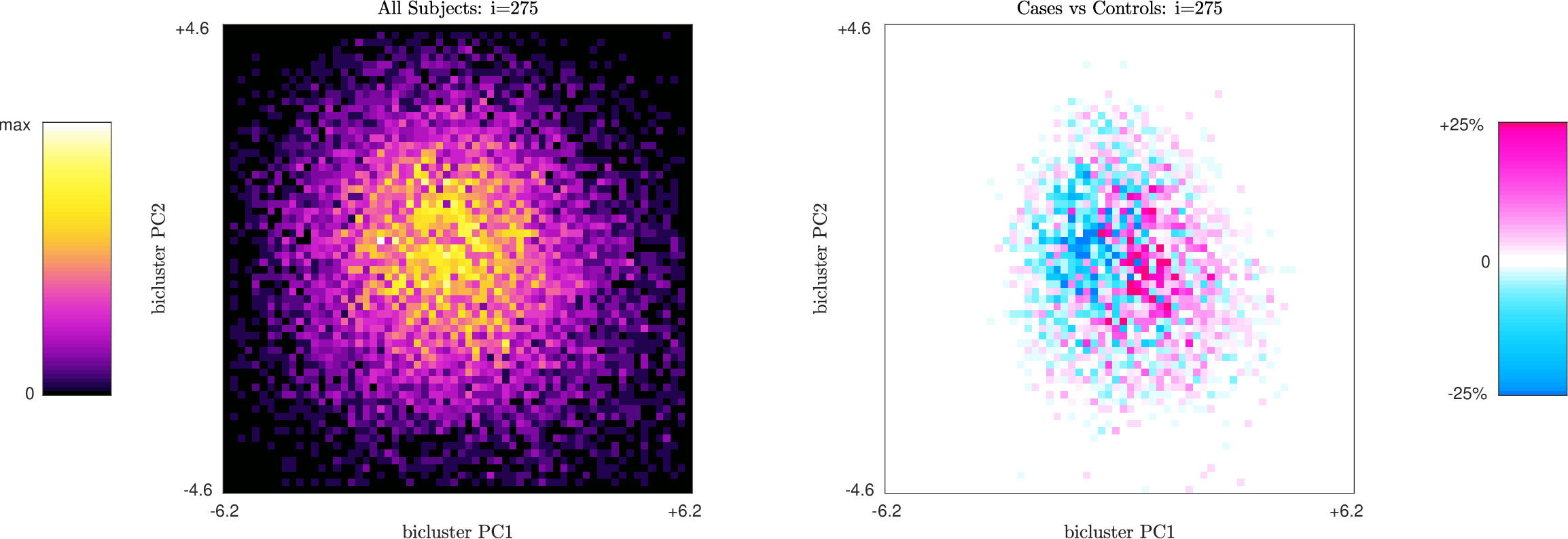}
  \caption{
    This figure is analogous to Fig \ref{fig:pca_heatmap_ver4_trn4_tst1_ni175}, except that we use iteration $i=275$.
    }
  \label{fig:pca_heatmap_ver4_trn4_tst1_ni275}
\end{figure}

\begin{figure}
  \centering
  \includegraphics[trim=20 0 0 0,clip,width=5.5in]{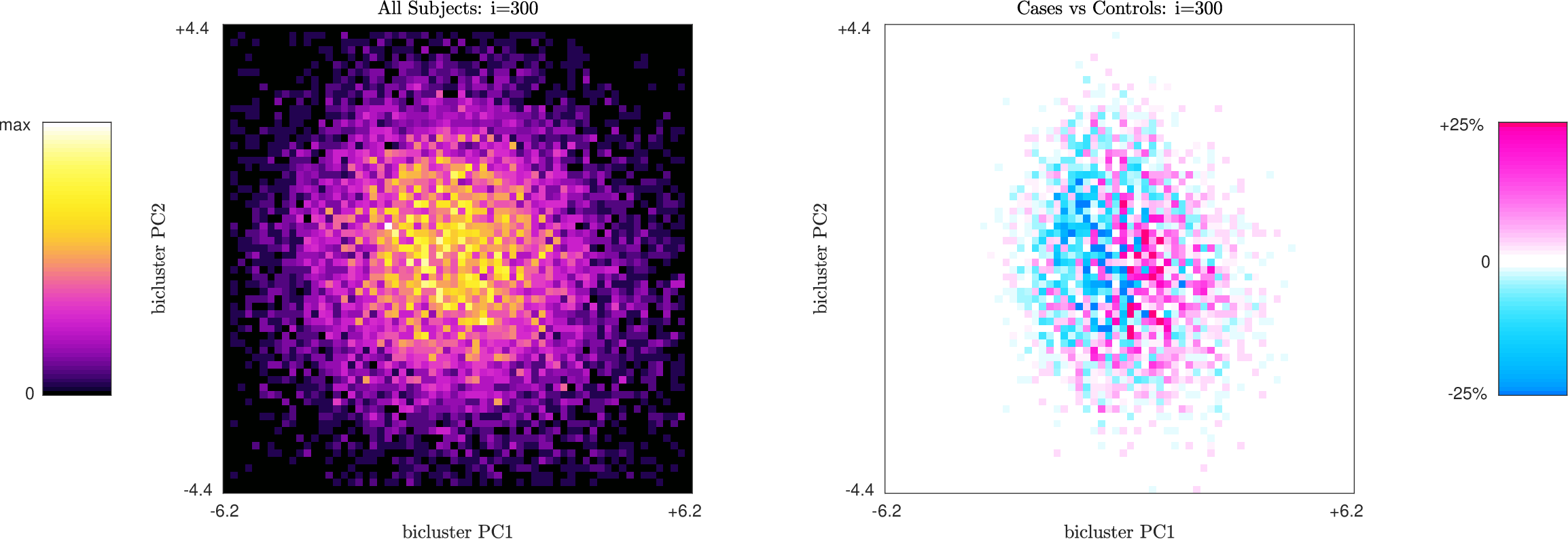}
  \caption{
    This figure is analogous to Fig \ref{fig:pca_heatmap_ver4_trn4_tst1_ni175}, except that we use iteration $i=300$.
    }
  \label{fig:pca_heatmap_ver4_trn4_tst1_ni300}
\end{figure}

\begin{figure}
  \centering
  \includegraphics[trim=20 0 0 0,clip,width=5.5in]{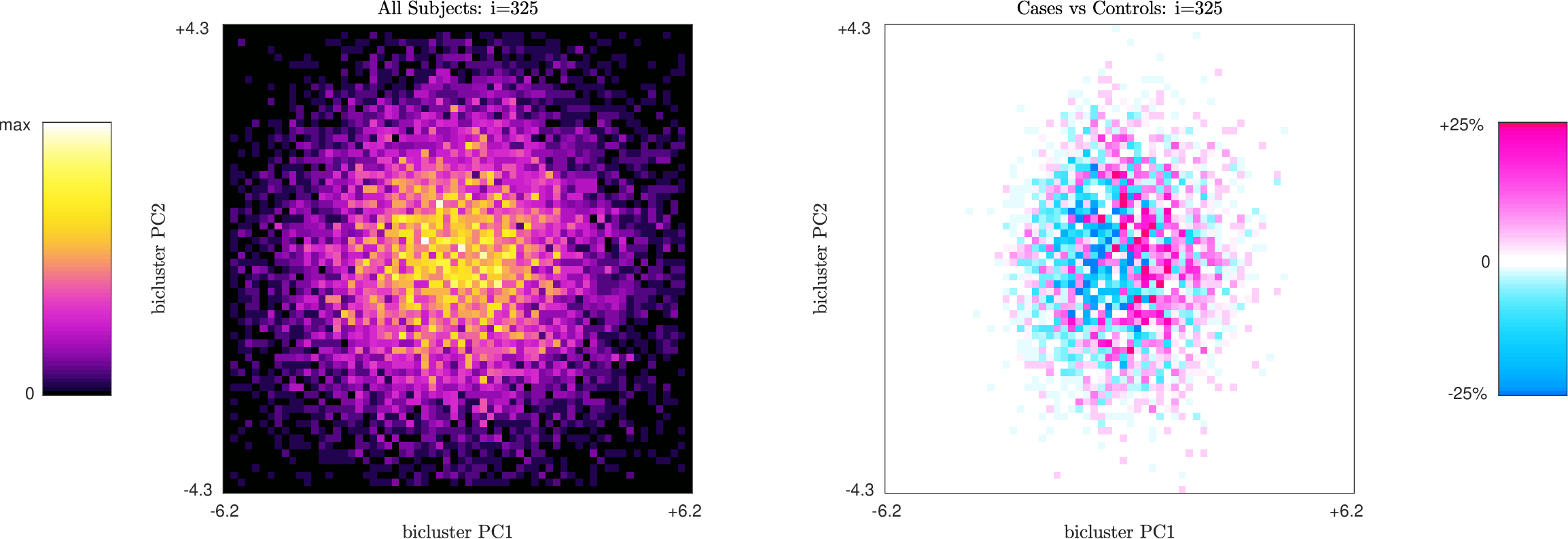}
  \caption{
    This figure is analogous to Fig \ref{fig:pca_heatmap_ver4_trn4_tst1_ni175}, except that we use iteration $i=325$.
    }
  \label{fig:pca_heatmap_ver4_trn4_tst1_ni325}
\end{figure}

\begin{figure}
  \centering
  \includegraphics[trim=20 0 0 0,clip,width=5.5in]{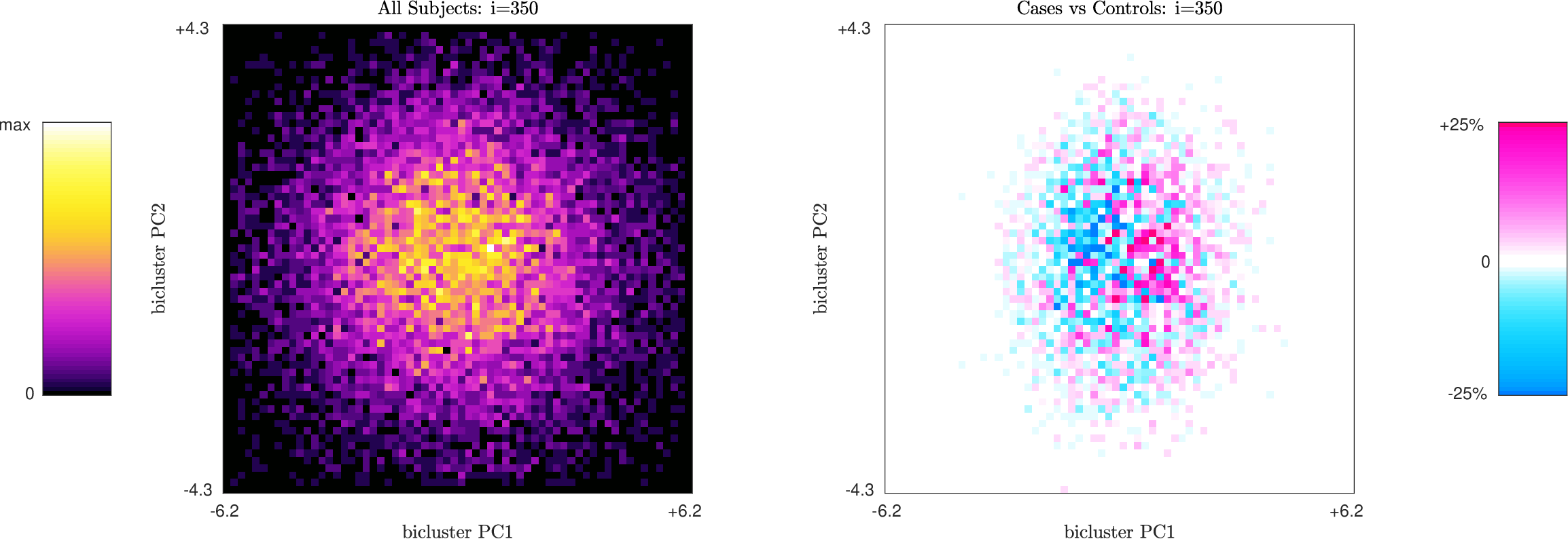}
  \caption{
    This figure is analogous to Fig \ref{fig:pca_heatmap_ver4_trn4_tst1_ni175}, except that we use iteration $i=350$.
    }
  \label{fig:pca_heatmap_ver4_trn4_tst1_ni350}
\end{figure}

Figs \ref{fig:pca_heatmap_ver4_trn4_tst1_ni175}-\ref{fig:pca_heatmap_ver4_trn4_tst1_ni350} show heatmaps of the distribution of cases- and control-subjects from arm-2, after projection onto the dominant two principal-components of the bicluster found in arm-1. The different figures correspond to different delineations of the bicluster, as determined by the iteration-index $i$.


\subsection*{Case-specific biclusters in arms 2, 3 and 4}

\begin{figure}
  \centering
  \includegraphics[width=5.5in]{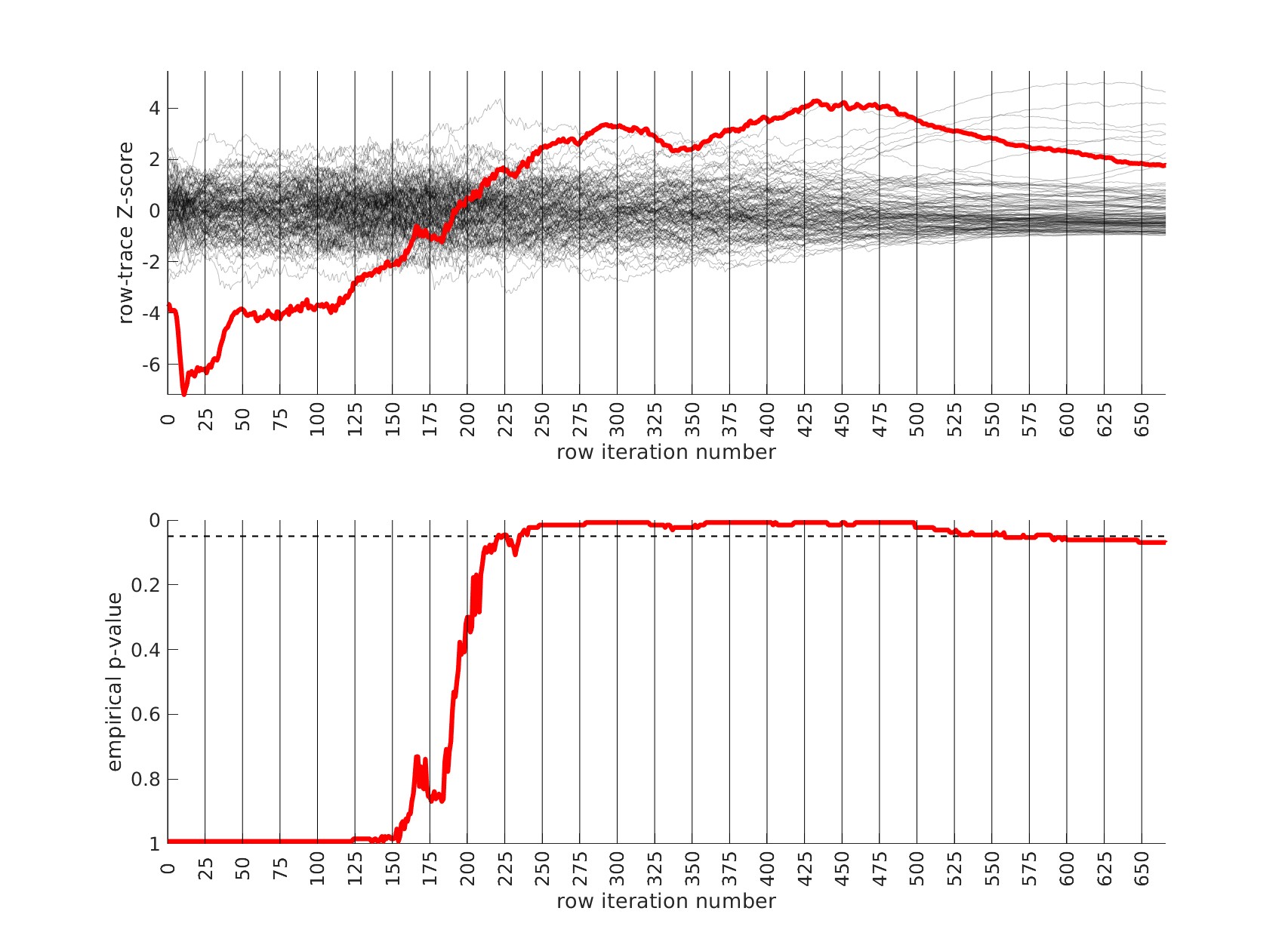}
  \caption{
    This figure is analogous to Fig \ref{fig:trace}, except that we show the traces resulting after we run our biclustering algorithm to search for case-specific biclusters using arm-2 as the discovery arm (again limited to SNPs with maf $\geq 0.25$).
    The overall p-value for the data (red-trace), estimated using the strategy in \cite{Rangan_2018}, is $p\sim 6.5/125 \sim 0.051$.
  }
  \label{fig:trace_cl1_maf01_dex_p25_D_m2r2_g004}
\end{figure}

\begin{figure}
  \centering
  \includegraphics[width=5.5in]{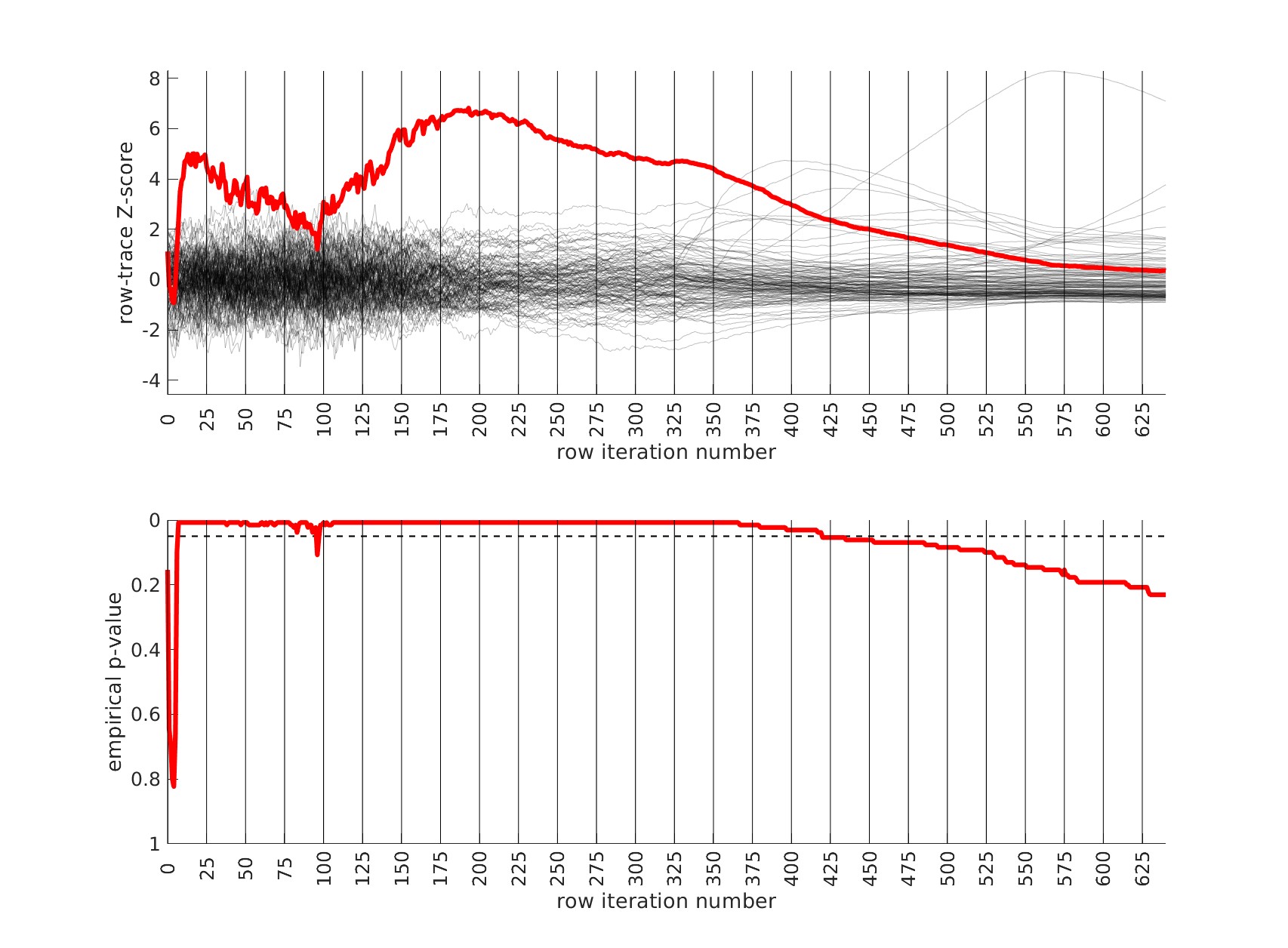}
  \caption{
    This figure is analogous to Figs \ref{fig:trace} and \ref{fig:trace_cl1_maf01_dex_p25_D_m2r2_g004}, except that we show the traces resulting after we run our biclustering algorithm using arm-3 as the discovery arm.
    The overall p-value for the data is $p\lesssim 1/64$.
  }
  \label{fig:trace_cl2_maf01_dex_p25_D_m2r2_g004}
\end{figure}

\begin{figure}
  \centering
  \includegraphics[width=5.5in]{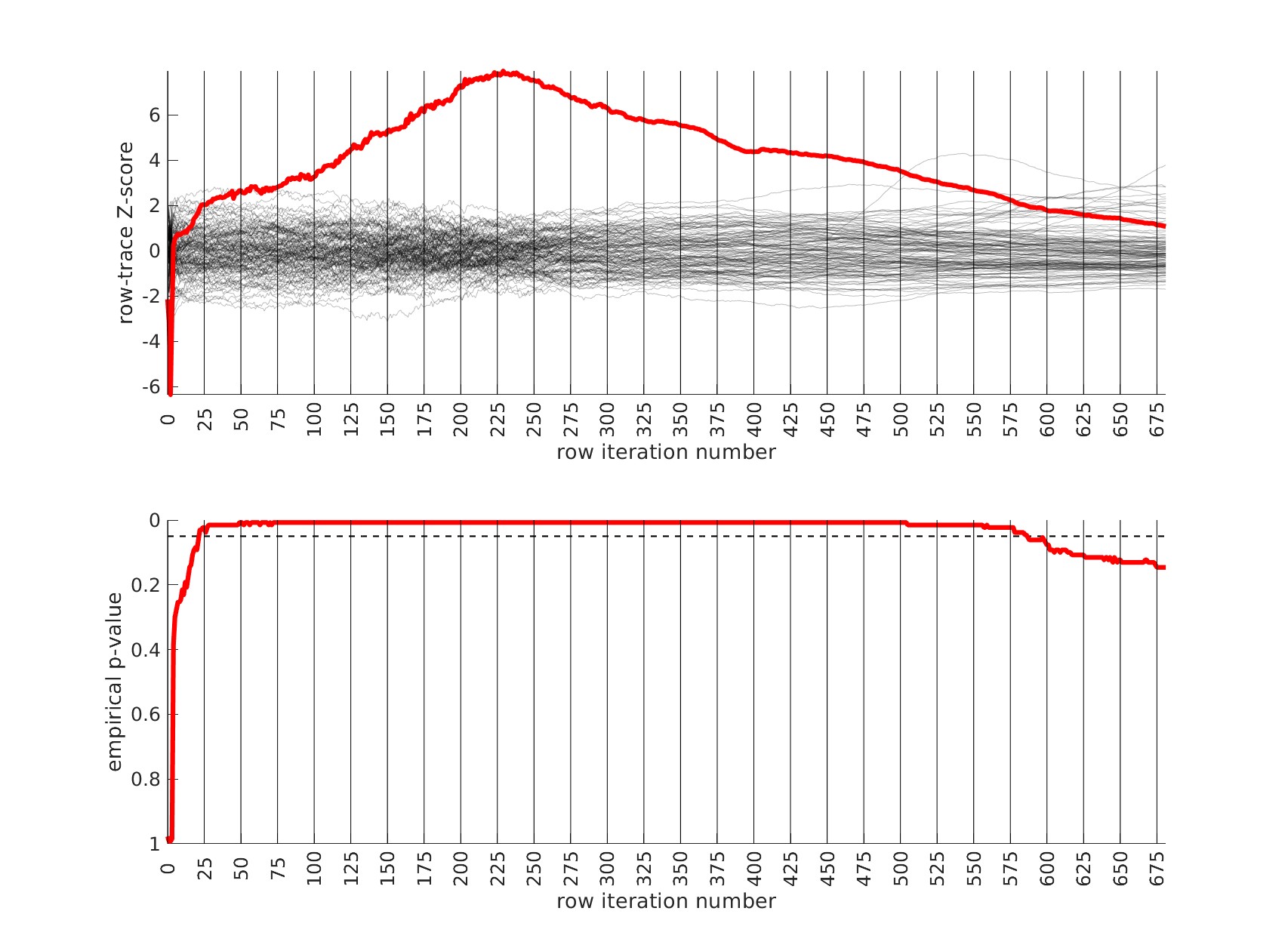}
  \caption{
    This figure is analogous to Figs \ref{fig:trace}, \ref{fig:trace_cl1_maf01_dex_p25_D_m2r2_g004} and \ref{fig:trace_cl2_maf01_dex_p25_D_m2r2_g004}, except that we use arm-3 as the discovery arm.
    The overall p-value for the data is $p\lesssim 1/128$.
  }
  \label{fig:trace_cl3_maf01_dex_p25_D_m2r2_g004}
\end{figure}

Figs \ref{fig:trace_cl1_maf01_dex_p25_D_m2r2_g004}-\ref{fig:trace_cl3_maf01_dex_p25_D_m2r2_g004} show the traces resulting after we use our biclustering algorithm to search for case-specific biclusters in arms 2-4.

\bibliographystyle{vancouver}
\bibliography{OConnell_Coombes_Review_2021_bibliography_20240928,extra_bib_20240928}

\end{document}